\documentclass[reprint,superscriptaddress,preprintnumbers,longbibliography,
amsmath,amssymb,aps,superscriptaddress,floatfix,tightenlines,pra
]{revtex4-2}
\usepackage{subfigure}
\usepackage{graphicx}
\usepackage{dcolumn}
\usepackage{bm}
\usepackage{color}
\usepackage[colorlinks=true,
citecolor=green,
linkcolor=red,
anchorcolor=black,
urlcolor=purple]{hyperref}
\usepackage{overpic}
\usepackage{epstopdf}
\usepackage{braket}
\usepackage{changes}
\definechangesauthor[name={Yun-Hao Shi},color=blue]{syh}
\definechangesauthor[name={Run-Qiu Yang},color=red]{yrq}

\begin{document}


\title{Quantum simulation of Hawking radiation and curved spacetime with a superconducting on-chip black hole}

\author{Yun-Hao Shi}
\thanks{These authors contributed equally to this work.}
\affiliation{Institute of Physics, Chinese Academy of Sciences, Beijing 100190, China}
\affiliation{School of Physical Sciences, University of Chinese Academy of Sciences, Beijing 100049, China}

\author{Run-Qiu Yang}
\thanks{These authors contributed equally to this work.}
\affiliation{Center for Joint Quantum Studies and Department of Physics, School of Science, Tianjin University, Tianjin 300350, China}%

\author{Zhongcheng Xiang}
\thanks{These authors contributed equally to this work.}
\affiliation{Institute of Physics, Chinese Academy of Sciences, Beijing 100190, China}

\author{Zi-Yong Ge}
\affiliation{Theoretical Quantum Physics Laboratory, RIKEN Cluster for Pioneering Research, Wako-shi, Saitama 351-0198, Japan}

\author{Hao Li}
\affiliation{Institute of Physics, Chinese Academy of Sciences, Beijing 100190, China}
\affiliation{School of Physics, Northwest University, Xi'an 710127, China}

\author{Yong-Yi Wang}
\affiliation{Institute of Physics, Chinese Academy of Sciences, Beijing 100190, China}
\affiliation{School of Physical Sciences, University of Chinese Academy of Sciences, Beijing 100049, China}

\author{Kaixuan Huang}
\affiliation{Beijing Academy of Quantum Information Sciences, Beijing 100193, China}%

\author{Ye Tian}
\affiliation{Institute of Physics, Chinese Academy of Sciences, Beijing 100190, China}

\author{Xiaohui Song}
\affiliation{Institute of Physics, Chinese Academy of Sciences, Beijing 100190, China}

\author{Dongning Zheng}
\email{dzheng@iphy.ac.cn}
\affiliation{Institute of Physics, Chinese Academy of Sciences, Beijing 100190, China}
\affiliation{School of Physical Sciences, University of Chinese Academy of Sciences, Beijing 100049, China}
\affiliation{Songshan Lake Materials Laboratory, Dongguan, Guangdong 523808, China}

\author{Kai Xu}
\email{kaixu@iphy.ac.cn}
\affiliation{Institute of Physics, Chinese Academy of Sciences, Beijing 100190, China}
\affiliation{School of Physical Sciences, University of Chinese Academy of Sciences, Beijing 100049, China}
\affiliation{Beijing Academy of Quantum Information Sciences, Beijing 100193, China}
\affiliation{Songshan Lake Materials Laboratory, Dongguan, Guangdong 523808, China}
\affiliation{CAS Center for Excellence in Topological Quantum Computation, University of Chinese Academy of Sciences, Beijing 100049, China}%

\author{Rong-Gen Cai}
\email{cairg@itp.ac.cn}
\affiliation{CAS Key Laboratory of Theoretical Physics, Institute of Theoretical Physics, Chinese Academy of Sciences, Beijing 100190, China}

\author{Heng Fan}
\email{hfan@iphy.ac.cn}
\affiliation{Institute of Physics, Chinese Academy of Sciences, Beijing 100190, China}
\affiliation{School of Physical Sciences, University of Chinese Academy of Sciences, Beijing 100049, China}
\affiliation{Beijing Academy of Quantum Information Sciences, Beijing 100193, China}
\affiliation{Songshan Lake Materials Laboratory, Dongguan, Guangdong 523808, China}
\affiliation{CAS Center for Excellence in Topological Quantum Computation, University of Chinese Academy of Sciences, Beijing 100049, China}%
\affiliation{Hefei National Laboratory, Hefei 230088, China}


\begin{abstract}

Hawking radiation is one of the quantum features of a black hole that can be understood as a quantum tunneling across the event horizon of the black hole, but it is quite difficult to directly observe the Hawking radiation of an astrophysical black hole. Here, we report a fermionic lattice-model-type realization of an analogue black hole by using a chain of 10 superconducting transmon qubits with interactions mediated by 9 transmon-type tunable couplers. The quantum walks of quasi-particle in the curved spacetime reflect the gravitational effect near the black hole, resulting in the behaviour of stimulated Hawking radiation, which is verified by the state tomography measurement of all 7 qubits outside the horizon. In addition, the dynamics of entanglement in the curved spacetime is directly measured. Our results would stimulate more interests to explore the related features of black holes using the programmable superconducting processor with tunable couplers.

\end{abstract}

\maketitle
\noindent
\textbf{\large{Introduction}}

In the classical picture, a particle falls into a black hole horizon and the horizon prevents the particle from turning back, then escape becomes impossible. However, taking into account quantum effects, the particle inside the black hole is doomed to gradually escape to the outside, leading to the Hawking radiation~\cite{Nature_Hawking1974}. The problem is that direct observation of such a quantum effect of a real black hole is difficult in astrophysics. For a black hole with solar mass, the associated Hawking temperature is only $\sim 10^{-8}$ K and the corresponding radiation probability is astronomically small. Given by this, various analogue systems were proposed to simulate a black hole and its physical effects in laboratories~\cite{PRL_Unruh1981}.
Over the past years, the theory of Hawking radiation has been tested in experiments based on various platforms engineered with analogue black holes,
such as using shallow water waves ~\cite{PRL_Unruh1981,PRD_Unruh1995,PRL_Weinfurtner2011,PRL_Weinfurtner2011,PRD_Michel2014,PRD_Euve2015,PRD_Coutant2016}, Bose-Einstein condensates (BEC)~\cite{NP_Steinhauer2016,PRL_Steinhauer2010,Nature_Steinhauer2019,PRL_Isoard2020,NP_Kolobov2021}, optical metamaterials and light~\cite{Science_Philbin2008,PRL_Drori2019,NPhoton_Sheng2013}, etc.

\begin{figure*}[htbp]
	\includegraphics[width=5.5in]{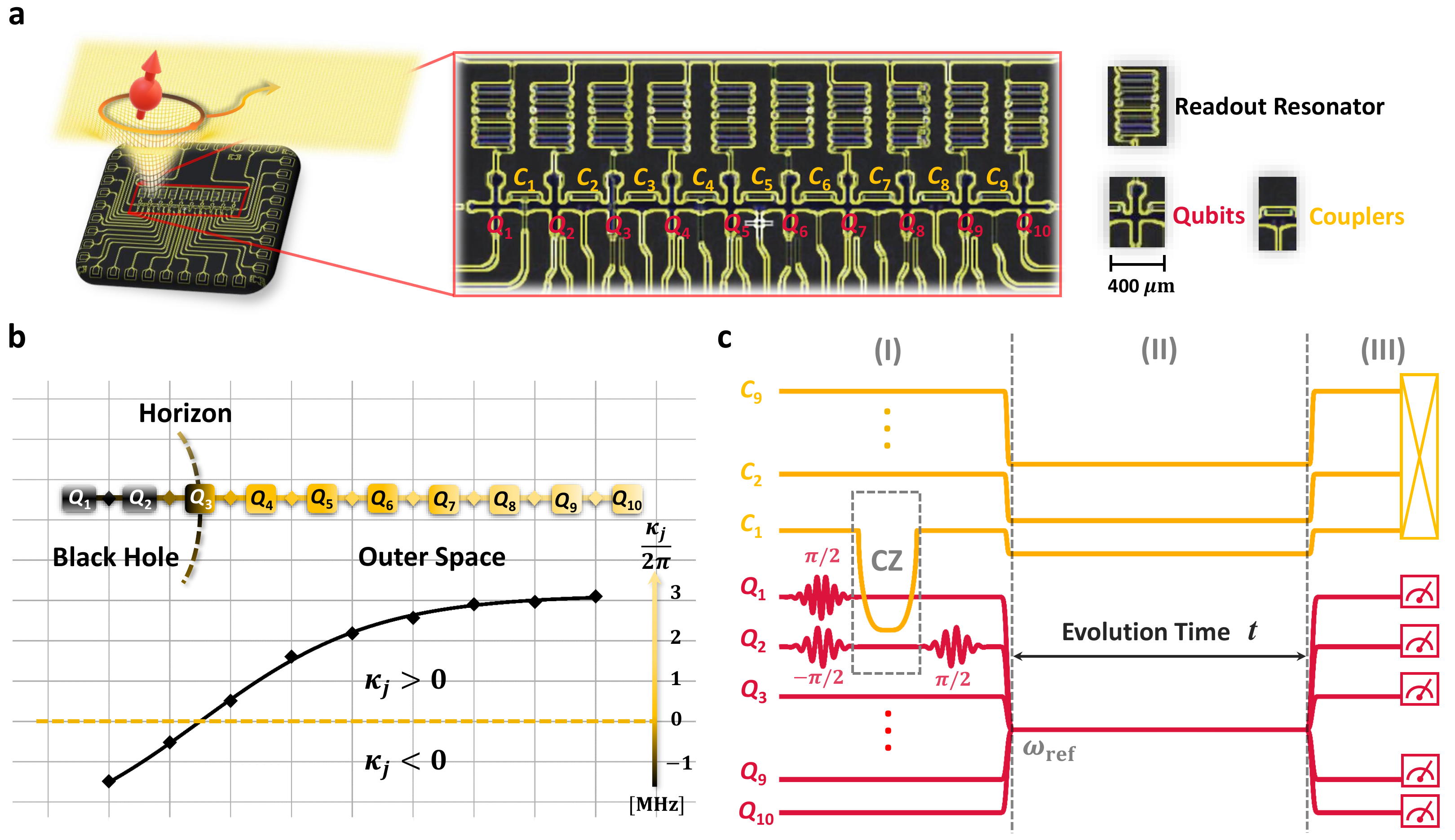}
	\caption{\textbf{On-chip analogue black hole.} \textbf{a}, False-color image of superconducting processor and schematic analogue black hole. Ten transmon qubits, $Q_1\sim Q_{10}$, shown as crosses, are integrated along a chain with nearest-neighbour couplings. Each nearest-neighbour two qubits are coupled via a coupler, $C_1\sim C_{9}$, realized by a transmon with only a flux bias line. All the transmons are frequency-tunable, but only the qubit has the XY control line and readout resonator. The schematic image represents the background of curved spacetime simulated by this superconducting chip. The red cartoon spin located at the upper-left denotes the evolution of one quasi-particle that is initially in the black hole and the outward-going radiation. \textbf{b}, Schematic representation of the site-dependent effective coupling strengths $\kappa_{j}$. In the experiment, the coupling $\kappa_{j}$ is designed according to Eq.~\eqref{Kappa}. There is a boundary analogous to the event horizon of a black hole, where the coupling changes its sign at site $Q_3$. Thus qubits $Q_1$ and $Q_2$ can be considered as the interior of the black hole, $Q_3$ is at the horizon, and $Q_4$ -- $Q_{10}$ are in the outside black hole. \textbf{c}, Experimental pulse sequence for observing dynamics of entanglement, which consists of three parts, i.e., (I) initialization, (II) evolution, and (III) measurement. For the initialization (I), we prepare an entangled Bell pair on $Q_1Q_2$ by combining several single-qubit pulses and a two-qubit control-phase (CZ) gate. At the left boundary of region (II), the curved (or flat) spacetime forms. Then the system will evolve according to the corresponding $\kappa_j$ in the Hamiltonian for a time $t$. In region (III), we perform the state tomography measurement. \label{fig:bh}}
\end{figure*}

On the other hand, the developments of superconducting processors enable us to simulate various intriguing problems of many-body systems, molecules, and to achieve quantum computational supremacy~\cite{Nature_Arute2019,RMP_Nori2014,PR_Nori2017,PRL_Wu2021}.
However, constructing an analogue black hole on a superconducting chip is still a challenge, which requires wide-range tunable and site-dependent couplings between qubits
to realize the curved spacetime~\cite{PRR_Yang2020}. Coincidentally, a recent architectural breakthrough of tunable couplers for superconducting circuit~\cite{PRApplied_Yan2018},
which has been exploited to implement fast and high-fidelity two-qubit gates~\cite{PRX_Sung2021,PRL_Xu2020,CPB_Xu2021,PRL_Collodo2020},
offers an opportunity to achieve specific coupling distribution analogous to the curved spacetime.
We develop such a superconducting processor integrated with a one-dimensional (1D) array of 10 qubits with interaction
couplings controlled by 9 tunable couplers, see Fig.~\ref{fig:bh}, which can realize both flat and curved spacetime backgrounds.
The quantum walks of quasi-particle excitations of superconducting qubits are performed to simulate the dynamics of particles in a black hole background, including dynamics of an entangled pair inside the horizon.
By using multi-qubit state tomography, Hawking radiation is measured which is in agreement with theoretical prediction.
This new constructed analogue black hole then facilitates further investigations of other related problems of the black hole.
\\
\\
\noindent
\textbf{\large{Results}}
\\
\textbf{Model and setup}
\\
To consider the effects of curved spacetime on quantum matters, we consider a (1+1)-D Dirac field, of which the Dirac equation is written as ($\hbar=c=1$)~\cite{PRD_Mann1991,PRL_Pedernales2018}
\begin{equation}
\mathrm{i}\gamma^{a}e^{\mu}_{(a)}\partial_{\mu}\psi+\frac{\mathrm{i}}{2}\gamma^{a}\frac{1}{\sqrt{-g}}\partial_{\mu}\left(\sqrt{-g}e^{\mu}_{(a)}\right)\psi-m\psi=0,
\end{equation}
where $g$ is the determinate of $g_{\mu\nu}$, the vielbein $e^{(a)}_{\mu}$ satisfies the orthonormal condition $e^{(a)}_{\mu}e^{\nu}_{(a)}=\delta^{\nu}_{\mu}$ and the $\gamma$-matrices in the two-dimensional case are chosen to be $\gamma=(\sigma_z,~\mathrm{i}\sigma_y)$. In the Eddington-Finkelstein coordinates $\{t, x\}$ and in the massless limit $m\rightarrow0$, such a Dirac field can be quantized into a discrete XY lattice model with site-dependent hopping couplings.
The effective Hamiltonian reads (see Supplementary Information and ref.~\cite{PRR_Yang2020})
\begin{equation}\label{H}
\hat{H}= -\sum_{j}\kappa_j(\hat{\sigma}^{+}_j\hat{\sigma}^{-}_{j+1}+\hat{\sigma}^{-}_j\hat{\sigma}^{+}_{j+1})
-\sum_{j}\mu_j\hat{\sigma}^{+}_j\hat{\sigma}^{-}_j,
\end{equation}
where $\hat{\sigma}^{+}_j$ ($\hat{\sigma}^{-}_j$) is the raising (lowering) operator of the $j$-th qubit, $\mu_j$ denotes the on-site potential, the site-dependent coupling $\kappa_{j}$ takes the form $\kappa_{j}\approx f((j-j_{\mathrm{h}}+1/2)d)/{4d}$
with $d$ being the lattice constant.
Here, the function $f(x)$ is related to spacetime metric, which is given in the Eddington-Finkelstein coordinates $ \{t , x\}$ as
$\mathrm{d} s^2 = f(x)\mathrm{d}t^2 - 2\mathrm{d}t\mathrm{d}x$ (see Methods and Supplementary Information). The spatial position $x$ is discretized as $x_j=(j-j_{\mathrm{h}})d$. Since the horizon locates at $f(x_{\mathrm{h}})=0$ with $f'(x_{\mathrm{h}})>0$, the horizon in our analogues model is then defined at site $j=j_{\mathrm{h}}$ where $f(x_{\mathrm{h}})=0$, but the sign of $\kappa_{j}$ is different on its two sides of the horizon resulting in a black hole spacetime  structure. One side of the horizon is considered as the interior of the black hole, while the opposite side represents the exterior of the black hole.

We perform the experiment to simulate the black hole using a superconducting processor with a chain of 10 qubits $Q_1$--$Q_{10}$,
which represents the Hamiltonian (\ref{H}), additionally with 9 tunable couplers interspersed between every two nearest-neighbour qubits, see Fig.~\ref{fig:bh}. The effective hopping coupling $\kappa_{j}$ between qubits $Q_j$ and $Q_{j+1}$ can be tuned arbitrarily via programming the frequency of the corresponding coupler $C_j$, see Methods. To describe the curved spacetime experimentally, we adjust the frequencies of all the couplers, and design the effective coupling distribution as
\begin{equation} \label{Kappa}
\kappa_{j} = \frac{\beta\tanh\!\left((j-j_{\mathrm{h}}+1/2)\eta d\right)}{4\eta d}
\end{equation}
with $j_{\mathrm{h}}=3$, $\eta d=0.35$ and $\beta/(2\pi)\approx4.39~\mathrm{MHz}$. Here we choose $f(x)=\beta \tanh(\eta x)/\eta$, where $\eta$ controls the scale of variation of $f$ over each lattice site, which has the dimension of $1/d$. One can verify that this function $f(x)$ gives us a nonzero Riemannian curvature tensor and so describes a 2-dimensional curved spacetime. As shown in Fig.~\ref{fig:bh}\textbf{b}, the coupling $\kappa_{j}$ goes monotonically from negative to positive from $Q_3$'s left to right side. In this way, the information of the static curved spacetime background is encoded into the site-dependent coupling distribution. Thus, the site $Q_3$ where the sign of the coupling reverses can be analogous to the event horizon of the black hole, the side of negative coupling ($Q_1$-$Q_2$) can be considered as the interior of the black hole, and $Q_4$--$Q_{10}$ are outside the horizon. For comparison, we also realize a uniform coupling distribution with $\kappa_j/(2\pi)\approx 2.94~\mathrm{MHz}$ to realize a flat spacetime. In fact, from the viewpoint of the lattice qubit model, the results will be equivalent if the function $\kappa$ is replaced by $|\kappa|$ both in the case of curved and flat spacetime. Since we here map the coupling to the components of metric, the continuity requires $\kappa$ changes the sign when passing through the analogue horizon.

\begin{figure*}[ht]
	\includegraphics[width=5.8in]{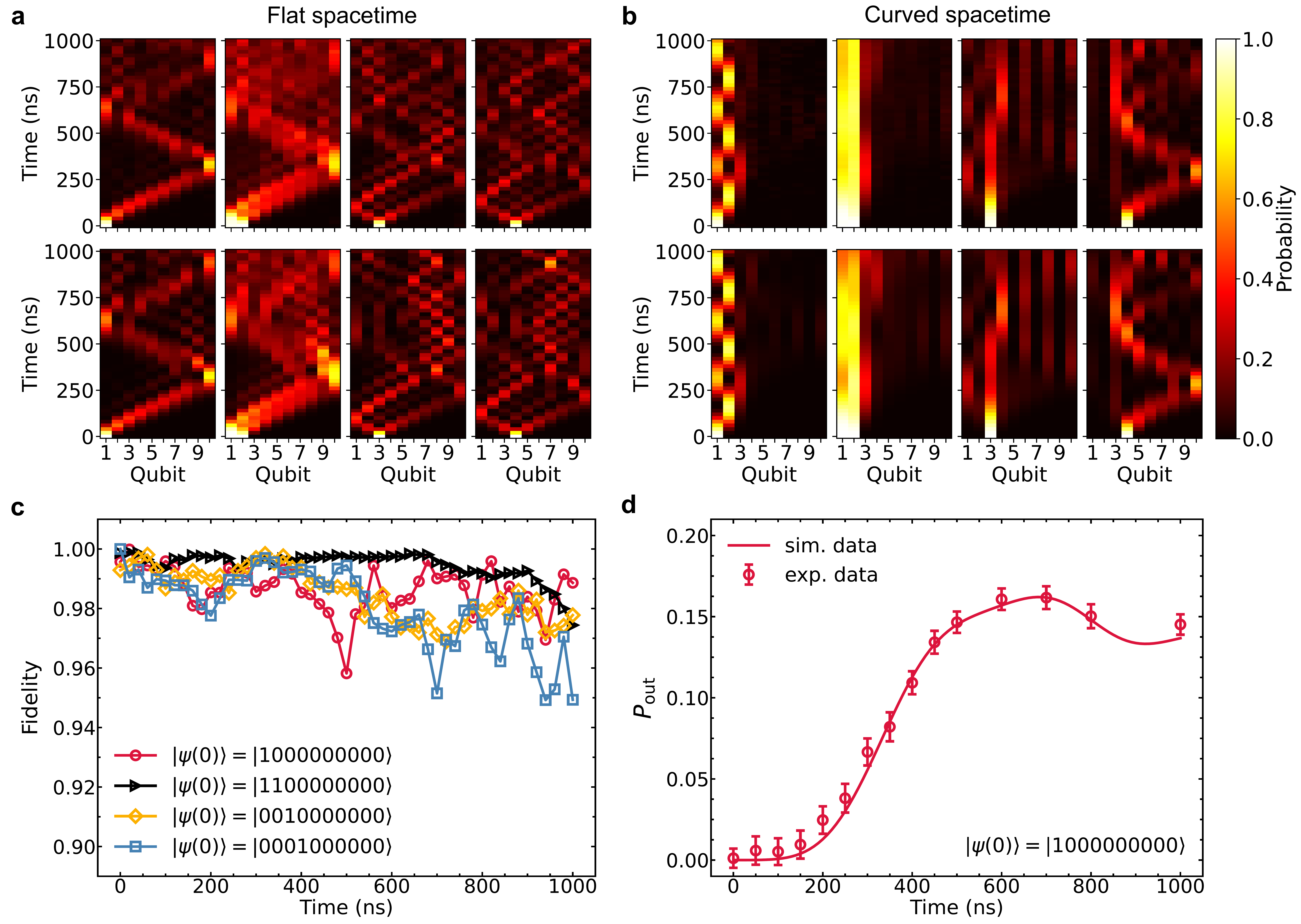}
	\caption{\textbf{Quantum walks in a 1D array of 10 superconducting qubits.} \textbf{a}, Results of the quantum walks in a flat spacetime for four different initial states, i.e., $\ket{\psi(0)}=\ket{1000000000}$, $\ket{1100000000}$, $\ket{0010000000}$ and $\ket{0001000000}$ with $\ket{0}$ representing the ground state of a qubit and $\ket{1}$ the excited state. The case of black hole spacetime is presented in \textbf{b}. The heatmap denotes the probabilities of excited-state for $Q_i$ in time. The horizontal axis is indexed as qubit number $i$, the vertical axis is the evolution time. Here we show both the numerical simulation and experimental data to compare the difference between the flat spacetime ($\kappa_j/(2\pi)\approx2.94$ MHz) and the curved spacetime ($\beta/(2\pi)\approx4.39$ MHz). \textbf{c}, The fidelity of quantum walks in the curved spacetime. \textbf{d}, The probability of finding a particle outside the horizon on qubits $\{Q_4Q_5Q_6Q_7Q_8Q_9Q_{10}\}$. Error bars are 1 SD calculated from all probability data of 50 repetitive experimental runs. \label{fig:walk}}
\end{figure*}

In the experiment, we first prepare an initial state $\ket{\psi(0)}$ with quasi-particle excitations, i.e., exciting qubits or creating
an entangled pair. The evolution of the initial state known as quantum walk will be governed by Schr\"odinger equation $\ket{\psi(t)}=e^{-\mathrm{i}\hat{H}t}\ket{\psi(0)}$ based on
1D programmable controlled Hamiltonian (\ref{H}). The dynamics of the prepared states then simulate the
behavior of quasi-particle in the studied (1+1)-dimensional spacetime with a designed flat or curved structure.
\\
\\
\textbf{Quantum walks in analogue curved spacetime}
\\
Figures~\ref{fig:walk}\textbf{a} and~\ref{fig:walk}\textbf{b} show the propagation of quasi-particles in flat and curved spacetimes, respectively. Here we initialize the system by preparing four different single-particle or two-particle states, including $\ket{\psi(0)}=\ket{1000000000}$, $\ket{1100000000}$, $\ket{0010000000}$ and $\ket{0001000000}$ with $\ket{0}$ and $\ket{1}$ being the eigenstates of $\hat{\sigma}^{+}_j\hat{\sigma}^{-}_j$. Once the initial state is prepared, we apply the rectangular Z pulses on all qubits to quench them in resonance at a reference frequency of $\omega_{\mathrm{ref}}/(2\pi)\approx5.1$ GHz. Meanwhile, the hopping coupling $\kappa_j$ between qubits is fixed as Eq.~\eqref{Kappa} (curved spacetime) or a constant (flat spacetime) by controlling couplers. After evolving for time $t$, all qubits are biased back to idle points for readout. The occupation of quasi-particle density distribution $p_j(t):=\bra{\psi(t)}\hat{\sigma}^{+}_j\hat{\sigma}^{-}_j\ket{\psi(t)}$ is measured by averaging 5000 repeated single-shot measurements, as shown
in Figs.~\ref{fig:walk}\textbf{a} and~\ref{fig:walk}\textbf{b}.

Figure.~\ref{fig:walk}\textbf{a} shows that the propagation of quasi-particle in the flat spacetime is unimpeded, corresponding to the result of conventional quantum walk with diffusive expansion~\cite{PRL_Childs2009,Science_Karski2009,Science_Yan2019,Science_Gong2021}. In contrast, the particle is mainly trapped in our on-chip black hole due to the analogue gravity around the horizon $Q_3$, as shown in Fig.~\ref{fig:walk}\textbf{b} with the initial state $\ket{\psi(0)}=\ket{1000000000}$ and $\ket{\psi(0)}=\ket{1100000000}$. Due to the infalling Eddington-Finkelstein coordinates we took, our model only simulates the outgoing modes of the particle (see Supplementary Information). Hence, the interior and exterior of black hole are equivalent so that the same phenomenon can be observed in that case where the particle is initially prepared in the exterior of black hole ($\ket{\psi(0)}=\ket{0001000000}$).

Here, we also present the result of the particle initialized at the horizon in Fig.~\ref{fig:walk}\textbf{b}, i.e., $\ket{\psi(0)}=\ket{0010000000}$. In the continuous curved spacetime, the particle initialized at the horizon is bound to the horizon forever due to the zero couplings on both sides of the horizon. However, in the finite-size lattice, the coupling strengths on both sides of the horizon are not strictly zero even though they are very small ($\approx0.54$ MHz). The particle seems to be localized at the horizon for a very short time, but it is doomed to escape from the constraints due to the finite-size effects. 

To show the accuracy of the experimental results of quantum walk in the curved spacetime, we present the fidelity $F(t)=\sum_{j=1}^{10}\sqrt{p_j(t)q_j(t)}$ between the measured and theoretical probability distributions $p_j(t)$ and $q_j(t)$ in Fig.~\ref{fig:walk}\textbf{c}.
The high fidelity, greater than 97$\%$ within 400 ns experiment time, implies that our experimental results are consistent with the theoretical predictions as also demonstrated
by the similarity between experimental data and numerical simulations. Note that in both cases of the flat and curved spacetimes the particle will be reflected when it arrives at the boundary $Q_1$ or $Q_{10}$.
\\
\\
\textbf{Observation of analogue Hawking radiation}
\\
Black holes emit thermal radiation leading to evaporation, known as Hawking radiation. However, its observation is a challenge even for an analogue black hole due to the accuracy of the experiment. The Hawking radiation of a black hole is spontaneous in nature. The first realization of spontaneous Hawking radiation in an analogue experiment was in BEC system~\cite{NP_Steinhauer2016}. Here we report an observation of analogue Hawking radiation on the superconducting quantum chip, which is also the first quantum realization of ``lattice black hole'' originally proposed by T. Jacobson more than twenty years ago~\cite{PRD_Corley1997,PRD_Jacobson1999}. 

For the initial state with a particle inside the horizon in our experiment, the evolution of the state
shows the propagation of particle that results in a nonzero density of state outside the horizon is equivalent to the Hawking radiation of the black hole.
Note that the Hawking radiation observed here is stimulated because we induce an excitation by flipping a qubit in $\ket{1}$.

Defining the probability of finding a particle outside the horizon as $P_{\mathrm{out}}=\sum_{j=4}^{10}p_j$,
Fig.~\ref{fig:walk}\textbf{d} shows a rising tendency of $P_{\mathrm{out}}$ in time. This result can be considered as an important signature of Hawking radiation for the analogue black hole~\cite{PRD_Unruh1995,PRL_Weinfurtner2011,PRD_Parentani1995,PRL_Drori2019}.

The theory of Hawking radiation points out that the probability of radiation satisfies a canonical blackbody spectrum,
\begin{equation}
P_{\mathrm{out}}(E)\propto e^{-\frac{E}{T_{\mathrm{H}}}},\label{P_out}
\end{equation}
where $E$ denotes the energy of particle outside the horizon, $T_{\mathrm{H}}/(2\pi)=g_{\mathrm{h}} /(4\pi^2)$ is defined as the effective temperature of the Hawking radiation, and $g_{\mathrm{h}}=\frac{1}{2}f'(x_{\mathrm{h}})=\beta/2$ represents the surface gravity of the black hole~\cite{PRR_Yang2020}. The derivation of Eq.~\eqref{P_out} can be constructed by using the picture of quantum tunneling to obtain the tunneling rate of particle~\cite{PRD_Damour1976,PRL_Parikh2000,JHEP_Arzano2005}. We use this picture in this work to understand Hawking radiation. Such a picture is equivalent to a field theoretical viewpoint of ``particle-antiparticle pairs" created around the horizon: the antiparticle (negative energy) falls into the black hole and annihilates with this particle inside the black hole, the particle outside the horizon is materialized and escapes into infinity (see Supplementary Information). Also, Eq.~\eqref{P_out} can be viewed as the detailed balance relation between the creation and annihilation of particle around the horizon in a thermal environment~\cite{RMP_Nori2012,NJP_Nori2012}.

The tunneling picture of Hawking radiation here is similar to the quantum fluid model of analogue horizon~\cite{PRL_Giovanazzi2005} with two differences in details. The first one is that the analogue horizon of ref. \cite{PRL_Giovanazzi2005} is created by transonic flow but we here create analogue horizon by inhomogeneous lattice hopping. The second is that the injected beam of \cite{PRL_Giovanazzi2005} is from the subsonic region (outside horizon) so that the reflected flow stands for the flow of Hawking radiation (classically the infalling beam should be swallowed by horizon completely and there is no reflected mode), but we here create a particle inside the horizon so the transmission flow is the Hawking radiation.

\begin{figure}[ht]
	\includegraphics[width=0.46\textwidth]{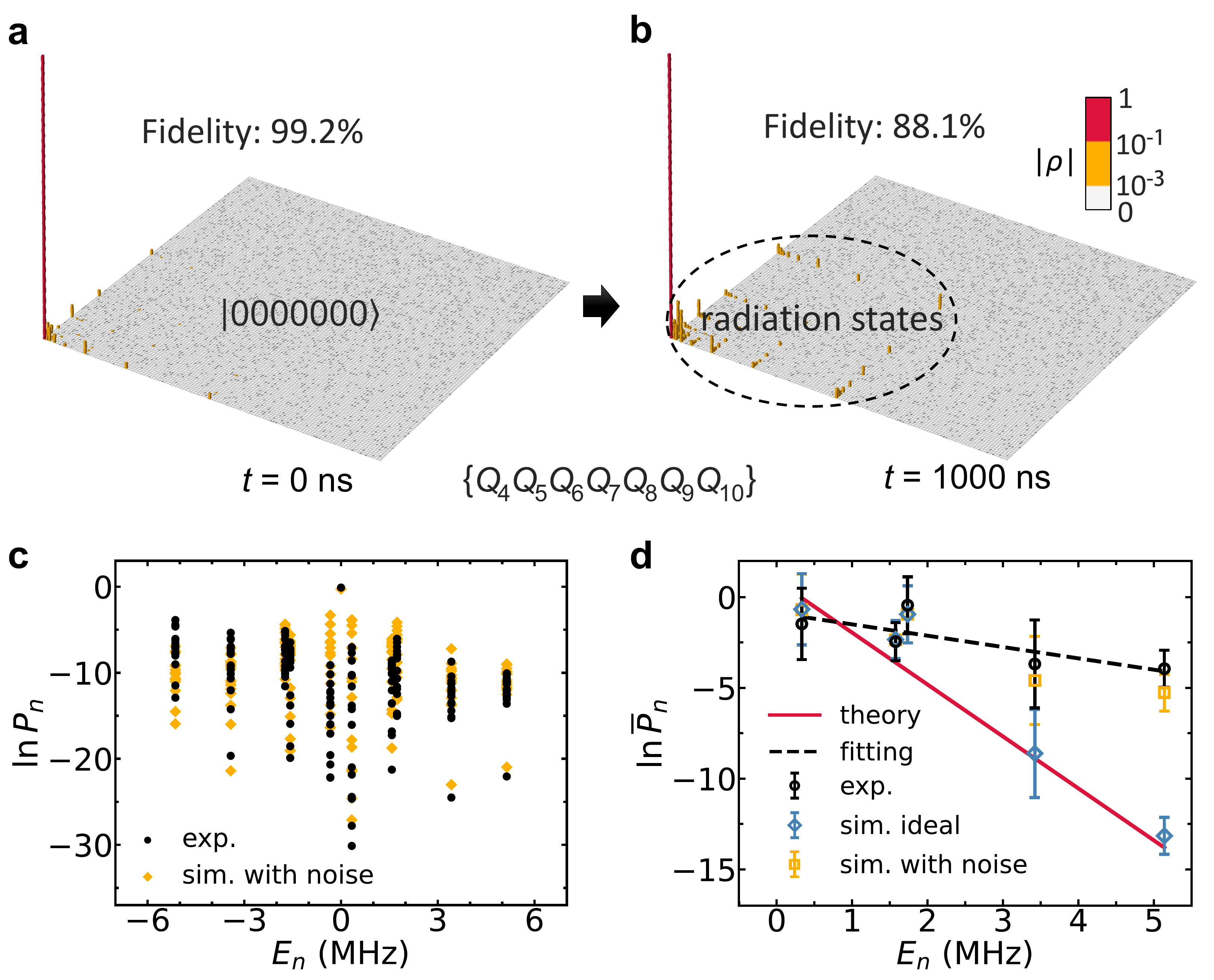}
	\caption{\textbf{Observation of analogue Hawking radiation.} \textbf{a}, The 7-qubit density matrix at $t=0$ ns. Initially, only $Q_1$ is prepared in $\ket{1}$ and all the qubits outside the horizon are almost in $\ket{0}$. \textbf{b}, The 7-qubit density matrix at $t=1000$ ns after the quench dynamics. Due to the Hawking radiation, radiation states can be detected with small probabilities. The fidelity between ideal and experimental density matrix at $t=0$ and $1000$ ns are $99.2\%$ and $88.1\%$, respectively. \textbf{c}, The logarithmic probability of finding a particle outside the horizon $P_n$ vs. its energy $E_n$. \textbf{d}, The logarithm of average radiation probability vs. the energy of particle when $E_n>0$. Error bars are 1 SD calculated from the tomography data at the same energy. The slope of the red line represents the reciprocal of Hawking temperature without noise, where the Hawking temperature here is given by $T_{\mathrm{H}}/(2\pi)=\beta/(8\pi^2)\approx0.35$ MHz or $\approx1.7\times10^{-5}$ K in Kelvin temperature. The experimental results are in agreement with the simulated data for low energy but diverge at high energy due to experiment noises. \label{fig:Hawking}}
\end{figure}

To obtain the radiation probabilities, we perform the quantum state tomography (QST) on the 7 qubits ($Q_4$--$Q_{10}$) outside the horizon at $t=0$ and $t=1000$ ns, such a final time is long enough so that the particle inside the black hole has finished its tunneling to the outside but the boundary effect is negligible to the results. Here, the initial state is $|\psi(0)\rangle=|1000000000\rangle$, i.e., a particle in the black hole has a certain position. When $t=0$ ns, no radiation can be detected and all the qubits outside the horizon are almost in $\ket{0}$, see Fig.~\ref{fig:Hawking}\textbf{a}. After a long time $t=1000$ ns, one may have a small chance to probe the particle outside the horizon, see Fig.~\ref{fig:Hawking}\textbf{b}. The corresponding probabilities of radiation can be extracted from the measured 7-qubit density matrix. Assuming that $|E_n\rangle$ is the $n$-th eigenenergy of total Hamiltonian and $\hat{\rho}_{\text{out}}$ is the density matrix outside obtained by QST, then the probability of finding a particle of energy $E_n$ outside the horizon can be obtained as $P_n=\langle E_n|\hat{\rho}_{\text{out}}|E_n\rangle$, see Methods. Although there are $2^{10}=1024$ eigenstates for 10-qubit Hamiltonian Eq.~\eqref{H} and the same number of $P_n$, the radiation states involve only 10 single-particle excited eigenstates due to the particle number conservation. As a consequence, only those $P_n$ that are corresponding to single-particle excited eigenstates have non-zero values, as shown in Fig.~\ref{fig:Hawking}\textbf{c}. Therefore, we take the average of $P_n$ with the same positive energy $E_n$ as $\bar{P}_n$ to describe the average probability of finding a particle outside with $E_n>0$. It can be expected that the relation between $\bar{P}_{n}$ and $E_{n}$ will agree with the theoretical prediction in Eq.~\eqref{P_out}. In Fig.~\ref{fig:Hawking}\textbf{d}, the simulated results show that the logarithm of the average radiation probability is approximately linear in energy with Hawking temperature $1.7\times10^{-5}$~K. The fitted Hawking temperature of experimental data is around $\sim 7.7\times 10^{-5}$~K, showing validity with the same order of magnitude. The deviation between experimental data and ideal simulation data is mainly caused by the evolution of the imperfect initial state. The fidelity between the imperfect initial state in the experiment and the ideal initial state is $99.2\%$, which may derive from the experimental noises including XY crosstalk, thermal excitation, leakage, etc. We substitute such an experimental state for the ideal initial state in the numerical simulation of Hawking radiation, then the results of numerical simulation agree with experimental results better.

Since the analogue Hawking radiation is characterized by the temperature, we then give an estimation of how large a black hole in our real universe could reproduce the same temperature. If we consider a Schwarzschild black hole in four-dimensional spacetime with the same Hawking temperature $T_{\mathrm{H}}$, its mass can be calculated by $M/M_{\mathrm{s}}=6.4\times10^{-8} \mathrm{K}/T_{\mathrm{H}}$~\cite{Nature_Hawking1974}, where $M_{\mathrm{s}}\approx2\times10^{30}$ kg is the solar mass. For the simulated black hole in our work, $M/M_{\mathrm{s}}\sim10^{-3}$, whereas the typical value reported in BEC system for this quantity can be $\sim10^2$~\cite{NP_Kolobov2021}.
This significant difference in magnitude is attributed to the scales of the setup in different experimental systems. In superconducting qubits, the coupling strength is usually on the order of MHz and thus the analogue surface gravity $g_{\mathrm{h}}$ is of the same magnitude, leading to $T_{\mathrm{H}}=g_{\mathrm{h}}/(2\pi)\sim10^{-5}$~K. Differently, the effective Hawking temperature of sonic black hole depends on the gradient of velocity at the analogue horizon. The BEC system and the shallow water wave system typically give us $T_{\mathrm{H}}\sim10^{-10}$~K~\cite{PRL_Steinhauer2010,Nature_Steinhauer2019, PRL_Isoard2020, NP_Kolobov2021} and $10^{-12}$~K~\cite{PRL_Weinfurtner2011}, respectively.
\\
\\
\textbf{Dynamics of an entangled pair in the analogue black hole}
\\
Hawking predicted that the entanglement entropy increases when a black hole forms and evaporates due to the Hawking radiation. Each Hawking particle is entangled with a partner particle in the black hole. Such kind of quantum feature plays a crucial role in studying black holes and quantum information~\cite{NRP_Maldacena2020}.

\begin{figure}[ht]
	\includegraphics[width=3.3in]{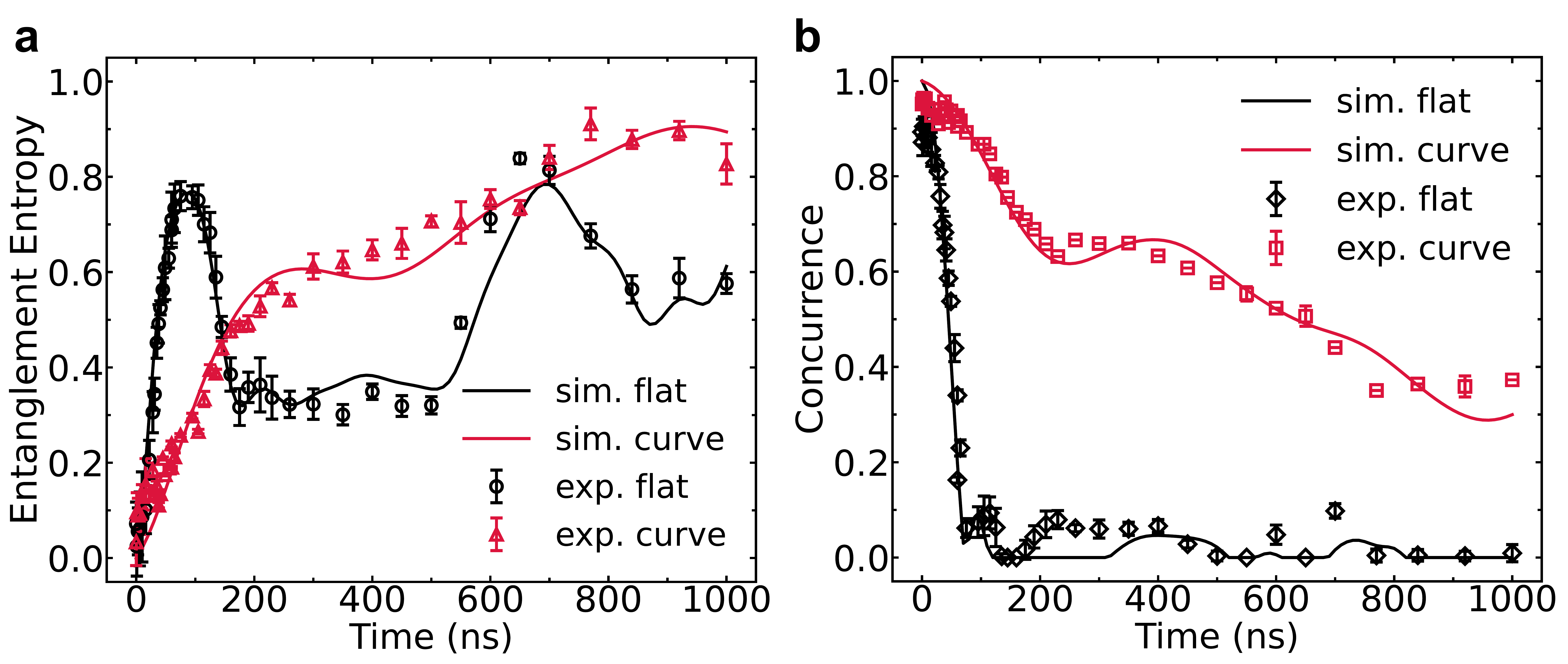}
	\caption{\textbf{Dynamics of entanglement in the analogue black hole}. \textbf{a}, The entanglement entropy vs. evolution time in different spacetimes. The entanglement entropy gradually increases with time in the curved spacetime. \textbf{b}, The concurrence (entanglement between the pair in black hole) vs. evolution time in different spacetimes. Error bars are 1 SD calculated from all tomography data of 10 repetitive experimental runs. The rapid decline of concurrence in the flat spacetime is observed, while the concurrence in the curved spacetime is protected due to the analogue gravity. The solid lines are the results of numerical simulation. Here we set the coupling for the flat spacetime to be a constant ($\kappa/(2\pi)\approx2.94$ MHz) and $\beta/(2\pi)\approx4.39$ MHz for the curved spacetime. \label{fig:entanglement}}
\end{figure}

To investigate the dynamical entanglement and non-local correlation both in flat spacetime and curved spacetime, we initially prepare an entangled pair $\ket{\psi_{\mathrm{in}}(0)}=(\ket{00}+\ket{11})/\sqrt{2}$ (Fig.~\ref{fig:bh}\textbf{c}). The mean fidelity of prepared entangled state is up to $99.15\%$. The dynamics of such an initial entangled state in flat or curved spacetime is observed by time-dependent QST measurement. We obtain the two-qubit density matrix $\hat{\rho}_{\mathrm{in}}(t)$ from the results of QST, and use it to compute the entanglement entropy and the concurrence (see Methods). In Fig.~\ref{fig:entanglement}\textbf{a}, the entanglement entropy in the case of curved spacetime progressively increases due to the Hawking radiation, while in the flat spacetime it has two wavefronts resulting from the quantum interference and reflection respectively~\cite{Science_Yan2019}. On the other hand, the concurrence decreases with time in both cases, reflecting the process of entanglement being lost into the environment. However, the speed of entanglement propagation is limited by the gravitational effects near the horizon, and thus the decrease in concurrence is slowed in the curved spacetime case compared to the flat spacetime case, as shown in Fig.~\ref{fig:entanglement}\textbf{b}.
\\
\\
\noindent
\textbf{\large{Discussion}}
\\
In summary, we have experimentally simulated a curved spacetime of black hole and observed an analogy of Hawking radiation in a superconducting processor with tunable couplers. An high-fidelity entangled pair is also prepared inside the horizon and the corresponding dynamics of entanglement is observed. Our results may stimulate more interests to explore the related features of black holes by using programmable superconducting processor with tunable couplers, and the techniques of calibrating and controlling coupler devices will pave the way for simulating intriguing physics with quantum many-body systems of different coupling distribution.

Our current results are a step in the direction of creating quantum systems with properties analogous to those of black holes. However, many more problems remain to be addressed in a complete simulation of quantum field theory in curved spacetime, both in theory and experiment. Theoretically, it is necessary to study different dimensional systems and investigate a comprehensive theory for mapping the various gravity fields into experimental realizable models. Experimentally, it is expected to expand the category of simulated Hamiltonians, extend the scale of qubits and enhance the control accuracy. In addition to pure analogue experiments, hybrid digital-analogue devices with substantial flexibility in near-term applications also need to be focused~\cite{Nature_Daley2022}. Last but not the least, we must return to the basic problems of quantum field theory and try to translate more fundamental questions, for example, how generic is the emergence of gravity or what happens to spacetime when quantum corrections are fairly important~\cite{NRP_Maldacena2020}.
\\
\\
\noindent
\textbf{\large{Methods}}
\\
\textbf{Metric of two-dimensional spacetime}. Consider a general two-dimensional spacetime background with a fixed static metric $g_{\mu\nu}$,  the metric in the Schwarzschild coordinates $(t_{\mathrm{s}},x)$ reads $	\mathrm{d}s^2=f(x)\mathrm{d}t_{\mathrm{s}}^2-f^{-1}(x)\mathrm{d}x^2$. To describe a black hole with nonzero temperature in 2-dimensional spacetime, we require that the function $f$ has a root at $x=x_{\mathrm{h}}$ with $f'(x_{\mathrm{h}})>0$ and $f(x)>0$ for $x>x_{\mathrm{h}}$ standing for the exterior of the black hole, while $f(x)<0$ for $x<x_{\mathrm{h}}$ for the interior. The horizon of black hole then locates at $x=x_{\mathrm{h}}$. For our purpose and experimental setups, we transform above metric into ``advanced Eddington-Finkelstein coordinates'' by the coordinates transformation $t=t_{\mathrm{s}}+\int{f^{-1}(x)\mathrm{d}x}$. In the coordinates $\{t,x\}$, the metric now becomes $	\mathrm{d} s^2 = f(x)\mathrm{d}t^2 - 2\mathrm{d}t\mathrm{d}x$. The differences between the ``time-orthogonal coordinates'' and ``advanced Eddington-Finkelstein coordinates'' can be found in Supplementary Note 1.
\\
\\
\textbf{Tunable effective couplings}. To construct both flat and curved spacetime background on a single superconducting quantum chip, we use tunable coupler device. The effective coupling between nearest-neighbour qubits derives from their direct capacitive coupling and the indirect virtual exchange coupling via the coupler in between, where the former is untunable and the latter depends on the frequency of the coupler, see Supplementary Note 3. To achieve accurate control of couplings, we develop an efficient and automatic calibration for multi-qubit devices with tunable couplers, see Supplementary Note 6. In the experiments, we apply fast flux-bias Z pulses on the couplers to adjust their frequencies, contributing to the effective coupling distribution. The site-dependent coupling distribution $\kappa_j$ as Eq.~\eqref{Kappa} corresponds to the curved spacetime ($\beta/(2\pi)\approx4.39$ MHz, Fig.~\ref{fig:bh}\textbf{b}), while a uniform coupling distribution ($\kappa_j/(2\pi)\approx2.94$ MHz) is related to the flat spacetime.
\\
\\
\textbf{Calculation of radiation probabilities}. We perform the 7-qubit QST in the observation of analogue Hawking radiation and obtain the density matrix outside the horizon in the 7-qubit Hilbert space. Then we set the states of the other three qubits to $\ket{0}$ and construct the density matrix in the 10-qubit Hilbert space $\hat{\rho}_{\mathrm{out}}$. The probability of finding a particle of energy $E_n$ outside the horizon $P_n$ is calculated as $P_n=\langle E_n|\hat{\rho}_{\text{out}}|E_n\rangle$.
\\
\\
\textbf{Measurement of entanglement}. As shown in Fig.~\ref{fig:bh}\textbf{c}, we prepare the initial entangled pair in the black hole by using two parallel rotations $\hat{R}_{\pi/2}^y$ and $\hat{R}_{-\pi/2}^y$, a CZ gate ($\hat{U}_{\mathrm{CZ}}=\mathrm{diag}(1,1,1,-1)$) and a single-qubit rotation $\hat{R}_{\pi/2}^y$ in sequence. The ideal initial state of the two qubits before the quench dynamics thus is $\ket{\psi_{\mathrm{in}}(0)}=\left(\hat{I}\otimes\hat{R}_{\pi/2}^y\right)\hat{U}_{\mathrm{CZ}}\left(\hat{R}_{\pi/2}^y\otimes\hat{R}_{-\pi/2}^y\right)\ket{00}=(\ket{00}+\ket{11})/\sqrt{2}$.

The state of the total system (the interior of black hole and the rest) is always a pure state during the quench dynamics. Thus, the entanglement entropy of the subsystems: $S(\hat{\rho}_{\mathrm{in}})=S(\hat{\rho}_{\mathrm{rest}})$, which quantifies the entanglement contained in this bipartite quantum system. In our experiment, the cost of measuring $\hat{\rho}_{\mathrm{rest}}$ is much higher than measuring $\hat{\rho}_{\mathrm{in}}$ due to the dimension of the Hilbert space. Therefore, we measure $\hat{\rho}_{\mathrm{in}}$ and calculate $S(\hat{\rho}_{\mathrm{in}})$ as the entanglement measure.


To characteristic the entanglement between the two qubits in the black hole, we use the well-defined measure: concurrence~\cite{PRL_Wootters1998}, which can be calculated as $E(\hat{\rho}_{\mathrm{in}})=\max\{0, \lambda_1-\lambda_2-\lambda_3-\lambda_4\}$ with $\lambda_i$ being the square roots of the eigenvalues of matrix $\hat{\rho}_{\mathrm{in}}\tilde{\hat{\rho}}_{\mathrm{in}}$ in decreasing order, where $\tilde{\hat{\rho}}_{\mathrm{in}}=(\hat{\sigma}_y\otimes\hat{\sigma}_y)\hat{\rho}_{\mathrm{in}}^{\ast}(\hat{\sigma}_y\otimes\hat{\sigma}_y)$ is the spin-flipped state of $\hat{\rho}_{\mathrm{in}}$ with ${\sigma}_y$ being Pauli matrix.
\\
\\
\textbf{\large{Data availability}}
\\
The source data underlying all figures are available at \url{https://doi.org/10.6084/m9.figshare.22802981}.~Other data are available from the corresponding author upon request.

\begin{thebibliography}{43}%
	\makeatletter
	\providecommand \@ifxundefined [1]{%
		\@ifx{#1\undefined}
	}%
	\providecommand \@ifnum [1]{%
		\ifnum #1\expandafter \@firstoftwo
		\else \expandafter \@secondoftwo
		\fi
	}%
	\providecommand \@ifx [1]{%
		\ifx #1\expandafter \@firstoftwo
		\else \expandafter \@secondoftwo
		\fi
	}%
	\providecommand \natexlab [1]{#1}%
	\providecommand \enquote  [1]{``#1''}%
	\providecommand \bibnamefont  [1]{#1}%
	\providecommand \bibfnamefont [1]{#1}%
	\providecommand \citenamefont [1]{#1}%
	\providecommand \href@noop [0]{\@secondoftwo}%
	\providecommand \href [0]{\begingroup \@sanitize@url \@href}%
	\providecommand \@href[1]{\@@startlink{#1}\@@href}%
	\providecommand \@@href[1]{\endgroup#1\@@endlink}%
	\providecommand \@sanitize@url [0]{\catcode `\\12\catcode `\$12\catcode
		`\&12\catcode `\#12\catcode `\^12\catcode `\_12\catcode `\%12\relax}%
	\providecommand \@@startlink[1]{}%
	\providecommand \@@endlink[0]{}%
	\providecommand \url  [0]{\begingroup\@sanitize@url \@url }%
	\providecommand \@url [1]{\endgroup\@href {#1}{\urlprefix }}%
	\providecommand \urlprefix  [0]{URL }%
	\providecommand \Eprint [0]{\href }%
	\providecommand \doibase [0]{https://doi.org/}%
	\providecommand \selectlanguage [0]{\@gobble}%
	\providecommand \bibinfo  [0]{\@secondoftwo}%
	\providecommand \bibfield  [0]{\@secondoftwo}%
	\providecommand \translation [1]{[#1]}%
	\providecommand \BibitemOpen [0]{}%
	\providecommand \bibitemStop [0]{}%
	\providecommand \bibitemNoStop [0]{.\EOS\space}%
	\providecommand \EOS [0]{\spacefactor3000\relax}%
	\providecommand \BibitemShut  [1]{\csname bibitem#1\endcsname}%
	\let\auto@bib@innerbib\@empty
	\bibitem [{\citenamefont {Hawking}(1974)}]{Nature_Hawking1974}%
	\BibitemOpen
	\bibfield  {author} {\bibinfo {author} {\bibfnamefont {S.~W.}\ \bibnamefont
			{Hawking}},\ }\bibfield  {title} {\bibinfo {title} {{Black hole
				explosions?}},\ }\href {https://doi.org/10.1038/248030a0} {\bibfield
		{journal} {\bibinfo  {journal} {Nature}\ }\textbf {\bibinfo {volume} {248}},\
		\bibinfo {pages} {30} (\bibinfo {year} {1974})}\BibitemShut {NoStop}%
	\bibitem [{\citenamefont {Unruh}(1981)}]{PRL_Unruh1981}%
	\BibitemOpen
	\bibfield  {author} {\bibinfo {author} {\bibfnamefont {W.~G.}\ \bibnamefont
			{Unruh}},\ }\bibfield  {title} {\bibinfo {title} {{Experimental Black-Hole
				Evaporation?}},\ }\href {https://doi.org/10.1103/PhysRevLett.46.1351}
	{\bibfield  {journal} {\bibinfo  {journal} {Phys. Rev. Lett.}\ }\textbf
		{\bibinfo {volume} {46}},\ \bibinfo {pages} {1351} (\bibinfo {year}
		{1981})}\BibitemShut {NoStop}%
	\bibitem [{\citenamefont {Unruh}(1995)}]{PRD_Unruh1995}%
	\BibitemOpen
	\bibfield  {author} {\bibinfo {author} {\bibfnamefont {W.~G.}\ \bibnamefont
			{Unruh}},\ }\bibfield  {title} {\bibinfo {title} {{Sonic analogue of black
				holes and the effects of high frequencies on black hole evaporation}},\
	}\href {https://doi.org/10.1103/PhysRevD.51.2827} {\bibfield  {journal}
		{\bibinfo  {journal} {Phys. Rev. D}\ }\textbf {\bibinfo {volume} {51}},\
		\bibinfo {pages} {2827} (\bibinfo {year} {1995})}\BibitemShut {NoStop}%
	\bibitem [{\citenamefont {Weinfurtner}\ \emph {et~al.}(2011)\citenamefont
		{Weinfurtner}, \citenamefont {Tedford}, \citenamefont {Penrice},
		\citenamefont {Unruh},\ and\ \citenamefont {Lawrence}}]{PRL_Weinfurtner2011}%
	\BibitemOpen
	\bibfield  {author} {\bibinfo {author} {\bibfnamefont {S.}~\bibnamefont
			{Weinfurtner}}, \bibinfo {author} {\bibfnamefont {E.~W.}\ \bibnamefont
			{Tedford}}, \bibinfo {author} {\bibfnamefont {M.~C.~J.}\ \bibnamefont
			{Penrice}}, \bibinfo {author} {\bibfnamefont {W.~G.}\ \bibnamefont {Unruh}},\
		and\ \bibinfo {author} {\bibfnamefont {G.~A.}\ \bibnamefont {Lawrence}},\
	}\bibfield  {title} {\bibinfo {title} {{Measurement of Stimulated Hawking
				Emission in an Analogue System}},\ }\href
	{https://doi.org/10.1103/PhysRevLett.106.021302} {\bibfield  {journal}
		{\bibinfo  {journal} {Phys. Rev. Lett.}\ }\textbf {\bibinfo {volume} {106}},\
		\bibinfo {pages} {021302} (\bibinfo {year} {2011})}\BibitemShut {NoStop}%
	\bibitem [{\citenamefont {Michel}\ and\ \citenamefont
		{Parentani}(2014)}]{PRD_Michel2014}%
	\BibitemOpen
	\bibfield  {author} {\bibinfo {author} {\bibfnamefont {F.}~\bibnamefont
			{Michel}}\ and\ \bibinfo {author} {\bibfnamefont {R.}~\bibnamefont
			{Parentani}},\ }\bibfield  {title} {\bibinfo {title} {{Probing the thermal
				character of analogue Hawking radiation for shallow water waves?}},\ }\href
	{https://doi.org/10.1103/PhysRevD.90.044033} {\bibfield  {journal} {\bibinfo
			{journal} {Phys. Rev. D}\ }\textbf {\bibinfo {volume} {90}},\ \bibinfo
		{pages} {044033} (\bibinfo {year} {2014})}\BibitemShut {NoStop}%
	\bibitem [{\citenamefont {Euv\'e}\ \emph {et~al.}(2015)\citenamefont {Euv\'e},
		\citenamefont {Michel}, \citenamefont {Parentani},\ and\ \citenamefont
		{Rousseaux}}]{PRD_Euve2015}%
	\BibitemOpen
	\bibfield  {author} {\bibinfo {author} {\bibfnamefont {L.-P.}\ \bibnamefont
			{Euv\'e}}, \bibinfo {author} {\bibfnamefont {F.}~\bibnamefont {Michel}},
		\bibinfo {author} {\bibfnamefont {R.}~\bibnamefont {Parentani}},\ and\
		\bibinfo {author} {\bibfnamefont {G.}~\bibnamefont {Rousseaux}},\ }\bibfield
	{title} {\bibinfo {title} {{Wave blocking and partial transmission in
				subcritical flows over an obstacle}},\ }\href
	{https://doi.org/10.1103/PhysRevD.91.024020} {\bibfield  {journal} {\bibinfo
			{journal} {Phys. Rev. D}\ }\textbf {\bibinfo {volume} {91}},\ \bibinfo
		{pages} {024020} (\bibinfo {year} {2015})}\BibitemShut {NoStop}%
	\bibitem [{\citenamefont {Coutant}\ and\ \citenamefont
		{Weinfurtner}(2016)}]{PRD_Coutant2016}%
	\BibitemOpen
	\bibfield  {author} {\bibinfo {author} {\bibfnamefont {A.}~\bibnamefont
			{Coutant}}\ and\ \bibinfo {author} {\bibfnamefont {S.}~\bibnamefont
			{Weinfurtner}},\ }\bibfield  {title} {\bibinfo {title} {{The imprint of the
				analogue Hawking effect in subcritical flows}},\ }\href
	{https://doi.org/10.1103/PhysRevD.94.064026} {\bibfield  {journal} {\bibinfo
			{journal} {Phys. Rev. D}\ }\textbf {\bibinfo {volume} {94}},\ \bibinfo
		{pages} {064026} (\bibinfo {year} {2016})}\BibitemShut {NoStop}%
	\bibitem [{\citenamefont {Steinhauer}(2016)}]{NP_Steinhauer2016}%
	\BibitemOpen
	\bibfield  {author} {\bibinfo {author} {\bibfnamefont {J.}~\bibnamefont
			{Steinhauer}},\ }\bibfield  {title} {\bibinfo {title} {{Observation of
				quantum Hawking radiation and its entanglement in an analogue black hole}},\
	}\href {https://doi.org/10.1038/nphys3863} {\bibfield  {journal} {\bibinfo
			{journal} {Nature Physics}\ }\textbf {\bibinfo {volume} {12}},\ \bibinfo
		{pages} {959} (\bibinfo {year} {2016})}\BibitemShut {NoStop}%
	\bibitem [{\citenamefont {Lahav}\ \emph {et~al.}(2010)\citenamefont {Lahav},
		\citenamefont {Itah}, \citenamefont {Blumkin}, \citenamefont {Gordon},
		\citenamefont {Rinott}, \citenamefont {Zayats},\ and\ \citenamefont
		{Steinhauer}}]{PRL_Steinhauer2010}%
	\BibitemOpen
	\bibfield  {author} {\bibinfo {author} {\bibfnamefont {O.}~\bibnamefont
			{Lahav}}, \bibinfo {author} {\bibfnamefont {A.}~\bibnamefont {Itah}},
		\bibinfo {author} {\bibfnamefont {A.}~\bibnamefont {Blumkin}}, \bibinfo
		{author} {\bibfnamefont {C.}~\bibnamefont {Gordon}}, \bibinfo {author}
		{\bibfnamefont {S.}~\bibnamefont {Rinott}}, \bibinfo {author} {\bibfnamefont
			{A.}~\bibnamefont {Zayats}},\ and\ \bibinfo {author} {\bibfnamefont
			{J.}~\bibnamefont {Steinhauer}},\ }\bibfield  {title} {\bibinfo {title}
		{{Realization of a Sonic Black Hole Analog in a Bose-Einstein Condensate}},\
	}\href {https://doi.org/10.1103/PhysRevLett.105.240401} {\bibfield  {journal}
		{\bibinfo  {journal} {Phys. Rev. Lett.}\ }\textbf {\bibinfo {volume} {105}},\
		\bibinfo {pages} {240401} (\bibinfo {year} {2010})}\BibitemShut {NoStop}%
	\bibitem [{\citenamefont {{Mu{\~{n}}oz de Nova}}\ \emph
		{et~al.}(2019)\citenamefont {{Mu{\~{n}}oz de Nova}}, \citenamefont
		{Golubkov}, \citenamefont {Kolobov},\ and\ \citenamefont
		{Steinhauer}}]{Nature_Steinhauer2019}%
	\BibitemOpen
	\bibfield  {author} {\bibinfo {author} {\bibfnamefont {J.~R.}\ \bibnamefont
			{{Mu{\~{n}}oz de Nova}}}, \bibinfo {author} {\bibfnamefont {K.}~\bibnamefont
			{Golubkov}}, \bibinfo {author} {\bibfnamefont {V.~I.}\ \bibnamefont
			{Kolobov}},\ and\ \bibinfo {author} {\bibfnamefont {J.}~\bibnamefont
			{Steinhauer}},\ }\bibfield  {title} {\bibinfo {title} {{Observation of
				thermal Hawking radiation and its temperature in an analogue black hole}},\
	}\href {https://doi.org/10.1038/s41586-019-1241-0} {\bibfield  {journal}
		{\bibinfo  {journal} {Nature}\ }\textbf {\bibinfo {volume} {569}},\ \bibinfo
		{pages} {688} (\bibinfo {year} {2019})}\BibitemShut {NoStop}%
	\bibitem [{\citenamefont {Isoard}\ and\ \citenamefont
		{Pavloff}(2020)}]{PRL_Isoard2020}%
	\BibitemOpen
	\bibfield  {author} {\bibinfo {author} {\bibfnamefont {M.}~\bibnamefont
			{Isoard}}\ and\ \bibinfo {author} {\bibfnamefont {N.}~\bibnamefont
			{Pavloff}},\ }\bibfield  {title} {\bibinfo {title} {{Departing from
				Thermality of Analogue Hawking Radiation in a Bose-Einstein Condensate}},\
	}\href {https://doi.org/10.1103/PhysRevLett.124.060401} {\bibfield  {journal}
		{\bibinfo  {journal} {Physical Review Letters}\ }\textbf {\bibinfo {volume}
			{124}},\ \bibinfo {pages} {060401} (\bibinfo {year} {2020})}\BibitemShut
	{NoStop}%
	\bibitem [{\citenamefont {Kolobov}\ \emph {et~al.}(2021)\citenamefont
		{Kolobov}, \citenamefont {Golubkov}, \citenamefont {{Mu{\~{n}}oz de Nova}},\
		and\ \citenamefont {Steinhauer}}]{NP_Kolobov2021}%
	\BibitemOpen
	\bibfield  {author} {\bibinfo {author} {\bibfnamefont {V.~I.}\ \bibnamefont
			{Kolobov}}, \bibinfo {author} {\bibfnamefont {K.}~\bibnamefont {Golubkov}},
		\bibinfo {author} {\bibfnamefont {J.~R.}\ \bibnamefont {{Mu{\~{n}}oz de
					Nova}}},\ and\ \bibinfo {author} {\bibfnamefont {J.}~\bibnamefont
			{Steinhauer}},\ }\bibfield  {title} {\bibinfo {title} {{Observation of
				stationary spontaneous Hawking radiation and the time evolution of an
				analogue black hole}},\ }\href {https://doi.org/10.1038/s41567-020-01076-0}
	{\bibfield  {journal} {\bibinfo  {journal} {Nature Physics}\ }\textbf
		{\bibinfo {volume} {17}},\ \bibinfo {pages} {362} (\bibinfo {year}
		{2021})}\BibitemShut {NoStop}%
	\bibitem [{\citenamefont {Philbin}\ \emph {et~al.}(2008)\citenamefont
		{Philbin}, \citenamefont {Kuklewicz}, \citenamefont {Robertson},
		\citenamefont {Hill}, \citenamefont {Konig},\ and\ \citenamefont
		{Leonhardt}}]{Science_Philbin2008}%
	\BibitemOpen
	\bibfield  {author} {\bibinfo {author} {\bibfnamefont {T.~G.}\ \bibnamefont
			{Philbin}}, \bibinfo {author} {\bibfnamefont {C.}~\bibnamefont {Kuklewicz}},
		\bibinfo {author} {\bibfnamefont {S.}~\bibnamefont {Robertson}}, \bibinfo
		{author} {\bibfnamefont {S.}~\bibnamefont {Hill}}, \bibinfo {author}
		{\bibfnamefont {F.}~\bibnamefont {Konig}},\ and\ \bibinfo {author}
		{\bibfnamefont {U.}~\bibnamefont {Leonhardt}},\ }\bibfield  {title} {\bibinfo
		{title} {{Fiber-Optical Analog of the Event Horizon}},\ }\href
	{https://doi.org/10.1126/science.1153625} {\bibfield  {journal} {\bibinfo
			{journal} {Science}\ }\textbf {\bibinfo {volume} {319}},\ \bibinfo {pages}
		{1367} (\bibinfo {year} {2008})}\BibitemShut {NoStop}%
	\bibitem [{\citenamefont {Drori}\ \emph {et~al.}(2019)\citenamefont {Drori},
		\citenamefont {Rosenberg}, \citenamefont {Bermudez}, \citenamefont
		{Silberberg},\ and\ \citenamefont {Leonhardt}}]{PRL_Drori2019}%
	\BibitemOpen
	\bibfield  {author} {\bibinfo {author} {\bibfnamefont {J.}~\bibnamefont
			{Drori}}, \bibinfo {author} {\bibfnamefont {Y.}~\bibnamefont {Rosenberg}},
		\bibinfo {author} {\bibfnamefont {D.}~\bibnamefont {Bermudez}}, \bibinfo
		{author} {\bibfnamefont {Y.}~\bibnamefont {Silberberg}},\ and\ \bibinfo
		{author} {\bibfnamefont {U.}~\bibnamefont {Leonhardt}},\ }\bibfield  {title}
	{\bibinfo {title} {{Observation of Stimulated Hawking Radiation in an Optical
				Analogue}},\ }\href {https://doi.org/10.1103/PhysRevLett.122.010404}
	{\bibfield  {journal} {\bibinfo  {journal} {Phys. Rev. Lett.}\ }\textbf
		{\bibinfo {volume} {122}},\ \bibinfo {pages} {010404} (\bibinfo {year}
		{2019})}\BibitemShut {NoStop}%
	\bibitem [{\citenamefont {Sheng}\ \emph {et~al.}(2013)\citenamefont {Sheng},
		\citenamefont {Liu}, \citenamefont {Wang}, \citenamefont {Zhu},\ and\
		\citenamefont {Genov}}]{NPhoton_Sheng2013}%
	\BibitemOpen
	\bibfield  {author} {\bibinfo {author} {\bibfnamefont {C.}~\bibnamefont
			{Sheng}}, \bibinfo {author} {\bibfnamefont {H.}~\bibnamefont {Liu}}, \bibinfo
		{author} {\bibfnamefont {Y.}~\bibnamefont {Wang}}, \bibinfo {author}
		{\bibfnamefont {S.~N.}\ \bibnamefont {Zhu}},\ and\ \bibinfo {author}
		{\bibfnamefont {D.~A.}\ \bibnamefont {Genov}},\ }\bibfield  {title} {\bibinfo
		{title} {{Trapping light by mimicking gravitational lensing}},\ }\href
	{https://doi.org/10.1038/nphoton.2013.247} {\bibfield  {journal} {\bibinfo
			{journal} {Nature Photonics}\ }\textbf {\bibinfo {volume} {7}},\ \bibinfo
		{pages} {902} (\bibinfo {year} {2013})}\BibitemShut {NoStop}%
	\bibitem [{\citenamefont {Arute}\ \emph {et~al.}(2019)\citenamefont {Arute},
		\citenamefont {Arya}, \citenamefont {Babbush}, \citenamefont {Bacon},
		\citenamefont {Bardin}, \citenamefont {Barends}, \citenamefont {Biswas},
		\citenamefont {Boixo}, \citenamefont {Brandao}, \citenamefont {Buell},
		\citenamefont {Burkett}, \citenamefont {Chen}, \citenamefont {Chen},
		\citenamefont {Chiaro}, \citenamefont {Collins}, \citenamefont {Courtney},
		\citenamefont {Dunsworth}, \citenamefont {Farhi}, \citenamefont {Foxen},
		\citenamefont {Fowler}, \citenamefont {Gidney}, \citenamefont {Giustina},
		\citenamefont {Graff}, \citenamefont {Guerin}, \citenamefont {Habegger},
		\citenamefont {Harrigan}, \citenamefont {Hartmann}, \citenamefont {Ho},
		\citenamefont {Hoffmann}, \citenamefont {Huang}, \citenamefont {Humble},
		\citenamefont {Isakov}, \citenamefont {Jeffrey}, \citenamefont {Jiang},
		\citenamefont {Kafri}, \citenamefont {Kechedzhi}, \citenamefont {Kelly},
		\citenamefont {Klimov}, \citenamefont {Knysh}, \citenamefont {Korotkov},
		\citenamefont {Kostritsa}, \citenamefont {Landhuis}, \citenamefont
		{Lindmark}, \citenamefont {Lucero}, \citenamefont {Lyakh}, \citenamefont
		{Mandr{\`{a}}}, \citenamefont {McClean}, \citenamefont {McEwen},
		\citenamefont {Megrant}, \citenamefont {Mi}, \citenamefont {Michielsen},
		\citenamefont {Mohseni}, \citenamefont {Mutus}, \citenamefont {Naaman},
		\citenamefont {Neeley}, \citenamefont {Neill}, \citenamefont {Niu},
		\citenamefont {Ostby}, \citenamefont {Petukhov}, \citenamefont {Platt},
		\citenamefont {Quintana}, \citenamefont {Rieffel}, \citenamefont {Roushan},
		\citenamefont {Rubin}, \citenamefont {Sank}, \citenamefont {Satzinger},
		\citenamefont {Smelyanskiy}, \citenamefont {Sung}, \citenamefont
		{Trevithick}, \citenamefont {Vainsencher}, \citenamefont {Villalonga},
		\citenamefont {White}, \citenamefont {Yao}, \citenamefont {Yeh},
		\citenamefont {Zalcman}, \citenamefont {Neven},\ and\ \citenamefont
		{Martinis}}]{Nature_Arute2019}%
	\BibitemOpen
	\bibfield  {author} {\bibinfo {author} {\bibfnamefont {F.}~\bibnamefont
			{Arute}}, \bibinfo {author} {\bibfnamefont {K.}~\bibnamefont {Arya}},
		\bibinfo {author} {\bibfnamefont {R.}~\bibnamefont {Babbush}}, \bibinfo
		{author} {\bibfnamefont {D.}~\bibnamefont {Bacon}}, \bibinfo {author}
		{\bibfnamefont {J.~C.}\ \bibnamefont {Bardin}}, \bibinfo {author}
		{\bibfnamefont {R.}~\bibnamefont {Barends}}, \bibinfo {author} {\bibfnamefont
			{R.}~\bibnamefont {Biswas}}, \bibinfo {author} {\bibfnamefont
			{S.}~\bibnamefont {Boixo}}, \bibinfo {author} {\bibfnamefont {F.~G. S.~L.}\
			\bibnamefont {Brandao}}, \bibinfo {author} {\bibfnamefont {D.~A.}\
			\bibnamefont {Buell}}, \bibinfo {author} {\bibfnamefont {B.}~\bibnamefont
			{Burkett}}, \bibinfo {author} {\bibfnamefont {Y.}~\bibnamefont {Chen}},
		\bibinfo {author} {\bibfnamefont {Z.}~\bibnamefont {Chen}}, \bibinfo {author}
		{\bibfnamefont {B.}~\bibnamefont {Chiaro}}, \bibinfo {author} {\bibfnamefont
			{R.}~\bibnamefont {Collins}}, \bibinfo {author} {\bibfnamefont
			{W.}~\bibnamefont {Courtney}}, \bibinfo {author} {\bibfnamefont
			{A.}~\bibnamefont {Dunsworth}}, \bibinfo {author} {\bibfnamefont
			{E.}~\bibnamefont {Farhi}}, \bibinfo {author} {\bibfnamefont
			{B.}~\bibnamefont {Foxen}}, \bibinfo {author} {\bibfnamefont
			{A.}~\bibnamefont {Fowler}}, \bibinfo {author} {\bibfnamefont
			{C.}~\bibnamefont {Gidney}}, \bibinfo {author} {\bibfnamefont
			{M.}~\bibnamefont {Giustina}}, \bibinfo {author} {\bibfnamefont
			{R.}~\bibnamefont {Graff}}, \bibinfo {author} {\bibfnamefont
			{K.}~\bibnamefont {Guerin}}, \bibinfo {author} {\bibfnamefont
			{S.}~\bibnamefont {Habegger}}, \bibinfo {author} {\bibfnamefont {M.~P.}\
			\bibnamefont {Harrigan}}, \bibinfo {author} {\bibfnamefont {M.~J.}\
			\bibnamefont {Hartmann}}, \bibinfo {author} {\bibfnamefont {A.}~\bibnamefont
			{Ho}}, \bibinfo {author} {\bibfnamefont {M.}~\bibnamefont {Hoffmann}},
		\bibinfo {author} {\bibfnamefont {T.}~\bibnamefont {Huang}}, \bibinfo
		{author} {\bibfnamefont {T.~S.}\ \bibnamefont {Humble}}, \bibinfo {author}
		{\bibfnamefont {S.~V.}\ \bibnamefont {Isakov}}, \bibinfo {author}
		{\bibfnamefont {E.}~\bibnamefont {Jeffrey}}, \bibinfo {author} {\bibfnamefont
			{Z.}~\bibnamefont {Jiang}}, \bibinfo {author} {\bibfnamefont
			{D.}~\bibnamefont {Kafri}}, \bibinfo {author} {\bibfnamefont
			{K.}~\bibnamefont {Kechedzhi}}, \bibinfo {author} {\bibfnamefont
			{J.}~\bibnamefont {Kelly}}, \bibinfo {author} {\bibfnamefont {P.~V.}\
			\bibnamefont {Klimov}}, \bibinfo {author} {\bibfnamefont {S.}~\bibnamefont
			{Knysh}}, \bibinfo {author} {\bibfnamefont {A.}~\bibnamefont {Korotkov}},
		\bibinfo {author} {\bibfnamefont {F.}~\bibnamefont {Kostritsa}}, \bibinfo
		{author} {\bibfnamefont {D.}~\bibnamefont {Landhuis}}, \bibinfo {author}
		{\bibfnamefont {M.}~\bibnamefont {Lindmark}}, \bibinfo {author}
		{\bibfnamefont {E.}~\bibnamefont {Lucero}}, \bibinfo {author} {\bibfnamefont
			{D.}~\bibnamefont {Lyakh}}, \bibinfo {author} {\bibfnamefont
			{S.}~\bibnamefont {Mandr{\`{a}}}}, \bibinfo {author} {\bibfnamefont {J.~R.}\
			\bibnamefont {McClean}}, \bibinfo {author} {\bibfnamefont {M.}~\bibnamefont
			{McEwen}}, \bibinfo {author} {\bibfnamefont {A.}~\bibnamefont {Megrant}},
		\bibinfo {author} {\bibfnamefont {X.}~\bibnamefont {Mi}}, \bibinfo {author}
		{\bibfnamefont {K.}~\bibnamefont {Michielsen}}, \bibinfo {author}
		{\bibfnamefont {M.}~\bibnamefont {Mohseni}}, \bibinfo {author} {\bibfnamefont
			{J.}~\bibnamefont {Mutus}}, \bibinfo {author} {\bibfnamefont
			{O.}~\bibnamefont {Naaman}}, \bibinfo {author} {\bibfnamefont
			{M.}~\bibnamefont {Neeley}}, \bibinfo {author} {\bibfnamefont
			{C.}~\bibnamefont {Neill}}, \bibinfo {author} {\bibfnamefont {M.~Y.}\
			\bibnamefont {Niu}}, \bibinfo {author} {\bibfnamefont {E.}~\bibnamefont
			{Ostby}}, \bibinfo {author} {\bibfnamefont {A.}~\bibnamefont {Petukhov}},
		\bibinfo {author} {\bibfnamefont {J.~C.}\ \bibnamefont {Platt}}, \bibinfo
		{author} {\bibfnamefont {C.}~\bibnamefont {Quintana}}, \bibinfo {author}
		{\bibfnamefont {E.~G.}\ \bibnamefont {Rieffel}}, \bibinfo {author}
		{\bibfnamefont {P.}~\bibnamefont {Roushan}}, \bibinfo {author} {\bibfnamefont
			{N.~C.}\ \bibnamefont {Rubin}}, \bibinfo {author} {\bibfnamefont
			{D.}~\bibnamefont {Sank}}, \bibinfo {author} {\bibfnamefont {K.~J.}\
			\bibnamefont {Satzinger}}, \bibinfo {author} {\bibfnamefont {V.}~\bibnamefont
			{Smelyanskiy}}, \bibinfo {author} {\bibfnamefont {K.~J.}\ \bibnamefont
			{Sung}}, \bibinfo {author} {\bibfnamefont {M.~D.}\ \bibnamefont
			{Trevithick}}, \bibinfo {author} {\bibfnamefont {A.}~\bibnamefont
			{Vainsencher}}, \bibinfo {author} {\bibfnamefont {B.}~\bibnamefont
			{Villalonga}}, \bibinfo {author} {\bibfnamefont {T.}~\bibnamefont {White}},
		\bibinfo {author} {\bibfnamefont {Z.~J.}\ \bibnamefont {Yao}}, \bibinfo
		{author} {\bibfnamefont {P.}~\bibnamefont {Yeh}}, \bibinfo {author}
		{\bibfnamefont {A.}~\bibnamefont {Zalcman}}, \bibinfo {author} {\bibfnamefont
			{H.}~\bibnamefont {Neven}},\ and\ \bibinfo {author} {\bibfnamefont {J.~M.}\
			\bibnamefont {Martinis}},\ }\bibfield  {title} {\bibinfo {title} {{Quantum
				supremacy using a programmable superconducting processor}},\ }\href
	{https://doi.org/10.1038/s41586-019-1666-5} {\bibfield  {journal} {\bibinfo
			{journal} {Nature}\ }\textbf {\bibinfo {volume} {574}},\ \bibinfo {pages}
		{505} (\bibinfo {year} {2019})}\BibitemShut {NoStop}%
	\bibitem [{\citenamefont {Georgescu}\ \emph {et~al.}(2014)\citenamefont
		{Georgescu}, \citenamefont {Ashhab},\ and\ \citenamefont
		{Nori}}]{RMP_Nori2014}%
	\BibitemOpen
	\bibfield  {author} {\bibinfo {author} {\bibfnamefont {I.~M.}\ \bibnamefont
			{Georgescu}}, \bibinfo {author} {\bibfnamefont {S.}~\bibnamefont {Ashhab}},\
		and\ \bibinfo {author} {\bibfnamefont {F.}~\bibnamefont {Nori}},\ }\bibfield
	{title} {\bibinfo {title} {{Quantum simulation}},\ }\href
	{https://doi.org/10.1103/RevModPhys.86.153} {\bibfield  {journal} {\bibinfo
			{journal} {Rev. Mod. Phys.}\ }\textbf {\bibinfo {volume} {86}},\ \bibinfo
		{pages} {153} (\bibinfo {year} {2014})}\BibitemShut {NoStop}%
	\bibitem [{\citenamefont {Gu}\ \emph {et~al.}(2017)\citenamefont {Gu},
		\citenamefont {Kockum}, \citenamefont {Miranowicz}, \citenamefont {xi~Liu},\
		and\ \citenamefont {Nori}}]{PR_Nori2017}%
	\BibitemOpen
	\bibfield  {author} {\bibinfo {author} {\bibfnamefont {X.}~\bibnamefont
			{Gu}}, \bibinfo {author} {\bibfnamefont {A.~F.}\ \bibnamefont {Kockum}},
		\bibinfo {author} {\bibfnamefont {A.}~\bibnamefont {Miranowicz}}, \bibinfo
		{author} {\bibfnamefont {Y.}~\bibnamefont {xi~Liu}},\ and\ \bibinfo {author}
		{\bibfnamefont {F.}~\bibnamefont {Nori}},\ }\bibfield  {title} {\bibinfo
		{title} {Microwave photonics with superconducting quantum circuits},\ }\href
	{https://doi.org/https://doi.org/10.1016/j.physrep.2017.10.002} {\bibfield
		{journal} {\bibinfo  {journal} {Physics Reports}\ }\textbf {\bibinfo {volume}
			{718-719}},\ \bibinfo {pages} {1} (\bibinfo {year} {2017})}\BibitemShut
	{NoStop}%
	\bibitem [{\citenamefont {Wu}\ \emph {et~al.}(2021)\citenamefont {Wu},
		\citenamefont {Bao}, \citenamefont {Cao}, \citenamefont {Chen}, \citenamefont
		{Chen}, \citenamefont {Chen}, \citenamefont {Chung}, \citenamefont {Deng},
		\citenamefont {Du}, \citenamefont {Fan}, \citenamefont {Gong}, \citenamefont
		{Guo}, \citenamefont {Guo}, \citenamefont {Guo}, \citenamefont {Han},
		\citenamefont {Hong}, \citenamefont {Huang}, \citenamefont {Huo},
		\citenamefont {Li}, \citenamefont {Li}, \citenamefont {Li}, \citenamefont
		{Li}, \citenamefont {Liang}, \citenamefont {Lin}, \citenamefont {Lin},
		\citenamefont {Qian}, \citenamefont {Qiao}, \citenamefont {Rong},
		\citenamefont {Su}, \citenamefont {Sun}, \citenamefont {Wang}, \citenamefont
		{Wang}, \citenamefont {Wu}, \citenamefont {Xu}, \citenamefont {Yan},
		\citenamefont {Yang}, \citenamefont {Yang}, \citenamefont {Ye}, \citenamefont
		{Yin}, \citenamefont {Ying}, \citenamefont {Yu}, \citenamefont {Zha},
		\citenamefont {Zhang}, \citenamefont {Zhang}, \citenamefont {Zhang},
		\citenamefont {Zhang}, \citenamefont {Zhao}, \citenamefont {Zhao},
		\citenamefont {Zhou}, \citenamefont {Zhu}, \citenamefont {Lu}, \citenamefont
		{Peng}, \citenamefont {Zhu},\ and\ \citenamefont {Pan}}]{PRL_Wu2021}%
	\BibitemOpen
	\bibfield  {author} {\bibinfo {author} {\bibfnamefont {Y.}~\bibnamefont
			{Wu}}, \bibinfo {author} {\bibfnamefont {W.-S.}\ \bibnamefont {Bao}},
		\bibinfo {author} {\bibfnamefont {S.}~\bibnamefont {Cao}}, \bibinfo {author}
		{\bibfnamefont {F.}~\bibnamefont {Chen}}, \bibinfo {author} {\bibfnamefont
			{M.-C.}\ \bibnamefont {Chen}}, \bibinfo {author} {\bibfnamefont
			{X.}~\bibnamefont {Chen}}, \bibinfo {author} {\bibfnamefont {T.-H.}\
			\bibnamefont {Chung}}, \bibinfo {author} {\bibfnamefont {H.}~\bibnamefont
			{Deng}}, \bibinfo {author} {\bibfnamefont {Y.}~\bibnamefont {Du}}, \bibinfo
		{author} {\bibfnamefont {D.}~\bibnamefont {Fan}}, \bibinfo {author}
		{\bibfnamefont {M.}~\bibnamefont {Gong}}, \bibinfo {author} {\bibfnamefont
			{C.}~\bibnamefont {Guo}}, \bibinfo {author} {\bibfnamefont {C.}~\bibnamefont
			{Guo}}, \bibinfo {author} {\bibfnamefont {S.}~\bibnamefont {Guo}}, \bibinfo
		{author} {\bibfnamefont {L.}~\bibnamefont {Han}}, \bibinfo {author}
		{\bibfnamefont {L.}~\bibnamefont {Hong}}, \bibinfo {author} {\bibfnamefont
			{H.-L.}\ \bibnamefont {Huang}}, \bibinfo {author} {\bibfnamefont {Y.-H.}\
			\bibnamefont {Huo}}, \bibinfo {author} {\bibfnamefont {L.}~\bibnamefont
			{Li}}, \bibinfo {author} {\bibfnamefont {N.}~\bibnamefont {Li}}, \bibinfo
		{author} {\bibfnamefont {S.}~\bibnamefont {Li}}, \bibinfo {author}
		{\bibfnamefont {Y.}~\bibnamefont {Li}}, \bibinfo {author} {\bibfnamefont
			{F.}~\bibnamefont {Liang}}, \bibinfo {author} {\bibfnamefont
			{C.}~\bibnamefont {Lin}}, \bibinfo {author} {\bibfnamefont {J.}~\bibnamefont
			{Lin}}, \bibinfo {author} {\bibfnamefont {H.}~\bibnamefont {Qian}}, \bibinfo
		{author} {\bibfnamefont {D.}~\bibnamefont {Qiao}}, \bibinfo {author}
		{\bibfnamefont {H.}~\bibnamefont {Rong}}, \bibinfo {author} {\bibfnamefont
			{H.}~\bibnamefont {Su}}, \bibinfo {author} {\bibfnamefont {L.}~\bibnamefont
			{Sun}}, \bibinfo {author} {\bibfnamefont {L.}~\bibnamefont {Wang}}, \bibinfo
		{author} {\bibfnamefont {S.}~\bibnamefont {Wang}}, \bibinfo {author}
		{\bibfnamefont {D.}~\bibnamefont {Wu}}, \bibinfo {author} {\bibfnamefont
			{Y.}~\bibnamefont {Xu}}, \bibinfo {author} {\bibfnamefont {K.}~\bibnamefont
			{Yan}}, \bibinfo {author} {\bibfnamefont {W.}~\bibnamefont {Yang}}, \bibinfo
		{author} {\bibfnamefont {Y.}~\bibnamefont {Yang}}, \bibinfo {author}
		{\bibfnamefont {Y.}~\bibnamefont {Ye}}, \bibinfo {author} {\bibfnamefont
			{J.}~\bibnamefont {Yin}}, \bibinfo {author} {\bibfnamefont {C.}~\bibnamefont
			{Ying}}, \bibinfo {author} {\bibfnamefont {J.}~\bibnamefont {Yu}}, \bibinfo
		{author} {\bibfnamefont {C.}~\bibnamefont {Zha}}, \bibinfo {author}
		{\bibfnamefont {C.}~\bibnamefont {Zhang}}, \bibinfo {author} {\bibfnamefont
			{H.}~\bibnamefont {Zhang}}, \bibinfo {author} {\bibfnamefont
			{K.}~\bibnamefont {Zhang}}, \bibinfo {author} {\bibfnamefont
			{Y.}~\bibnamefont {Zhang}}, \bibinfo {author} {\bibfnamefont
			{H.}~\bibnamefont {Zhao}}, \bibinfo {author} {\bibfnamefont {Y.}~\bibnamefont
			{Zhao}}, \bibinfo {author} {\bibfnamefont {L.}~\bibnamefont {Zhou}}, \bibinfo
		{author} {\bibfnamefont {Q.}~\bibnamefont {Zhu}}, \bibinfo {author}
		{\bibfnamefont {C.-Y.}\ \bibnamefont {Lu}}, \bibinfo {author} {\bibfnamefont
			{C.-Z.}\ \bibnamefont {Peng}}, \bibinfo {author} {\bibfnamefont
			{X.}~\bibnamefont {Zhu}},\ and\ \bibinfo {author} {\bibfnamefont {J.-W.}\
			\bibnamefont {Pan}},\ }\bibfield  {title} {\bibinfo {title} {{Strong Quantum
				Computational Advantage Using a Superconducting Quantum Processor}},\ }\href
	{https://doi.org/10.1103/PhysRevLett.127.180501} {\bibfield  {journal}
		{\bibinfo  {journal} {Physical Review Letters}\ }\textbf {\bibinfo {volume}
			{127}},\ \bibinfo {pages} {180501} (\bibinfo {year} {2021})}\BibitemShut
	{NoStop}%
	\bibitem [{\citenamefont {Yang}\ \emph {et~al.}(2020)\citenamefont {Yang},
		\citenamefont {Liu}, \citenamefont {Zhu}, \citenamefont {Luo},\ and\
		\citenamefont {Cai}}]{PRR_Yang2020}%
	\BibitemOpen
	\bibfield  {author} {\bibinfo {author} {\bibfnamefont {R.-Q.}\ \bibnamefont
			{Yang}}, \bibinfo {author} {\bibfnamefont {H.}~\bibnamefont {Liu}}, \bibinfo
		{author} {\bibfnamefont {S.}~\bibnamefont {Zhu}}, \bibinfo {author}
		{\bibfnamefont {L.}~\bibnamefont {Luo}},\ and\ \bibinfo {author}
		{\bibfnamefont {R.-G.}\ \bibnamefont {Cai}},\ }\bibfield  {title} {\bibinfo
		{title} {{Simulating quantum field theory in curved spacetime with quantum
				many-body systems}},\ }\href
	{https://doi.org/10.1103/PhysRevResearch.2.023107} {\bibfield  {journal}
		{\bibinfo  {journal} {Phys. Rev. Research}\ }\textbf {\bibinfo {volume}
			{2}},\ \bibinfo {pages} {023107} (\bibinfo {year} {2020})}\BibitemShut
	{NoStop}%
	\bibitem [{\citenamefont {Yan}\ \emph {et~al.}(2018)\citenamefont {Yan},
		\citenamefont {Krantz}, \citenamefont {Sung}, \citenamefont {Kjaergaard},
		\citenamefont {Campbell}, \citenamefont {Orlando}, \citenamefont
		{Gustavsson},\ and\ \citenamefont {Oliver}}]{PRApplied_Yan2018}%
	\BibitemOpen
	\bibfield  {author} {\bibinfo {author} {\bibfnamefont {F.}~\bibnamefont
			{Yan}}, \bibinfo {author} {\bibfnamefont {P.}~\bibnamefont {Krantz}},
		\bibinfo {author} {\bibfnamefont {Y.}~\bibnamefont {Sung}}, \bibinfo {author}
		{\bibfnamefont {M.}~\bibnamefont {Kjaergaard}}, \bibinfo {author}
		{\bibfnamefont {D.~L.}\ \bibnamefont {Campbell}}, \bibinfo {author}
		{\bibfnamefont {T.~P.}\ \bibnamefont {Orlando}}, \bibinfo {author}
		{\bibfnamefont {S.}~\bibnamefont {Gustavsson}},\ and\ \bibinfo {author}
		{\bibfnamefont {W.~D.}\ \bibnamefont {Oliver}},\ }\bibfield  {title}
	{\bibinfo {title} {{Tunable Coupling Scheme for Implementing High-Fidelity
				Two-Qubit Gates}},\ }\href {https://doi.org/10.1103/PhysRevApplied.10.054062}
	{\bibfield  {journal} {\bibinfo  {journal} {Phys. Rev. Applied}\ }\textbf
		{\bibinfo {volume} {10}},\ \bibinfo {pages} {054062} (\bibinfo {year}
		{2018})}\BibitemShut {NoStop}%
	\bibitem [{\citenamefont {Sung}\ \emph {et~al.}(2021)\citenamefont {Sung},
		\citenamefont {Ding}, \citenamefont {Braum\"uller}, \citenamefont
		{Veps\"al\"ainen}, \citenamefont {Kannan}, \citenamefont {Kjaergaard},
		\citenamefont {Greene}, \citenamefont {Samach}, \citenamefont {McNally},
		\citenamefont {Kim}, \citenamefont {Melville}, \citenamefont {Niedzielski},
		\citenamefont {Schwartz}, \citenamefont {Yoder}, \citenamefont {Orlando},
		\citenamefont {Gustavsson},\ and\ \citenamefont {Oliver}}]{PRX_Sung2021}%
	\BibitemOpen
	\bibfield  {author} {\bibinfo {author} {\bibfnamefont {Y.}~\bibnamefont
			{Sung}}, \bibinfo {author} {\bibfnamefont {L.}~\bibnamefont {Ding}}, \bibinfo
		{author} {\bibfnamefont {J.}~\bibnamefont {Braum\"uller}}, \bibinfo {author}
		{\bibfnamefont {A.}~\bibnamefont {Veps\"al\"ainen}}, \bibinfo {author}
		{\bibfnamefont {B.}~\bibnamefont {Kannan}}, \bibinfo {author} {\bibfnamefont
			{M.}~\bibnamefont {Kjaergaard}}, \bibinfo {author} {\bibfnamefont
			{A.}~\bibnamefont {Greene}}, \bibinfo {author} {\bibfnamefont {G.~O.}\
			\bibnamefont {Samach}}, \bibinfo {author} {\bibfnamefont {C.}~\bibnamefont
			{McNally}}, \bibinfo {author} {\bibfnamefont {D.}~\bibnamefont {Kim}},
		\bibinfo {author} {\bibfnamefont {A.}~\bibnamefont {Melville}}, \bibinfo
		{author} {\bibfnamefont {B.~M.}\ \bibnamefont {Niedzielski}}, \bibinfo
		{author} {\bibfnamefont {M.~E.}\ \bibnamefont {Schwartz}}, \bibinfo {author}
		{\bibfnamefont {J.~L.}\ \bibnamefont {Yoder}}, \bibinfo {author}
		{\bibfnamefont {T.~P.}\ \bibnamefont {Orlando}}, \bibinfo {author}
		{\bibfnamefont {S.}~\bibnamefont {Gustavsson}},\ and\ \bibinfo {author}
		{\bibfnamefont {W.~D.}\ \bibnamefont {Oliver}},\ }\bibfield  {title}
	{\bibinfo {title} {{Realization of High-Fidelity CZ and $ZZ$-Free iSWAP Gates
				with a Tunable Coupler}},\ }\href
	{https://doi.org/10.1103/PhysRevX.11.021058} {\bibfield  {journal} {\bibinfo
			{journal} {Phys. Rev. X}\ }\textbf {\bibinfo {volume} {11}},\ \bibinfo
		{pages} {021058} (\bibinfo {year} {2021})}\BibitemShut {NoStop}%
	\bibitem [{\citenamefont {Xu}\ \emph {et~al.}(2020)\citenamefont {Xu},
		\citenamefont {Chu}, \citenamefont {Yuan}, \citenamefont {Qiu}, \citenamefont
		{Zhou}, \citenamefont {Zhang}, \citenamefont {Tan}, \citenamefont {Yu},
		\citenamefont {Liu}, \citenamefont {Li}, \citenamefont {Yan},\ and\
		\citenamefont {Yu}}]{PRL_Xu2020}%
	\BibitemOpen
	\bibfield  {author} {\bibinfo {author} {\bibfnamefont {Y.}~\bibnamefont
			{Xu}}, \bibinfo {author} {\bibfnamefont {J.}~\bibnamefont {Chu}}, \bibinfo
		{author} {\bibfnamefont {J.}~\bibnamefont {Yuan}}, \bibinfo {author}
		{\bibfnamefont {J.}~\bibnamefont {Qiu}}, \bibinfo {author} {\bibfnamefont
			{Y.}~\bibnamefont {Zhou}}, \bibinfo {author} {\bibfnamefont {L.}~\bibnamefont
			{Zhang}}, \bibinfo {author} {\bibfnamefont {X.}~\bibnamefont {Tan}}, \bibinfo
		{author} {\bibfnamefont {Y.}~\bibnamefont {Yu}}, \bibinfo {author}
		{\bibfnamefont {S.}~\bibnamefont {Liu}}, \bibinfo {author} {\bibfnamefont
			{J.}~\bibnamefont {Li}}, \bibinfo {author} {\bibfnamefont {F.}~\bibnamefont
			{Yan}},\ and\ \bibinfo {author} {\bibfnamefont {D.}~\bibnamefont {Yu}},\
	}\bibfield  {title} {\bibinfo {title} {{High-Fidelity, High-Scalability
				Two-Qubit Gate Scheme for Superconducting Qubits}},\ }\href
	{https://doi.org/10.1103/PhysRevLett.125.240503} {\bibfield  {journal}
		{\bibinfo  {journal} {Phys. Rev. Lett.}\ }\textbf {\bibinfo {volume} {125}},\
		\bibinfo {pages} {240503} (\bibinfo {year} {2020})}\BibitemShut {NoStop}%
	\bibitem [{\citenamefont {Xu}\ \emph {et~al.}(2021)\citenamefont {Xu},
		\citenamefont {Liu}, \citenamefont {Li}, \citenamefont {Han}, \citenamefont
		{Zhang}, \citenamefont {Linghu}, \citenamefont {Li}, \citenamefont {Chen},
		\citenamefont {Yang}, \citenamefont {Wang}, \citenamefont {Ma}, \citenamefont
		{Xue}, \citenamefont {Jin},\ and\ \citenamefont {Yu}}]{CPB_Xu2021}%
	\BibitemOpen
	\bibfield  {author} {\bibinfo {author} {\bibfnamefont {H.}~\bibnamefont
			{Xu}}, \bibinfo {author} {\bibfnamefont {W.}~\bibnamefont {Liu}}, \bibinfo
		{author} {\bibfnamefont {Z.}~\bibnamefont {Li}}, \bibinfo {author}
		{\bibfnamefont {J.}~\bibnamefont {Han}}, \bibinfo {author} {\bibfnamefont
			{J.}~\bibnamefont {Zhang}}, \bibinfo {author} {\bibfnamefont
			{K.}~\bibnamefont {Linghu}}, \bibinfo {author} {\bibfnamefont
			{Y.}~\bibnamefont {Li}}, \bibinfo {author} {\bibfnamefont {M.}~\bibnamefont
			{Chen}}, \bibinfo {author} {\bibfnamefont {Z.}~\bibnamefont {Yang}}, \bibinfo
		{author} {\bibfnamefont {J.}~\bibnamefont {Wang}}, \bibinfo {author}
		{\bibfnamefont {T.}~\bibnamefont {Ma}}, \bibinfo {author} {\bibfnamefont
			{G.}~\bibnamefont {Xue}}, \bibinfo {author} {\bibfnamefont {Y.}~\bibnamefont
			{Jin}},\ and\ \bibinfo {author} {\bibfnamefont {H.}~\bibnamefont {Yu}},\
	}\bibfield  {title} {\bibinfo {title} {{Realization of adiabatic and diabatic
				CZ gates in superconducting qubits coupled with a tunable coupler}},\ }\href
	{https://doi.org/10.1088/1674-1056/abf03a} {\bibfield  {journal} {\bibinfo
			{journal} {Chinese Physics B}\ }\textbf {\bibinfo {volume} {30}},\ \bibinfo
		{pages} {044212} (\bibinfo {year} {2021})}\BibitemShut {NoStop}%
	\bibitem [{\citenamefont {Collodo}\ \emph {et~al.}(2020)\citenamefont
		{Collodo}, \citenamefont {Herrmann}, \citenamefont {Lacroix}, \citenamefont
		{Andersen}, \citenamefont {Remm}, \citenamefont {Lazar}, \citenamefont
		{Besse}, \citenamefont {Walter}, \citenamefont {Wallraff},\ and\
		\citenamefont {Eichler}}]{PRL_Collodo2020}%
	\BibitemOpen
	\bibfield  {author} {\bibinfo {author} {\bibfnamefont {M.~C.}\ \bibnamefont
			{Collodo}}, \bibinfo {author} {\bibfnamefont {J.}~\bibnamefont {Herrmann}},
		\bibinfo {author} {\bibfnamefont {N.}~\bibnamefont {Lacroix}}, \bibinfo
		{author} {\bibfnamefont {C.~K.}\ \bibnamefont {Andersen}}, \bibinfo {author}
		{\bibfnamefont {A.}~\bibnamefont {Remm}}, \bibinfo {author} {\bibfnamefont
			{S.}~\bibnamefont {Lazar}}, \bibinfo {author} {\bibfnamefont {J.-C.}\
			\bibnamefont {Besse}}, \bibinfo {author} {\bibfnamefont {T.}~\bibnamefont
			{Walter}}, \bibinfo {author} {\bibfnamefont {A.}~\bibnamefont {Wallraff}},\
		and\ \bibinfo {author} {\bibfnamefont {C.}~\bibnamefont {Eichler}},\
	}\bibfield  {title} {\bibinfo {title} {{Implementation of Conditional Phase
				Gates Based on Tunable $ZZ$ Interactions}},\ }\href
	{https://doi.org/10.1103/PhysRevLett.125.240502} {\bibfield  {journal}
		{\bibinfo  {journal} {Phys. Rev. Lett.}\ }\textbf {\bibinfo {volume} {125}},\
		\bibinfo {pages} {240502} (\bibinfo {year} {2020})}\BibitemShut {NoStop}%
	\bibitem [{\citenamefont {Mann}\ \emph {et~al.}(1991)\citenamefont {Mann},
		\citenamefont {Morsink}, \citenamefont {Sikkema},\ and\ \citenamefont
		{Steele}}]{PRD_Mann1991}%
	\BibitemOpen
	\bibfield  {author} {\bibinfo {author} {\bibfnamefont {R.~B.}\ \bibnamefont
			{Mann}}, \bibinfo {author} {\bibfnamefont {S.~M.}\ \bibnamefont {Morsink}},
		\bibinfo {author} {\bibfnamefont {A.~E.}\ \bibnamefont {Sikkema}},\ and\
		\bibinfo {author} {\bibfnamefont {T.~G.}\ \bibnamefont {Steele}},\ }\bibfield
	{title} {\bibinfo {title} {{Semiclassical gravity in 1+1 dimensions}},\
	}\href {https://doi.org/10.1103/PhysRevD.43.3948} {\bibfield  {journal}
		{\bibinfo  {journal} {Phys. Rev. D}\ }\textbf {\bibinfo {volume} {43}},\
		\bibinfo {pages} {3948} (\bibinfo {year} {1991})}\BibitemShut {NoStop}%
	\bibitem [{\citenamefont {Pedernales}\ \emph {et~al.}(2018)\citenamefont
		{Pedernales}, \citenamefont {Beau}, \citenamefont {Pittman}, \citenamefont
		{Egusquiza}, \citenamefont {Lamata}, \citenamefont {Solano},\ and\
		\citenamefont {del Campo}}]{PRL_Pedernales2018}%
	\BibitemOpen
	\bibfield  {author} {\bibinfo {author} {\bibfnamefont {J.~S.}\ \bibnamefont
			{Pedernales}}, \bibinfo {author} {\bibfnamefont {M.}~\bibnamefont {Beau}},
		\bibinfo {author} {\bibfnamefont {S.~M.}\ \bibnamefont {Pittman}}, \bibinfo
		{author} {\bibfnamefont {I.~L.}\ \bibnamefont {Egusquiza}}, \bibinfo {author}
		{\bibfnamefont {L.}~\bibnamefont {Lamata}}, \bibinfo {author} {\bibfnamefont
			{E.}~\bibnamefont {Solano}},\ and\ \bibinfo {author} {\bibfnamefont
			{A.}~\bibnamefont {del Campo}},\ }\bibfield  {title} {\bibinfo {title}
		{{Dirac Equation in ($1+1$)-Dimensional Curved Spacetime and the Multiphoton
				Quantum Rabi Model}},\ }\href
	{https://doi.org/10.1103/PhysRevLett.120.160403} {\bibfield  {journal}
		{\bibinfo  {journal} {Phys. Rev. Lett.}\ }\textbf {\bibinfo {volume} {120}},\
		\bibinfo {pages} {160403} (\bibinfo {year} {2018})}\BibitemShut {NoStop}%
	\bibitem [{\citenamefont {Childs}(2009)}]{PRL_Childs2009}%
	\BibitemOpen
	\bibfield  {author} {\bibinfo {author} {\bibfnamefont {A.~M.}\ \bibnamefont
			{Childs}},\ }\bibfield  {title} {\bibinfo {title} {{Universal Computation by
				Quantum Walk}},\ }\href {https://doi.org/10.1103/PhysRevLett.102.180501}
	{\bibfield  {journal} {\bibinfo  {journal} {Physical Review Letters}\
		}\textbf {\bibinfo {volume} {102}},\ \bibinfo {pages} {180501} (\bibinfo
		{year} {2009})}\BibitemShut {NoStop}%
	\bibitem [{\citenamefont {Karski}\ \emph {et~al.}(2009)\citenamefont {Karski},
		\citenamefont {F{\"{o}}rster}, \citenamefont {Choi}, \citenamefont {Steffen},
		\citenamefont {Alt}, \citenamefont {Meschede},\ and\ \citenamefont
		{Widera}}]{Science_Karski2009}%
	\BibitemOpen
	\bibfield  {author} {\bibinfo {author} {\bibfnamefont {M.}~\bibnamefont
			{Karski}}, \bibinfo {author} {\bibfnamefont {L.}~\bibnamefont
			{F{\"{o}}rster}}, \bibinfo {author} {\bibfnamefont {J.-M.}\ \bibnamefont
			{Choi}}, \bibinfo {author} {\bibfnamefont {A.}~\bibnamefont {Steffen}},
		\bibinfo {author} {\bibfnamefont {W.}~\bibnamefont {Alt}}, \bibinfo {author}
		{\bibfnamefont {D.}~\bibnamefont {Meschede}},\ and\ \bibinfo {author}
		{\bibfnamefont {A.}~\bibnamefont {Widera}},\ }\bibfield  {title} {\bibinfo
		{title} {{Quantum Walk in Position Space with Single Optically Trapped
				Atoms}},\ }\href {https://doi.org/10.1126/science.1174436} {\bibfield
		{journal} {\bibinfo  {journal} {Science}\ }\textbf {\bibinfo {volume}
			{325}},\ \bibinfo {pages} {174} (\bibinfo {year} {2009})}\BibitemShut
	{NoStop}%
	\bibitem [{\citenamefont {Yan}\ \emph {et~al.}(2019)\citenamefont {Yan},
		\citenamefont {Zhang}, \citenamefont {Gong}, \citenamefont {Wu},
		\citenamefont {Zheng}, \citenamefont {Li}, \citenamefont {Wang},
		\citenamefont {Liang}, \citenamefont {Lin}, \citenamefont {Xu}, \citenamefont
		{Guo}, \citenamefont {Sun}, \citenamefont {Peng}, \citenamefont {Xia},
		\citenamefont {Deng}, \citenamefont {Rong}, \citenamefont {You},
		\citenamefont {Nori}, \citenamefont {Fan}, \citenamefont {Zhu},\ and\
		\citenamefont {Pan}}]{Science_Yan2019}%
	\BibitemOpen
	\bibfield  {author} {\bibinfo {author} {\bibfnamefont {Z.}~\bibnamefont
			{Yan}}, \bibinfo {author} {\bibfnamefont {Y.-R.}\ \bibnamefont {Zhang}},
		\bibinfo {author} {\bibfnamefont {M.}~\bibnamefont {Gong}}, \bibinfo {author}
		{\bibfnamefont {Y.}~\bibnamefont {Wu}}, \bibinfo {author} {\bibfnamefont
			{Y.}~\bibnamefont {Zheng}}, \bibinfo {author} {\bibfnamefont
			{S.}~\bibnamefont {Li}}, \bibinfo {author} {\bibfnamefont {C.}~\bibnamefont
			{Wang}}, \bibinfo {author} {\bibfnamefont {F.}~\bibnamefont {Liang}},
		\bibinfo {author} {\bibfnamefont {J.}~\bibnamefont {Lin}}, \bibinfo {author}
		{\bibfnamefont {Y.}~\bibnamefont {Xu}}, \bibinfo {author} {\bibfnamefont
			{C.}~\bibnamefont {Guo}}, \bibinfo {author} {\bibfnamefont {L.}~\bibnamefont
			{Sun}}, \bibinfo {author} {\bibfnamefont {C.-Z.}\ \bibnamefont {Peng}},
		\bibinfo {author} {\bibfnamefont {K.}~\bibnamefont {Xia}}, \bibinfo {author}
		{\bibfnamefont {H.}~\bibnamefont {Deng}}, \bibinfo {author} {\bibfnamefont
			{H.}~\bibnamefont {Rong}}, \bibinfo {author} {\bibfnamefont {J.~Q.}\
			\bibnamefont {You}}, \bibinfo {author} {\bibfnamefont {F.}~\bibnamefont
			{Nori}}, \bibinfo {author} {\bibfnamefont {H.}~\bibnamefont {Fan}}, \bibinfo
		{author} {\bibfnamefont {X.}~\bibnamefont {Zhu}},\ and\ \bibinfo {author}
		{\bibfnamefont {J.-W.}\ \bibnamefont {Pan}},\ }\bibfield  {title} {\bibinfo
		{title} {{Strongly correlated quantum walks with a 12-qubit superconducting
				processor}},\ }\href {https://doi.org/10.1126/science.aaw1611} {\bibfield
		{journal} {\bibinfo  {journal} {Science}\ }\textbf {\bibinfo {volume}
			{364}},\ \bibinfo {pages} {753} (\bibinfo {year} {2019})}\BibitemShut
	{NoStop}%
	\bibitem [{\citenamefont {Gong}\ \emph {et~al.}(2021)\citenamefont {Gong},
		\citenamefont {Wang}, \citenamefont {Zha}, \citenamefont {Chen},
		\citenamefont {Huang}, \citenamefont {Wu}, \citenamefont {Zhu}, \citenamefont
		{Zhao}, \citenamefont {Li}, \citenamefont {Guo}, \citenamefont {Qian},
		\citenamefont {Ye}, \citenamefont {Chen}, \citenamefont {Ying}, \citenamefont
		{Yu}, \citenamefont {Fan}, \citenamefont {Wu}, \citenamefont {Su},
		\citenamefont {Deng}, \citenamefont {Rong}, \citenamefont {Zhang},
		\citenamefont {Cao}, \citenamefont {Lin}, \citenamefont {Xu}, \citenamefont
		{Sun}, \citenamefont {Guo}, \citenamefont {Li}, \citenamefont {Liang},
		\citenamefont {Bastidas}, \citenamefont {Nemoto}, \citenamefont {Munro},
		\citenamefont {Huo}, \citenamefont {Lu}, \citenamefont {Peng}, \citenamefont
		{Zhu},\ and\ \citenamefont {Pan}}]{Science_Gong2021}%
	\BibitemOpen
	\bibfield  {author} {\bibinfo {author} {\bibfnamefont {M.}~\bibnamefont
			{Gong}}, \bibinfo {author} {\bibfnamefont {S.}~\bibnamefont {Wang}}, \bibinfo
		{author} {\bibfnamefont {C.}~\bibnamefont {Zha}}, \bibinfo {author}
		{\bibfnamefont {M.-C.}\ \bibnamefont {Chen}}, \bibinfo {author}
		{\bibfnamefont {H.-L.}\ \bibnamefont {Huang}}, \bibinfo {author}
		{\bibfnamefont {Y.}~\bibnamefont {Wu}}, \bibinfo {author} {\bibfnamefont
			{Q.}~\bibnamefont {Zhu}}, \bibinfo {author} {\bibfnamefont {Y.}~\bibnamefont
			{Zhao}}, \bibinfo {author} {\bibfnamefont {S.}~\bibnamefont {Li}}, \bibinfo
		{author} {\bibfnamefont {S.}~\bibnamefont {Guo}}, \bibinfo {author}
		{\bibfnamefont {H.}~\bibnamefont {Qian}}, \bibinfo {author} {\bibfnamefont
			{Y.}~\bibnamefont {Ye}}, \bibinfo {author} {\bibfnamefont {F.}~\bibnamefont
			{Chen}}, \bibinfo {author} {\bibfnamefont {C.}~\bibnamefont {Ying}}, \bibinfo
		{author} {\bibfnamefont {J.}~\bibnamefont {Yu}}, \bibinfo {author}
		{\bibfnamefont {D.}~\bibnamefont {Fan}}, \bibinfo {author} {\bibfnamefont
			{D.}~\bibnamefont {Wu}}, \bibinfo {author} {\bibfnamefont {H.}~\bibnamefont
			{Su}}, \bibinfo {author} {\bibfnamefont {H.}~\bibnamefont {Deng}}, \bibinfo
		{author} {\bibfnamefont {H.}~\bibnamefont {Rong}}, \bibinfo {author}
		{\bibfnamefont {K.}~\bibnamefont {Zhang}}, \bibinfo {author} {\bibfnamefont
			{S.}~\bibnamefont {Cao}}, \bibinfo {author} {\bibfnamefont {J.}~\bibnamefont
			{Lin}}, \bibinfo {author} {\bibfnamefont {Y.}~\bibnamefont {Xu}}, \bibinfo
		{author} {\bibfnamefont {L.}~\bibnamefont {Sun}}, \bibinfo {author}
		{\bibfnamefont {C.}~\bibnamefont {Guo}}, \bibinfo {author} {\bibfnamefont
			{N.}~\bibnamefont {Li}}, \bibinfo {author} {\bibfnamefont {F.}~\bibnamefont
			{Liang}}, \bibinfo {author} {\bibfnamefont {V.~M.}\ \bibnamefont {Bastidas}},
		\bibinfo {author} {\bibfnamefont {K.}~\bibnamefont {Nemoto}}, \bibinfo
		{author} {\bibfnamefont {W.~J.}\ \bibnamefont {Munro}}, \bibinfo {author}
		{\bibfnamefont {Y.-H.}\ \bibnamefont {Huo}}, \bibinfo {author} {\bibfnamefont
			{C.-Y.}\ \bibnamefont {Lu}}, \bibinfo {author} {\bibfnamefont {C.-Z.}\
			\bibnamefont {Peng}}, \bibinfo {author} {\bibfnamefont {X.}~\bibnamefont
			{Zhu}},\ and\ \bibinfo {author} {\bibfnamefont {J.-W.}\ \bibnamefont {Pan}},\
	}\bibfield  {title} {\bibinfo {title} {{Quantum walks on a programmable
				two-dimensional 62-qubit superconducting processor}},\ }\href
	{https://doi.org/10.1126/science.abg7812} {\bibfield  {journal} {\bibinfo
			{journal} {Science}\ }\textbf {\bibinfo {volume} {372}},\ \bibinfo {pages}
		{948} (\bibinfo {year} {2021})}\BibitemShut {NoStop}%
	\bibitem [{\citenamefont {Corley}\ and\ \citenamefont
		{Jacobson}(1998)}]{PRD_Corley1997}%
	\BibitemOpen
	\bibfield  {author} {\bibinfo {author} {\bibfnamefont {S.}~\bibnamefont
			{Corley}}\ and\ \bibinfo {author} {\bibfnamefont {T.}~\bibnamefont
			{Jacobson}},\ }\bibfield  {title} {\bibinfo {title} {Lattice black holes},\
	}\href {https://doi.org/10.1103/PhysRevD.57.6269} {\bibfield  {journal}
		{\bibinfo  {journal} {Phys. Rev. D}\ }\textbf {\bibinfo {volume} {57}},\
		\bibinfo {pages} {6269} (\bibinfo {year} {1998})}\BibitemShut {NoStop}%
	\bibitem [{\citenamefont {Jacobson}\ and\ \citenamefont
		{Mattingly}(1999)}]{PRD_Jacobson1999}%
	\BibitemOpen
	\bibfield  {author} {\bibinfo {author} {\bibfnamefont {T.}~\bibnamefont
			{Jacobson}}\ and\ \bibinfo {author} {\bibfnamefont {D.}~\bibnamefont
			{Mattingly}},\ }\bibfield  {title} {\bibinfo {title} {Hawking radiation on a
			falling lattice},\ }\href {https://doi.org/10.1103/PhysRevD.61.024017}
	{\bibfield  {journal} {\bibinfo  {journal} {Phys. Rev. D}\ }\textbf {\bibinfo
			{volume} {61}},\ \bibinfo {pages} {024017} (\bibinfo {year}
		{1999})}\BibitemShut {NoStop}%
	\bibitem [{\citenamefont {Brout}\ \emph {et~al.}(1995)\citenamefont {Brout},
		\citenamefont {Massar}, \citenamefont {Parentani},\ and\ \citenamefont
		{Spindel}}]{PRD_Parentani1995}%
	\BibitemOpen
	\bibfield  {author} {\bibinfo {author} {\bibfnamefont {R.}~\bibnamefont
			{Brout}}, \bibinfo {author} {\bibfnamefont {S.}~\bibnamefont {Massar}},
		\bibinfo {author} {\bibfnamefont {R.}~\bibnamefont {Parentani}},\ and\
		\bibinfo {author} {\bibfnamefont {P.}~\bibnamefont {Spindel}},\ }\bibfield
	{title} {\bibinfo {title} {{Hawking radiation without trans-Planckian
				frequencies}},\ }\href {https://doi.org/10.1103/PhysRevD.52.4559} {\bibfield
		{journal} {\bibinfo  {journal} {Phys. Rev. D}\ }\textbf {\bibinfo {volume}
			{52}},\ \bibinfo {pages} {4559} (\bibinfo {year} {1995})}\BibitemShut
	{NoStop}%
	\bibitem [{\citenamefont {Damour}\ and\ \citenamefont
		{Ruffini}(1976)}]{PRD_Damour1976}%
	\BibitemOpen
	\bibfield  {author} {\bibinfo {author} {\bibfnamefont {T.}~\bibnamefont
			{Damour}}\ and\ \bibinfo {author} {\bibfnamefont {R.}~\bibnamefont
			{Ruffini}},\ }\bibfield  {title} {\bibinfo {title} {{Black-hole evaporation
				in the Klein-Sauter-Heisenberg-Euler formalism}},\ }\href
	{https://doi.org/10.1103/PhysRevD.14.332} {\bibfield  {journal} {\bibinfo
			{journal} {Phys. Rev. D}\ }\textbf {\bibinfo {volume} {14}},\ \bibinfo
		{pages} {332} (\bibinfo {year} {1976})}\BibitemShut {NoStop}%
	\bibitem [{\citenamefont {Parikh}\ and\ \citenamefont
		{Wilczek}(2000)}]{PRL_Parikh2000}%
	\BibitemOpen
	\bibfield  {author} {\bibinfo {author} {\bibfnamefont {M.~K.}\ \bibnamefont
			{Parikh}}\ and\ \bibinfo {author} {\bibfnamefont {F.}~\bibnamefont
			{Wilczek}},\ }\bibfield  {title} {\bibinfo {title} {{Hawking Radiation As
				Tunneling}},\ }\href {https://doi.org/10.1103/PhysRevLett.85.5042} {\bibfield
		{journal} {\bibinfo  {journal} {Phys. Rev. Lett.}\ }\textbf {\bibinfo
			{volume} {85}},\ \bibinfo {pages} {5042} (\bibinfo {year}
		{2000})}\BibitemShut {NoStop}%
	\bibitem [{\citenamefont {Arzano}\ \emph {et~al.}(2005)\citenamefont {Arzano},
		\citenamefont {Medved},\ and\ \citenamefont {Vagenas}}]{JHEP_Arzano2005}%
	\BibitemOpen
	\bibfield  {author} {\bibinfo {author} {\bibfnamefont {M.}~\bibnamefont
			{Arzano}}, \bibinfo {author} {\bibfnamefont {A.~J.~M.}\ \bibnamefont
			{Medved}},\ and\ \bibinfo {author} {\bibfnamefont {E.~C.}\ \bibnamefont
			{Vagenas}},\ }\bibfield  {title} {\bibinfo {title} {{Hawking radiation as
				tunneling through the quantum horizon}},\ }\href
	{https://doi.org/10.1088/1126-6708/2005/09/037} {\bibfield  {journal}
		{\bibinfo  {journal} {Journal of High Energy Physics}\ }\textbf {\bibinfo
			{volume} {9}},\ \bibinfo {pages} {037} (\bibinfo {year} {2005})}\BibitemShut
	{NoStop}%
	\bibitem [{\citenamefont {Nation}\ \emph
		{et~al.}(2012{\natexlab{a}})\citenamefont {Nation}, \citenamefont
		{Johansson}, \citenamefont {Blencowe},\ and\ \citenamefont
		{Nori}}]{RMP_Nori2012}%
	\BibitemOpen
	\bibfield  {author} {\bibinfo {author} {\bibfnamefont {P.~D.}\ \bibnamefont
			{Nation}}, \bibinfo {author} {\bibfnamefont {J.~R.}\ \bibnamefont
			{Johansson}}, \bibinfo {author} {\bibfnamefont {M.~P.}\ \bibnamefont
			{Blencowe}},\ and\ \bibinfo {author} {\bibfnamefont {F.}~\bibnamefont
			{Nori}},\ }\bibfield  {title} {\bibinfo {title} {Colloquium: Stimulating
			uncertainty: Amplifying the quantum vacuum with superconducting circuits},\
	}\href {https://doi.org/10.1103/RevModPhys.84.1} {\bibfield  {journal}
		{\bibinfo  {journal} {Rev. Mod. Phys.}\ }\textbf {\bibinfo {volume} {84}},\
		\bibinfo {pages} {1} (\bibinfo {year} {2012}{\natexlab{a}})}\BibitemShut
	{NoStop}%
	\bibitem [{\citenamefont {Nation}\ \emph
		{et~al.}(2012{\natexlab{b}})\citenamefont {Nation}, \citenamefont
		{Blencowe},\ and\ \citenamefont {Nori}}]{NJP_Nori2012}%
	\BibitemOpen
	\bibfield  {author} {\bibinfo {author} {\bibfnamefont {P.~D.}\ \bibnamefont
			{Nation}}, \bibinfo {author} {\bibfnamefont {M.~P.}\ \bibnamefont
			{Blencowe}},\ and\ \bibinfo {author} {\bibfnamefont {F.}~\bibnamefont
			{Nori}},\ }\bibfield  {title} {\bibinfo {title} {Non-equilibrium landauer
			transport model for hawking radiation from a black hole},\ }\href
	{https://doi.org/10.1088/1367-2630/14/3/033013} {\bibfield  {journal}
		{\bibinfo  {journal} {New Journal of Physics}\ }\textbf {\bibinfo {volume}
			{14}},\ \bibinfo {pages} {033013} (\bibinfo {year}
		{2012}{\natexlab{b}})}\BibitemShut {NoStop}%
	\bibitem [{\citenamefont {Giovanazzi}(2005)}]{PRL_Giovanazzi2005}%
	\BibitemOpen
	\bibfield  {author} {\bibinfo {author} {\bibfnamefont {S.}~\bibnamefont
			{Giovanazzi}},\ }\bibfield  {title} {\bibinfo {title} {Hawking radiation in
			sonic black holes},\ }\href {https://doi.org/10.1103/PhysRevLett.94.061302}
	{\bibfield  {journal} {\bibinfo  {journal} {Phys. Rev. Lett.}\ }\textbf
		{\bibinfo {volume} {94}},\ \bibinfo {pages} {061302} (\bibinfo {year}
		{2005})}\BibitemShut {NoStop}%
	\bibitem [{\citenamefont {Maldacena}(2020)}]{NRP_Maldacena2020}%
	\BibitemOpen
	\bibfield  {author} {\bibinfo {author} {\bibfnamefont {J.}~\bibnamefont
			{Maldacena}},\ }\bibfield  {title} {\bibinfo {title} {{Black holes and
				quantum information}},\ }\href {https://doi.org/10.1038/s42254-019-0146-z}
	{\bibfield  {journal} {\bibinfo  {journal} {Nature Reviews Physics}\ }\textbf
		{\bibinfo {volume} {2}},\ \bibinfo {pages} {123} (\bibinfo {year}
		{2020})}\BibitemShut {NoStop}%
	\bibitem [{\citenamefont {Daley}\ \emph {et~al.}(2022)\citenamefont {Daley},
		\citenamefont {Bloch}, \citenamefont {Kokail}, \citenamefont {Flannigan},
		\citenamefont {Pearson}, \citenamefont {Troyer},\ and\ \citenamefont
		{Zoller}}]{Nature_Daley2022}%
	\BibitemOpen
	\bibfield  {author} {\bibinfo {author} {\bibfnamefont {A.~J.}\ \bibnamefont
			{Daley}}, \bibinfo {author} {\bibfnamefont {I.}~\bibnamefont {Bloch}},
		\bibinfo {author} {\bibfnamefont {C.}~\bibnamefont {Kokail}}, \bibinfo
		{author} {\bibfnamefont {S.}~\bibnamefont {Flannigan}}, \bibinfo {author}
		{\bibfnamefont {N.}~\bibnamefont {Pearson}}, \bibinfo {author} {\bibfnamefont
			{M.}~\bibnamefont {Troyer}},\ and\ \bibinfo {author} {\bibfnamefont
			{P.}~\bibnamefont {Zoller}},\ }\bibfield  {title} {\bibinfo {title}
		{{Practical quantum advantage in quantum simulation}},\ }\href
	{https://doi.org/10.1038/s41586-022-04940-6} {\bibfield  {journal} {\bibinfo
			{journal} {Nature}\ }\textbf {\bibinfo {volume} {607}},\ \bibinfo {pages}
		{667} (\bibinfo {year} {2022})}\BibitemShut {NoStop}%
	\bibitem [{\citenamefont {Wootters}(1998)}]{PRL_Wootters1998}%
	\BibitemOpen
	\bibfield  {author} {\bibinfo {author} {\bibfnamefont {W.~K.}\ \bibnamefont
			{Wootters}},\ }\bibfield  {title} {\bibinfo {title} {{Entanglement of
				Formation of an Arbitrary State of Two Qubits}},\ }\href
	{https://doi.org/10.1103/PhysRevLett.80.2245} {\bibfield  {journal} {\bibinfo
			{journal} {Phys. Rev. Lett.}\ }\textbf {\bibinfo {volume} {80}},\ \bibinfo
		{pages} {2245} (\bibinfo {year} {1998})}\BibitemShut {NoStop}%
\end{thebibliography}
\providecommand{\noopsort}[1]{}\providecommand{\singleletter}[1]{#1}%

\noindent
\textbf{\large{Acknowledgements}}
\\
This work was supported by the Synergetic Extreme Condition User Facility (SECUF). Devices were made at the Nanofabrication Facilities at Institute of Physics in Beijing. Z.X., D.Z., K.X. and H.F. are supported by the National Natural Science Foundation of China (Grant Nos. 92265207, 92065114, T2121001, 11934018 and 11904393), Innovation Program for Quantum Science and Technology (No. 2-6), the Key-Area Research and Development Program of Guangdong Province, China (Grant No. 2020B0303030001), the Strategic Priority Research Program of Chinese Academy of Sciences (Grant No. XDB28000000), Scientific Instrument Developing Project of Chinese Academy of Sciences (Grant No. YJKYYQ20200041), and Beijing Natural Science Foundation (Grant No. Z200009). R.-Q.Y. acknowledges the support of the National Natural Science Foundation of China (Grant No. 12005155).
\\
\\
\textbf{\large{Author contributions}}
\\
Y.-H.S. performed the experiment with the assistance of K.X.; R.-G.C. and H.F. conceived the idea; R.-Q.Y. and R.-G.C. provided theoretical support; K.X., Y.-H.S., R.-Q.Y., Z.-Y.G. and Y.-Y.W. performed the numerical simulation; Z.X. fabricated the device with the help of D.Z. and X.S.; Y.-H.S., H.L. and K.H. helped the experimental setup supervised by K.X. with the assistance of Y.T.; H.F. supervised the whole project; Y.-H.S., R.-Q.Y., Y.-Y.W., Z.-Y.G., R.-G.C. and H.F. cowrote the manuscript, and all authors contributed to the discussions of the results and and development of the manuscript.
\\
\\
\textbf{\large{Competing interests}}
\\
The authors declare no competing interests.

\end{document}



\title{Supplementary Information for ``Quantum simulation of Hawking radiation and curved spacetime with a superconducting on-chip black hole''}

\author{Yun-Hao Shi}
\thanks{These authors contributed equally to this work.}
\affiliation{Institute of Physics, Chinese Academy of Sciences, Beijing 100190, China}
\affiliation{School of Physical Sciences, University of Chinese Academy of Sciences, Beijing 100049, China}

\author{Run-Qiu Yang}
\thanks{These authors contributed equally to this work.}
\affiliation{Center for Joint Quantum Studies and Department of Physics, School of Science, Tianjin University, Tianjin 300350, China}%

\author{Zhongcheng Xiang}
\thanks{These authors contributed equally to this work.}
\affiliation{Institute of Physics, Chinese Academy of Sciences, Beijing 100190, China}

\author{Zi-Yong Ge}
\affiliation{Theoretical Quantum Physics Laboratory, RIKEN Cluster for Pioneering Research, Wako-shi, Saitama 351-0198, Japan}

\author{Hao Li}
\affiliation{Institute of Physics, Chinese Academy of Sciences, Beijing 100190, China}
\affiliation{School of Physics, Northwest University, Xi'an 710127, China}

\author{Yong-Yi Wang}
\affiliation{Institute of Physics, Chinese Academy of Sciences, Beijing 100190, China}
\affiliation{School of Physical Sciences, University of Chinese Academy of Sciences, Beijing 100049, China}

\author{Kaixuan Huang}
\affiliation{Beijing Academy of Quantum Information Sciences, Beijing 100193, China}%

\author{Ye Tian}
\affiliation{Institute of Physics, Chinese Academy of Sciences, Beijing 100190, China}

\author{Xiaohui Song}
\affiliation{Institute of Physics, Chinese Academy of Sciences, Beijing 100190, China}

\author{Dongning Zheng}
\email{dzheng@iphy.ac.cn}
\affiliation{Institute of Physics, Chinese Academy of Sciences, Beijing 100190, China}
\affiliation{School of Physical Sciences, University of Chinese Academy of Sciences, Beijing 100049, China}
\affiliation{Songshan Lake Materials Laboratory, Dongguan, Guangdong 523808, China}

\author{Kai Xu}
\email{kaixu@iphy.ac.cn}
\affiliation{Institute of Physics, Chinese Academy of Sciences, Beijing 100190, China}
\affiliation{School of Physical Sciences, University of Chinese Academy of Sciences, Beijing 100049, China}
\affiliation{Beijing Academy of Quantum Information Sciences, Beijing 100193, China}
\affiliation{Songshan Lake Materials Laboratory, Dongguan, Guangdong 523808, China}
\affiliation{CAS Center for Excellence in Topological Quantum Computation, University of Chinese Academy of Sciences, Beijing 100049, China}%

\author{Rong-Gen Cai}
\email{cairg@itp.ac.cn}
\affiliation{CAS Key Laboratory of Theoretical Physics, Institute of Theoretical Physics, Chinese Academy of Sciences, Beijing 100190, China}

\author{Heng Fan}
\email{hfan@iphy.ac.cn}
\affiliation{Institute of Physics, Chinese Academy of Sciences, Beijing 100190, China}
\affiliation{School of Physical Sciences, University of Chinese Academy of Sciences, Beijing 100049, China}
\affiliation{Beijing Academy of Quantum Information Sciences, Beijing 100193, China}
\affiliation{Songshan Lake Materials Laboratory, Dongguan, Guangdong 523808, China}
\affiliation{CAS Center for Excellence in Topological Quantum Computation, University of Chinese Academy of Sciences, Beijing 100049, China}%
\affiliation{Hefei National Laboratory, Hefei 230088, China}

\maketitle

\beginsupplement
\section{Introduction on Eddington-Finkelstein coordinates}\label{intro-bh}
In this paper, the spacetime geometry is present by advanced Eddington-Finkelstein coordinates (AEFC) $\{t,x\}$. Though the coordinate ``$t$'' plays the role of ``time'' in this system, there are a few differences compared with the usual time coordinate. In this appendix, we will give a basic introduction.

One simple way to obtain an intuition of advanced Eddington-Finkelstein coordinates $\{t,x\}$ is to consider the wave propagating in flat spacetime. Let us consider a Minkowski spacetime. The usual Minkowski coordinates (MC) is $\{t_{\mathrm{m}},x\}$, of which the metric reads
\begin{equation}\label{minkwg}
	\mathrm{d} s^2=\mathrm{d} t_{\mathrm{m}}^2-\mathrm{d} x^2
\end{equation}
The massless scalar field then will satisfy
\begin{equation}\label{wavescalar1}
	\partial_{t_{\mathrm{m}}}^2\phi-\partial_x^2\phi=0
\end{equation}
The solution of this equation, in general, is given by following ``traveling wave solution'',
\begin{equation}\label{soluphi1}
	\phi(t_{\mathrm{m}},x)=\phi_1(t_{\mathrm{m}}+x)+\phi_2(t_{\mathrm{m}}-x)\,,
\end{equation}
where $\phi_1(t_{\mathrm{m}}+x)$ stands for the advanced solution  and $\phi_2(t_{\mathrm{m}}-x)$ stands for the outgoing solution.

To covert the advanced Eddington-Finkelstein coordinates $\{t,x\}$, we consider a coordinates transformation
\begin{equation}\label{coorditrans1}
	t=t_{\mathrm{m}}+x\,,
\end{equation}
and so the metric then reads
\begin{equation}\label{minkwg2}
	\mathrm{d} s^2=\mathrm{d} t^2-2\mathrm{d} t\mathrm{d} x\,.
\end{equation}
It needs to note that, though $\mathrm{d} t_{\mathrm{m}}\neq\mathrm{d} t$, we still have
\begin{equation}\label{dphidv}
	\partial_{t_{\mathrm{m}}}h|_{\text{MC}}=\partial_th|_{\text{AEFC}}\,,
\end{equation}
for an arbitrary function $h$, i.e. the time derivatives are the same as that in the usual Minkowski coordinates and advanced Eddington-Finkelstein coordinates. Thus, the growth rate of a quantity in usual Minkowski coordinates can also be computed according to the time derivative of advanced Eddington-Finkelstein coordinates. On the contrary, if one wants to compute the spatial derivative, then the two coordinate systems will have different results
\begin{equation}\label{dphidv2}
	\partial_xh|_{\text{MC}}\neq\partial_xh|_{\text{AEFC}}\,
\end{equation}
in general since the derivative in left-side fixes $t_m$ but the derivative of right-side fixes $t=t_m+x$.

The propagators of wave are also very different in these two coordinate systems. From Eq.~\eqref{soluphi1}, one sees that the ``traveling wave solution'' in advanced Eddington-Finkelstein coordinates reads
\begin{equation}\label{soluphi2}
	\phi(t,x)=\phi_1(t)+\phi_2(t-2x)\,,
\end{equation}
This can also be obtained from the wave equation $2\partial_t\partial_x\phi+2\partial_x^2\phi=0$. It is a little surprising that the advanced wave $\phi_1$ has no propagator! In fact \textit{the infalling mode now becomes a boundary condition rather than a propagator}. If we impose a boundary condition
\begin{equation}\label{bd1}
	\phi(t,\pm\infty)=0\,,
\end{equation}
then we have $\phi_1(t)=0$ and there is only outgoing mode. Thus, the advanced Eddington-Finkelstein coordinates with boundary condition~\eqref{bd1} can only represent the propagator of outgoing modes. In other words, the advanced Eddington-Finkelstein coordinates play the role of a selector to choose only outgoing modes.

Though we assume that the spacetime is flat and the matter is a scalar field in the above discussion, the basic physical picture will still be true if we consider a curved 2-dimensional spacetime and Dirac field. In Fig. 2 of our main text, one can see that our model only simulates the outgoing modes, just as we expected in the above discussion. In general, the wave in gravitational fields contains both advanced modes and outgoing modes. The Hawking radiation is an energy flux towards infinity, i.e. carried by outgoing modes. This is one reason why this paper uses advanced Eddington-Finkelstein coordinates to study Hawking radiation.

\section{More explanation on the tunneling picture of Hawking radiation}\label{hawrad}
\begin{figure}
	\includegraphics[width=0.38\textwidth]{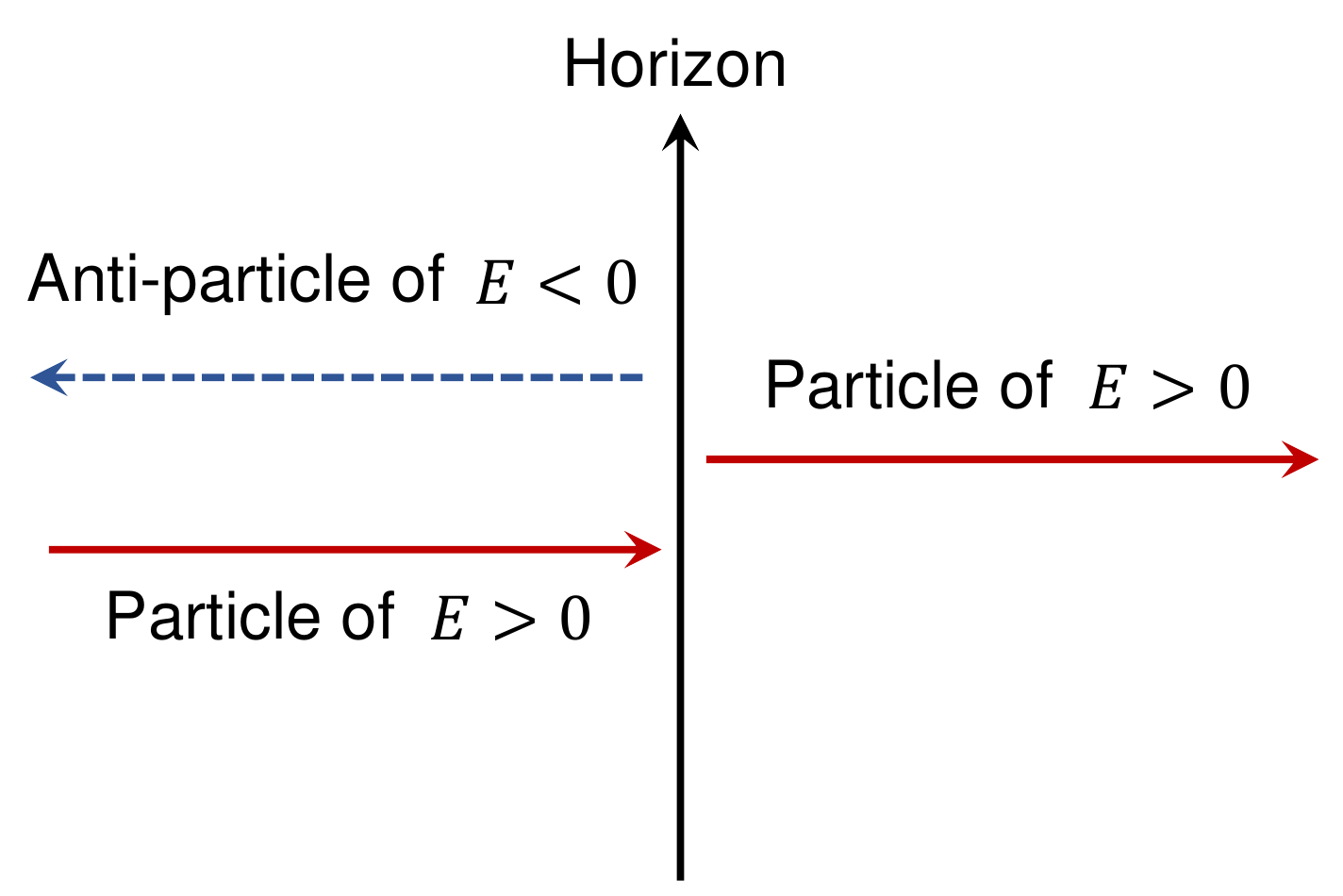}
	\caption{The anti-particle flow of negative energy infalling toward the interior of the black hole can always be interpreted as a particle flow of positive energy outgoing from the interior. }\label{fig:particle1}
\end{figure}
This paper uses the picture of ``quantum tunneling'' to understand the Hawking radiation~\cite{PRD_Damour1976,PRL_Parikh2000,JHEP_Arzano2005}. Though this is also a widespread picture to understand Hawking radiation in the community of black hole physics, it may not be familiar to other readers. Here we make a brief introduction to this picture.

At first glance, the tunneling picture is very different from the picture of ``pair creation'' outside the horizon. However, they are equivalent in physics. Based on the picture of ``pair creation'' in Hawking radiation, ``particle-antiparticle pairs'' can be created around the horizon. The antiparticle (negative energy) falls into the black hole and annihilates with positive energy particle inside the black hole, the particle outside the horizon is materialized and escapes into infinity. Note that the pair creation/annihilation is a virtual process, and the really materialized result is that the original particle inside the black hole disappears but an identical particle appears outside the horizon. The anti-particle of negative energy infalling the interior of the black hole can always be interpreted as a particle of positive energy outgoing from the interior, see schematic diagram Fig.~\ref{fig:particle1}. This leads to an equivalent picture to understand Hawking radiation via quantum tunneling: the particle inside the horizon escapes to the outside by quantum tunneling. Thus, the ``tunneling picture'' and ``pair creation picture'' are just two different pictures to understand the same physical phenomenon. Note that this ``tunneling picture'' does not violate causality since the spectrum is thermal and no information is carried.

\begin{figure}[b]
	\includegraphics[width=0.44\textwidth]{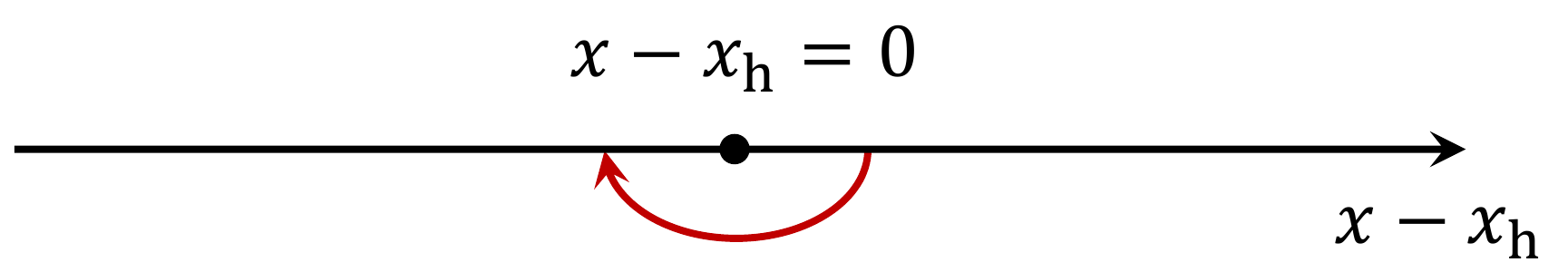}
	\caption{The wavefunction are connected in the complex plane $\ln|x_{\mathrm{h}}-x|\rightarrow\ln|x_{\mathrm{h}}-x|-i\pi$ when $x$ runs from outside horizon into inside horizon, which yields $|x_{\mathrm{h}}-x|^{i\omega/g_{\mathrm{h}}}\rightarrow|x_{\mathrm{h}}-x|^{i\omega/g_{\mathrm{h}}}e^{\pi \omega/g_{\mathrm{h}}}$.}\label{fig:particle2}
\end{figure}

Let us explain in detail how to use this tunneling picture to obtain the spectrum of radiation and corresponding temperature. For the spacetime with a black hole, the metric in the Schwarzschild coordinates $\{t_{\mathrm{s}}, x\}$ is given by $	\mathrm{d}s^2=f(x)\mathrm{d}t_{\mathrm{s}}^2-f^{-1}(x)\mathrm{d}x^2$.
We consider an outgoing mode with positive energy (corresponding to the observers of infinity) of a massless scalar or Dirac field, which can be written as ($\hbar=1$)
\begin{equation}\label{outmodes}
	\Phi(t_{\mathrm{s}},x)\propto\exp\left[-i\omega\left(t_{\mathrm{s}}-\int \frac{\mathrm{d} x}{f(x)}\right)\right].
\end{equation}
By using the Eddington-Finkelstein coordinates we have
\begin{equation}\label{outmodes}
	\Phi(t,x)\propto\exp\left[-i\omega\left(t-2\int \frac{\mathrm{d} x}{f(x)}\right)\right].
\end{equation}
Since $f(x)$ has a root at $f(x_{\mathrm{h}})$, we separate integration into two parts,
\begin{equation}\label{outmodes2}
	\int \frac{\mathrm{d} x}{f(x)}=F(x)+\int\frac{\mathrm{d} x}{2g_{\mathrm{h}}(x-x_{\mathrm{h}})}=F(x)+\frac1{2g_{\mathrm{h}}}\ln|x-x_{\mathrm{h}}|,
\end{equation}
where
\begin{equation}\label{outmodes3}
	F(x)=\int \left[\frac1{f(x)}-\frac1{2g_{\mathrm{h}}(x-x_{\mathrm{h}})}\right]\mathrm{d} x
\end{equation}
is regular at $x=x_{\mathrm{h}}$. Here $g_{\mathrm{h}}=f'(x_{\mathrm{h}})/2$ is the surface gravity. Thus, the outgoing positive mode then reads
\begin{equation}\label{outmodes4}
	\Phi(t,x)\propto e^{-i\omega[t-2F(x)]}|x-x_{\mathrm{h}}|^{i\omega/g_{\mathrm{h}}}.
\end{equation}
\begin{figure*}[ht]
	\includegraphics[width=6.1in]{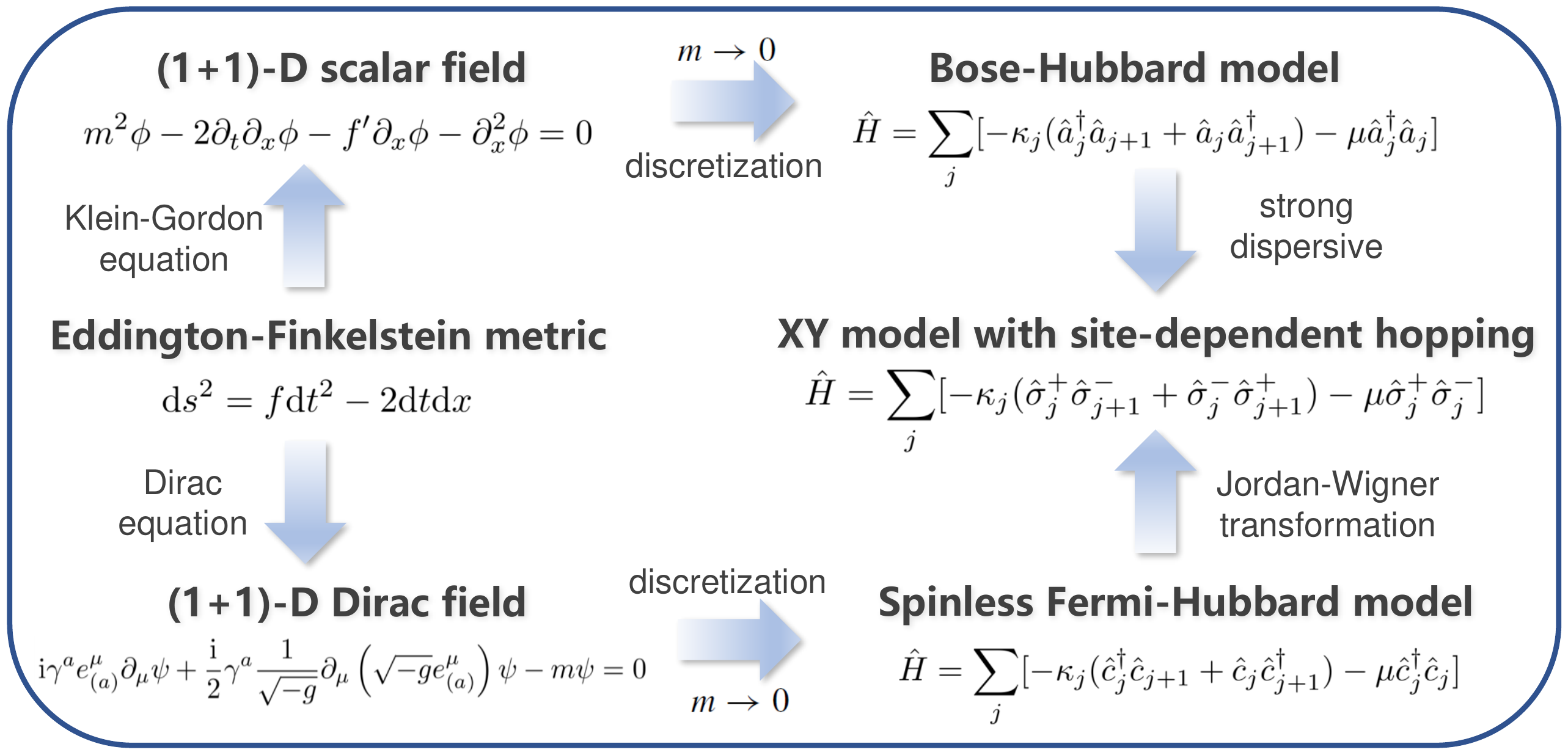}
	\caption{The theoretical framework of simulating quantum field theory in (1+1)-D curved spacetime with quantum many-body systems. The ``time" $t$ in the infalling Eddington-Finkelstein metric is given by $t=t_{\mathrm{s}}+\int{f^{-1}\mathrm{d}x}$ where $\{t_{\mathrm{s}},x\}$ are the Schwarzschild coordinates. In (1+1)-D configurations, the massless scalar field $\phi$ and the massless Dirac field $\psi$ can be discretized into the Bose-Hubbard model and the spinless Fermi-Hubbard model respectively, where the static curved spacetime background is encoded into the site-dependent hopping coupling distribution $\kappa_j$ satisfying Eq.~\eqref{eq:kappa_condition}. Here the on-site potential $\mu$ is an arbitrary constant. Note that both the Bose-Hubbard model and the spinless Fermi-Hubbard model can be transformed into the XY model, which implies their equivalence in (1+1)-D spacetime. This schematic diagram briefly summarizes the theory in ref.~\cite{PRR_Yang2020}. \label{fig:frame_theory} }
\end{figure*}
This outgoing mode has an infinite number of oscillations as $x\rightarrow x_{\mathrm{h}}$ and therefore cannot be straightforwardly extended to the inner region from the region outside the horizon. As argued in ref.~\cite{PRD_Damour1976}, we can use analytic continuation to connect two branches in a complex plane: the wave function $\Phi$ describing a particle state (positive frequencies) can be analytically continued to a complex plane (see Fig.~\ref{fig:particle2}). Then we obtain
\begin{equation}\label{outmodes5}
	\Phi(t,x)=\left\{
	\begin{split}
		&e^{-i\omega[t-2F(x)]}|x-x_{\mathrm{h}}|^{i\omega/g_{\mathrm{h}}},~~~x>x_{\mathrm{h}}\\
		&e^{-i\omega[t-2F(x)]}|x_{\mathrm{h}}-x|^{i\omega/g_{\mathrm{h}}}e^{\pi \omega/g_{\mathrm{h}}},~~~x<x_{\mathrm{h}}\,.
	\end{split}
	\right.
\end{equation}
This gives us the tunneling rate
\begin{equation}\label{tunnlingrate1}
	P=e^{-2\pi\omega/g_{\mathrm{h}}},
\end{equation}
which is identical to the detailed balance relation for transition rates in a thermal environment~\cite{RMP_Nori2012}. Note that the tunneling rate \eqref{tunnlingrate1} stands for the rate of a single particle. It is possible that the tunneling of multiple particles happens simultaneously. For bosons we have
\begin{equation}\label{bosons1}
	\begin{split}
		\text{particle number: }&~0~~~1~~~~2~~~~3~~~\cdots\\
		\text{probability: } &~1~~~P~~P^2~~P^3~~\cdots
	\end{split}
\end{equation}
Thus the occupation number of energy $\omega$ reads
\begin{equation}\label{occpn1}
	n(\omega)=\frac{\sum_{k=0}^\infty kP^k}{\sum_{k=0}^\infty P^k}=\frac1{e^{2\pi\omega/g_{\mathrm{h}}}-1}\,.
\end{equation}
This gives us the expected distribution of bosons and the temperature reads $T=g_{\mathrm{h}}/(2\pi)$ as predicted by Hawking. For fermions, if there is no other internal degree of freedom, the Pauli exclusion principle implies that there is  at most one particle of same energy. Thus, Eq.~\eqref{bosons1} is replaced by
\begin{equation}\label{bosons2}
	\begin{split}
		\text{particle number }n:&~0~~~1\\
		\text{probability:} &~1~~~P
	\end{split}
\end{equation}
Thus the occupation number of energy $\omega$ reads
\begin{equation}\label{occpn2}
	n(\omega)=\frac{\sum_{k=0}^1 kP^k}{\sum_{k=0}^1 P^k}=\frac1{e^{2\pi \omega/g_{\mathrm{h}}}+1}\,.
\end{equation}
This gives us the expected distribution of fermions and the temperature still reads $T=g_{\mathrm{h}}/(2\pi)$.

In addition, the composite system of the interior of the black hole and the exterior is isolated. The interior and the exterior exchange energy and particles via the horizon. Hence, the occupations Eq.~\eqref{occpn1} and ~\eqref{occpn2} can be viewed as the statistical averages of grand canonical distributions concerning bosons and fermions, respectively.

\section{\label{sec:Model} Model and Hamiltonian}

Our experiment is performed on a superconducting quantum processor which consists of 10 transmon qubits ($Q_1\sim Q_{10}$) and 9 transmon-type couplers ($C_1\sim C_{9}$). Each qubit and coupler are frequency-tunable, but only qubits have the XY control line and the readout resonator. The total Hamiltonian of this type of all-transmon system, with $N$ qubits and $N-1$ couplers, can be expressed as
\begin{eqnarray}
	\hat{H} &=& \hat{H}_{Q} \!+\! \hat{H}_{C} \!+\! \hat{H}_{{Q}-{Q}} \!+\! \hat{H}_{{Q}-{C}} \!+\! \hat{H}_{{C}-{C}}, \\
	\hat{H}_{Q}/\hbar &=& \sum_{j=1}^{N}\omega_{q_j}\hat{b}^{\dagger}_{q_j}\hat{b}_{q_j} + \frac{\alpha_{q_j}}{2}\hat{b}^{\dagger}_{q_j}\hat{b}^{\dagger}_{q_j}\hat{b}_{q_j}\hat{b}_{q_j},\\
	\hat{H}_{C}/\hbar &=& \sum_{j=1}^{N-1}\omega_{c_j}\hat{b}^{\dagger}_{c_j}\hat{b}_{c_j} + \frac{\alpha_{c_j}}{2}\hat{b}^{\dagger}_{c_j}\hat{b}^{\dagger}_{c_j}\hat{b}_{c_j}\hat{b}_{c_j},\\
	\hat{H}_{{Q}-{Q}}/\hbar &=& \sum_{j=1}^{N-1} g_{q_j,q_{j+1}}(\hat{b}^{\dagger}_{q_j}\hat{b}_{q_{j+1}}+\hat{b}_{q_j}\hat{b}^{\dagger}_{q_{j+1}}), \\
	\hat{H}_{{C}-{C}}/\hbar &=& \sum_{j=1}^{N-1} g_{c_j,c_{j+1}}(\hat{b}^{\dagger}_{c_j}\hat{b}_{c_{j+1}}+\hat{b}_{c_j}\hat{b}^{\dagger}_{c_{j+1}}), \\
	\hat{H}_{{Q}-{C}}/\hbar &=& \sum_{j=1}^{N-1} g_{q_j,c_j}(\hat{b}^{\dagger}_{q_j}\hat{b}_{c_j}+\hat{b}_{q_j}\hat{b}^{\dagger}_{c_j})\nonumber\\
	&~&\quad~+g_{q_{j+1},c_j}(\hat{b}^{\dagger}_{q_{j+1}}\hat{b}_{c_j}+\hat{b}_{q_{j+1}}\hat{b}^{\dagger}_{c_{j}}),
\end{eqnarray}
where $\hbar$ is the reduced Planck constant (for convenience $\hbar$ will be assumed to be 1 in the following), $\hat{b}_{q_j}(\hat{b}_{c_j})$ and $\hat{b}^{\dagger}_{q_j}(\hat{b}^{\dagger}_{c_j})$ denote the annihilation and creation operators of the $j$-th qubit (coupler), respectively. The corresponding frequencies and anharmonicities are $\omega_{q_j}(\omega_{c_j})$ and $\alpha_{q_j}(\alpha_{c_j})$. Every pair of two neighbouring qubits and their middle coupler are coupled through exchange-type interactions with coupling strengths $g_{q_j,c_j}$, $g_{q_{j+1},c_j}$ and $g_{q_j,q_{j+1}}$. Here, the total Hamiltonian has three parts, including qubit-qubit interaction $\hat{H}_{{Q}-{Q}}$, coupler-coupler interaction $\hat{H}_{{C}-{C}}$ and qubit-coupler interaction $\hat{H}_{{Q}-{C}}$. The total system is equivalent to a 19-qubit Bose-Hubbard model.

In our experiment, the strong dispersive condition $g_{q_j,c_j}\ll|\Delta_{q_j,c_j}|$ is satisfied, where $\Delta_{q_j,c_j}=\omega_{q_j}-\omega_{c_j}$ is the frequency detuning. By virtue of the so-called Schrieffer-Wolff transformation
\begin{eqnarray}
	\hat{U} & = & \exp\Big[\sum_{j=1}^{N-1}\frac{g_{q_j,c_j}}{\Delta_{q_j,c_j}}(\hat{b}^{\dagger}_{q_j}\hat{b}_{c_j}+\hat{b}_{q_j}\hat{b}^{\dagger}_{c_j})\nonumber\\
	& ~ &\label{eq:SWT}\qquad\qquad+\frac{g_{q_{j+1},c_j}}{\Delta_{q_{j+1},c_j}}(\hat{b}^{\dagger}_{q_{j+1}}\hat{b}_{c_j}+\hat{b}_{q_{j+1}}\hat{b}^{\dagger}_{c_{j}})\Big],
\end{eqnarray}
one can obtain the effective qubits Hamiltonian
\begin{eqnarray}
	\hat{\widetilde{H}} & = & \hat{U}\hat{H}\hat{U}^{\dagger} = \hat{\widetilde{H}}_{Q} + \hat{\widetilde{H}}_{{Q}-{Q}} \nonumber\\
	& = & \sum_{j=1}^{N}\widetilde{\omega}_{q_j}\hat{b}^{\dagger}_{q_j}\hat{b}_{q_j} + \frac{\alpha_{q_j}}{2}\hat{b}^{\dagger}_{q_j}\hat{b}^{\dagger}_{q_j}\hat{b}_{q_j}\hat{b}_{q_j}\nonumber\\
	& ~ &+\sum_{j=1}^{N-1} \widetilde{g}_{q_j,q_{j+1}}(\hat{b}^{\dagger}_{q_j}\hat{b}_{q_{j+1}}+\hat{b}_{q_j}\hat{b}^{\dagger}_{q_{j+1}}).
	\label{eq:effective_H}
\end{eqnarray}
with the corresponding dressed frequency
\begin{equation}
	\label{eq:effective_freq}\widetilde{\omega}_{q_j} =\begin{cases}
		& \omega_{q_j} + \frac{g^2_{q_j,c_j}}{\Delta_{q_j,c_j}},\qquad\qquad\qquad j=1 \\
		& \omega_{q_j} + \frac{g^2_{q_j,c_j}}{\Delta_{q_j,c_j}} + \frac{g^2_{q_j,c_{j+1}}}{\Delta_{q_j,c_{j+1}}},~1<j<N \\
		& \omega_{q_j} + \frac{g^2_{q_j,c_{j-1}}}{\Delta_{q_j,c_{j-1}}},\qquad\qquad\quad j=N
	\end{cases}
\end{equation}
and effective coupling strength
\begin{equation}
	\label{eq:effective_g}\widetilde{g}_{q_j,q_{j+1}} = g_{q_j,q_{j+1}} + \frac{g_{q_j,c_{j}}g_{q_{j+1},c_{j}}}{\Lambda^{c_j}_{q_j,q_{j+1}}},
\end{equation}
where $\Lambda^{c_j}_{q_j,q_{j+1}} = 2/\big(1/{\Delta_{q_j,c_j}}+1/{\Delta_{q_{j+1},c_j}}\big)$ is the harmonic mean of the frequencies detuning between the $j$-th coupler and its nearest neighbor qubits. Eq.~(\ref{eq:effective_g}) implies that the effective qubit-qubit coupling is derived from their direct capacitive coupling and the indirect virtual exchange coupling via the coupler in between. If the frequency of coupler is above the frequencies of qubits, $\Lambda^{c_j}_{q_j,q_{j+1}} < 0$ holds so that the effective coupling $\widetilde{g}_{q_j,q_{j+1}}$ can be tuned from positive to negative monotonically with gradually decreasing the frequency of coupler. Experimentally, we use the arbitrary waveform generator (AWG) to generate various fast-bias voltages applied to the corresponding couplers. These pulses on the Z control lines change the frequencies of couplers and then make it possible for the superconducting circuit with tunable couplers to engineer an arbitrary coupling distribution.

With $\widetilde{g}_{q_j,q_{j+1}}\ll\alpha_{q_j}$, the effective Hamiltonian Eq.~\eqref{eq:effective_H} can be rewritten as a site-dependent XY model:
\begin{equation}\label{eq:xymodel}
	\hat{\widetilde{H}} = -\sum_{j=1}^{N-1}\kappa_j(\hat{\sigma}^{+}_j\hat{\sigma}^{-}_{j+1}+\hat{\sigma}^{-}_j\hat{\sigma}^{+}_{j+1})
	-\sum_{j=1}^{N}\mu_j\hat{\sigma}^{+}_j\hat{\sigma}^{-}_j,
\end{equation}
where $\hat{\sigma}^{+}_j$ ($\hat{\sigma}^{-}_j$) is the raising (lowering) operator of the $j$-th qubit. Here we choose
\begin{equation}
	\kappa_j=-\widetilde{g}_{q_j,q_{j+1}},\quad\mu_j=-\widetilde{\omega}_{q_j}.
\end{equation}
For the Hamiltonian Eq.~\eqref{eq:xymodel}, one can map the spin variables to spinless fermion operators by introducing the Jordan-Wigner transformation~\cite{AP_Lieb1961}: $\hat{\sigma}^{+}_j = \hat{c}^{\dagger}_j\exp\left\{\mathrm{i}\pi\sum_{l=1}^{j-1}\hat{c}^{\dagger}_j\hat{c}_j\right\}$ and $\hat{\sigma}^{-}_j = \exp\left\{-\mathrm{i}\pi\sum_{l=1}^{j-1}\hat{c}^{\dagger}_j\hat{c}_j\right\}\hat{c}_j$, where the operators $\hat{c}^{\dagger}$ and $\hat{c}_j$ satisfy the commutation relations of fermions, i.e., $\{\hat{c}_j, \hat{c}_k\}=\{\hat{c}^{\dagger}_j, \hat{c}^{\dagger}_k\}=0$ and $\{\hat{c}_j, \hat{c}^{\dagger}_k\}=\delta_{jk}$. Hence, the effective Hamiltonian is mapped into a spinless fermion lattice model as
\begin{equation}\label{eq:fermion}
	\hat{\widetilde{H}} = - \sum_{j=1}^{N-1}\kappa_j(\hat{c}^{\dagger}_j\hat{c}_{j+1}+\hat{c}_j\hat{c}^{\dagger}_{j+1})
	-\sum_{j=1}^{N}\mu_j\hat{c}^{\dagger}_j\hat{c}_j.
\end{equation}

\section{\label{sec:Correspondence} Correspondence with two-dimensional curved spacetime}

Considering the Heisenberg equation $\mathrm{i}\frac{\mathrm{d}}{\mathrm{d}t}\hat{c}_j=[\hat{c}_j, \hat{\widetilde{H}}]$, the evolution equation for the operator $\hat{c}_j$ can be given by
\begin{equation}\label{eq:Heisenberg}
	\mathrm{i}\frac{\mathrm{d}}{\mathrm{d}t}\hat{c}_j = -\kappa_j\hat{c}_{j+1}-\kappa_{j-1}\hat{c}_{j-1}-\mu\hat{c}_j.
\end{equation}
By introducing a variable transformation $\hat{\widetilde{c}}_j(t)=(-\mathrm{i})^je^{-\mathrm{i}\mu t}\hat{c}_j$, we obtain
\begin{equation}\label{eq:Heisenberg_v2}
	\frac{\mathrm{d}}{\mathrm{d}t}\hat{\widetilde{c}}_j(t) = -\kappa_j\hat{\widetilde{c}}_{j+1}(t)+\kappa_{j-1}\hat{\widetilde{c}}_{j-1}(t).
\end{equation}
Here $\hat{\widetilde{c}}_j(t)$ can be viewed as a quantized operator of a discrete field $\varphi_j(t)$, and the spatial position can be discretized as $x=x_j=jd-x_{\mathrm{h}}$, where $x_{\mathrm{h}}=j_{\mathrm{h}}d$ ($j\in\mathbb{Z^+}$) and $d$ denotes the lattice constant.
Note the factor $(-i)^j$ is important to obtain the correct Heisenberg equation. The similar trick is widely used in artificial lattices to simulate quantum fields in flat or curved spacetimes~\cite{PRB_Longhi2010,PRL_Dreisow2010,AP_Koke2016}.

Now let us recover the continuous field $\varphi(t,x)$. If we define a function $f$ that is dependent of the spatial position $x_j$ and substitute $\kappa_{j}$ as,
\begin{equation}
	\kappa_{j}=\frac{f(x_{j+1})+f(x_{j})}{8d},
\end{equation}
according to Eq.~\eqref{eq:Heisenberg_v2}, $\varphi(t,x_j)\rightarrow\hat{\widetilde{c}}_j(t)/\sqrt{d}$ will obey the following relation in the continuum limit,
\begin{widetext}
	\begin{eqnarray}
		\frac{\partial}{\partial t}\varphi(t,x) &=& -\frac{f(x_{j+1}) +f(x_{j})}{8d}\varphi(t,x_{j+1})+\frac{f(x_{j})
			+f(x_{j-1})}{8d}\varphi(t,x_{j-1})\nonumber\\
		&=&-\frac{f(x_j)}{4}\cdot\frac{\varphi(t,x_{j+1})-\varphi(t,x_{j-1})}{2d}-\frac{1}{4}\cdot\frac{f(x_{j+1})\varphi(t,x_{j+1})-f(x_{j-1})\varphi(t,x_{j-1})}{2d}\nonumber\\
		&=&-\frac{f(x)}{4}\cdot\frac{\partial}{\partial x}\varphi(t,x)-\frac{1}{4}\cdot\frac{\partial}{\partial x}\left[f(x)\varphi(t,x)\right]\nonumber\\
		&=&-\frac{f(x)}{2}\cdot\frac{\partial}{\partial x}\varphi(t,x)-\frac{f'(x)}{4}\varphi(t,x).\label{eq:evolution}
	\end{eqnarray}
\end{widetext}
In fact, Eq.~\eqref{eq:evolution} can be considered as a special case of Dirac equation in the massless limit $m\rightarrow0$ if we decompose the Dirac field operator into $\psi=\frac{1}{\sqrt{2}}(\xi+\varphi, \xi-\varphi)^{\mathrm{T}}$. In the light of refs.~\cite{PRL_Pedernales2018,PRL_Mann1991}, the Dirac equation in (1+1)-dimensional curved spacetime with the metric $g_{\mu\nu}$ is written as
\begin{equation}\label{eq:Dirac}
	\mathrm{i}\gamma^{a}e^{\mu}_{(a)}\partial_{\mu}\psi+\frac{\mathrm{i}}{2}\gamma^{a}\frac{1}{\sqrt{-g}}\partial_{\mu}\left(\sqrt{-g}e^{\mu}_{(a)}\right)\psi-m\psi=0,
\end{equation}
where the $\gamma$-matrices in the two-dimensional case are $\gamma^{0}=\sigma_z$ and $\gamma^{1}=\mathrm{i}\sigma_y$, and the dyad is chosen as
\begin{equation}
	e^{\mu}_{(a)}=\left(
	\begin{array}{cc}
		-1 & 1 \\
		\frac{1-f}{2} & \frac{1+f}{2} \\
	\end{array}
	\right),
\end{equation}
which satisfies the orthonormal condition $e^{(a)}_{\mu}e^{\nu}_{(a)}=\delta^{\nu}_{\mu}$.
Thus, Eq.~\eqref{eq:Dirac} can be decomposed into two independent equations,
\begin{equation}\label{eq:Dirac_v2}
	\partial_t\varphi=-\frac{f}{2}\partial_x\varphi-\frac{f'}{4}\varphi+\frac{\mathrm{i}}{2}m\xi,~\partial_x\xi=-\mathrm{i}m\varphi.
\end{equation}
In the massless limit $m\rightarrow0$, one can find that Eq.~\eqref{eq:Dirac_v2} is in accord with Eq.~\eqref{eq:evolution}. Hence, what the effective Hamiltonian Eq.~\eqref{eq:fermion} describes is equivalent to a two-dimensional static curved spacetime governed by the massless Dirac equation if we set $\kappa_{j}$ as
\begin{equation}\label{eq:kappa_condition}
	\kappa_{j}=\frac{f(x_{j+1})+f(x_{j})}{8d}\approx\frac{f(x_j+d/2)}{4d}.
\end{equation}
There is only one single nondegenerate horizon $x_{\mathrm{h}}$ so that $f(x_{\mathrm{h}})=0$ and $f(x)>0$ when $x>x_\mathrm{h}$ and
\begin{equation}
	g_\mathrm{h}=\frac{1}{2}f'(x_\mathrm{h})>0,
\end{equation}
where $g_\mathrm{h}$ is the surface gravity of the horizon, which gives the Hawking temperature $T_\mathrm{H}=g_\mathrm{h}/(2\pi)$. In the main text, we set $f(x)=\beta\tanh{\eta x}/\eta$ with corresponding Hawking temperature $T_\mathrm{H}=\beta/(4\pi)$
and
\begin{equation}\label{eq:kappa_j}
\kappa_{j} = \frac{\beta\tanh\!\left((j-j_{\mathrm{h}}+1/2)\eta d\right)}{4\eta d},
\end{equation}
where $\eta$ controls the scale of variation of $f$ over each lattice site, which has the dimension of $1/d$. Here, we fix $j_{\mathrm{h}}=3$, $\beta/(2\pi)=4.39$~MHz, and $\eta d=0.35$ in the analogue curved spacetime experiments.

What we have shown above is the correspondence between XY model and the (1+1)-D Dirac field. The case of scalar field governed by Klein-Gordon equation is similar. For a complete presentation, one can refer to the earlier theoretical work~\cite{PRR_Yang2020}. Here we briefly summarize the theoretical framework, as shown in Fig.~\ref{fig:frame_theory}.

\begin{figure}[ht]
	\includegraphics[width=3.0in]{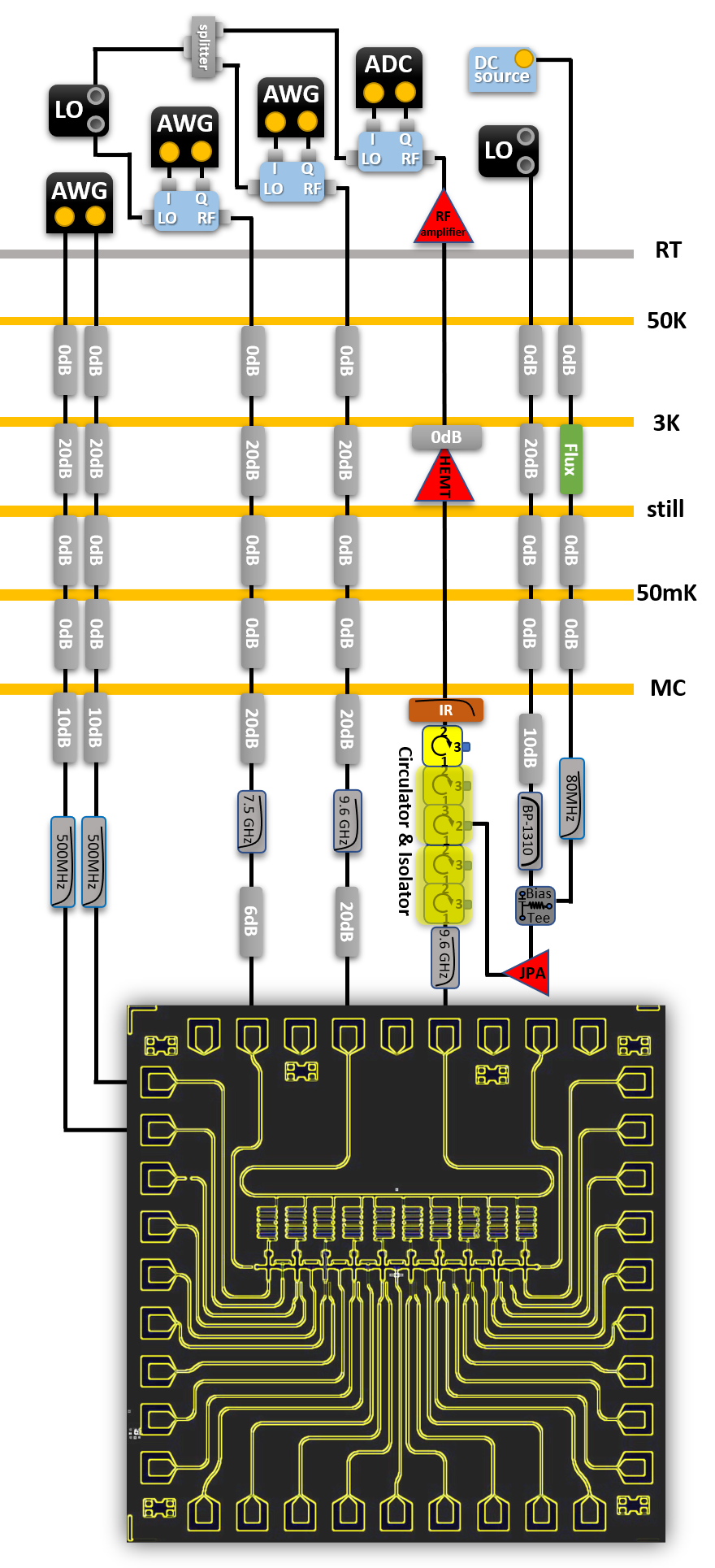}
	\caption{ A schematic diagram of the experimental system and partial wiring information.}\label{fig:wiring}
\end{figure}

\section{\label{sec:Setup} Experimental Setup and Device parameters}

Our superconducting quantum processor is placed in a BlueFors dilution refrigerator. The base temperature of the mixing chamber (MC) is about 9 mK. All qubits share one readout line equipped with a Josephson parametric amplifier (JPA) and a high-electron-mobility transistor (HEMT). Pulse on the readout transmission line is first generated as a mixture of local oscillation (LO) and the envelopes from an arbitrary waveform generator (AWG) and then demodulated by an analog digital converter (ADC). In this experiment, we replace the DC bias with a long Z square pulse generated by AWGs. Both XY and Z control signals are programmed in advance before being uploaded to AWGs. A schematic diagram of experiment setup is given in Fig.~\ref{fig:wiring}.

The device parameters are briefly shown in Table~\ref{tab:paras}. All the parameters are characterized by various relatively efficient and automatic methods, especially the parameters concerning couplers. Details of those methods in our experiment will be introduced in the following.

\begin{table*}
	
	\begin{ruledtabular}
		\begin{tabular}{ccccccccccc}
			qubit & $Q_1$ & $Q_2$ & $Q_3$ & $Q_4$ & $Q_5$ & $Q_6$ & $Q_7$ & $Q_8$ & $Q_9$ & $Q_{10}$ \\[.05cm] \hline
			\\[-.2cm]
			$\widetilde{\omega}^{01}_{q_j}/2\pi$ (GHz) & 5.300 & 4.760 & 5.330 & 4.805 & 5.278 & 4.830 & 5.231 & 4.705 & 5.180 & 4.655 \\[.09cm]
			$\widetilde{\omega}^r_{q_j}/2\pi$ (GHz) & 6.667 & 6.687 & 6.708 & 6.733 & 6.749 & 6.772 & 6.792 & 6.815 & 6.840 & 6.857 \\[.09cm]
			$E_{\mathrm{C}}/2\pi$ (MHz) & 195.8 & 194.5 & 195.4 & 198.4 & 197.0 & 196.2 & 195.7 & 201.8 & 199.7 & 203.2\\[.09cm]
			$E_{\mathrm{JJ}}/2\pi$ (GHz) & 20.69 & 19.88 & 19.78 & 18.55 & 19.30 & 19.52 & 19.39 & 17.66 & 18.64 & 18.00 \\[.09cm]
			$F_{0,q_j}$ & 0.965 & 0.958 & 0.974 & 0.938 & 0.956 & 0.940 & 0.982 & 0.952 & 0.949 & 0.943 \\[.09cm]
			$F_{1,q_j}$ & 0.885 & 0.901 & 0.907 & 0.887 & 0.885 & 0.910 & 0.871 & 0.875 & 0.904 & 0.906 \\[.09cm]
			$T_{1,q_j}$ ($\mu$s) & 24.9 & 20.1 & 18.6 & 27.8 & 27.9 & 25.4 & 24.1 & 23.5 & 37.4 & 27.7 \\[.09cm]
			$T^{\ast}_{2,q_j}$ ($\mu$s) & 4.2 & 1.5 & 5.4 & 2.0 & 5.9 & 1.9 & 5.2 & 2.7 & 5.9 & 2.0 \\[.09cm]
			$T^{\mathrm{Echo}}_{2,q_j}$ ($\mu$s) & 7.2 & 3.4 & 17.9 & 4.4 & 10.7 & 4.6 & 11.3 & 5.3 & 9.0 & 4.1 \\[.09cm]
			
			\hline\hline
			\\[-.3cm]
			coupler & $C_1$ & $C_2$ & $C_3$ & $C_4$ & $C_5$ & $C_6$ & $C_7$ & $C_8$ & $C_9$ & $~$\\[.05cm] \hline
			\\[-.2cm]
			$g_{q_j,c_j}/2\pi$ (MHz) & 98.07 & 84.13 & 98.88 & 85.96 & 96.46 & 87.50 & 96.00 & 85.63 & 96.02 & ~ \\[.09cm]
			$g_{q_{j+1},c_j}/2\pi$ (MHz) & 85.72 & 96.68 & 88.64 & 97.46 & 85.36 & 96.78 & 85.89 & 97.40 & 83.36 & ~ \\[.09cm]
			$g_{q_{j},q_{j+1}}/2\pi$ (MHz) & 10.41 & 10.06 & 10.05 & 9.73 & 9.56 & 9.97 & 9.55 & 9.64 & 9.82 & ~ \\[.09cm]
		\end{tabular}
	\end{ruledtabular}
	\caption{ List of device parameters. Here, $\widetilde{\omega}^{01}_{q_j(c_j)}$ is $\ket{0}\!\!\rightarrow\!\!\ket{1}$ transition frequency of the $j$-th qubit (coupler) with the corresponding readout frequency $\widetilde{\omega}^r_{q_j}$. $E_{\mathrm{C}}$ and $E_{\mathrm{JJ}}$ denote the charging energy and the Josephson energy. $F_{0,q_j}$ and $F_{1,q_j}$ are measure fidelities of $\ket{0}$ and $\ket{1}$, respectively. $T_{1,q_j}$ represents the energy relaxation time of $q_j$ at the idle point. The dephasing time $T^{\ast}_{2,q_j}$ is characterized by the Ramsey fringe experiment, while $T^{\mathrm{Echo}}_{2,q_j}$ is measured by spin echo sequence with an inserted $\pi$ pulse. The coupling strengths of exchange-type interactions between qubits and the corresponding coupler are $g_{q_{j},c_j}$ and $g_{q_{j+1},c_j}$, and the direct coupling of qubits is $g_{q_{j},q_{j+1}}$.}\label{tab:paras}
\end{table*}

\section{\label{sec:cali_auto} Efficient and automatic calibration for multi-qubit devices with tunable couplers}

Before carrying out our experiment for simulating an analogue black hole, we need to calibrate all 10 qubits and find the useful parameters of 9 couplers. This is far more difficult and time-consuming than calibrating a typical 10-qubit sample without tunable couplers. In order to measure and characterize device parameters more efficiently, we adopt an automatic calibration technology based on a combination of physical models and optimization methods.

\subsection{\label{subsec:cali_auto1} Spectrum of qubit and frequency calibration}

\begin{figure}[b]
	\includegraphics[width=3.0in]{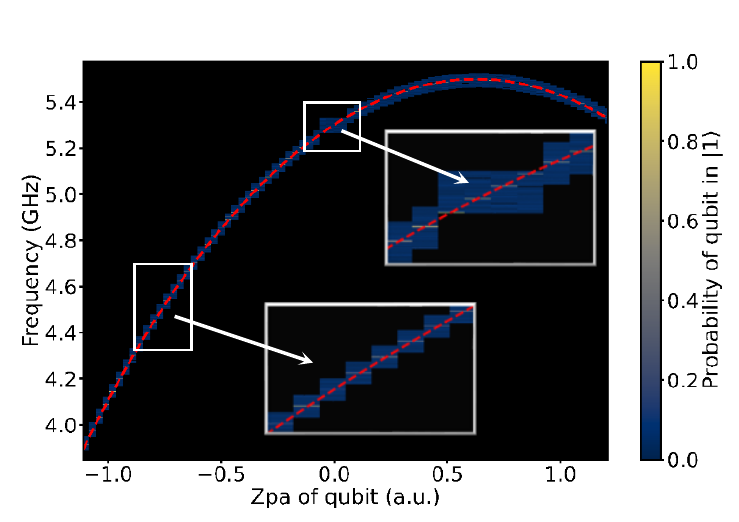}
	\caption{Experimental data of qubit automatical spectroscopy measurement. Here we take the spectrum of $Q_1$ as an example. The black area is unscanned in order to save time. We first scan a small square area (about four columns data) around Zpa $=$ 0 and then use polynomial curves to fit the peaks of these data. The corresponding polynomial fitting coefficients will help predict the next peak of the qubit spectrum. By constantly measuring, fitting and predicting, we obtain the experimental data of the qubit spectrum with a wide range. The mapping between qubit Zpa and its frequency can be obtained by fitting the experimental data based on Eq.~\eqref{eq:zpa2freq}.}\label{fig:spec_q1}
\end{figure}

First and foremost, all the qubits are individually brought up through the standard single-qubit calibration (from identifying the readout resonator frequency to calibrating $\pi$ pulse). If a qubit is brought up at a certain frequency, we need to perform a two-dimensional spectroscopy measurement to extract the mapping between Z-pulse amplitude (hereinafter referred as Zpa, each unit of 200 mV) of qubit bias and its frequency (such as Fig.~\ref{fig:spec_q1}) and this will contribute to automatically calibrate all the qubits together.

According to our all-transmon sample, each transmon consists of two parallel SIS-type Josephson junctions connected by a loop which is in series with a capacitor. The critical currents of two junctions are $I_{c_1}$ and $I_{c_2}$ and $E_{\mathrm{C}}$ denotes the charging energy of capacitor. By using the perturbation theory, the transition frequency can be approximately written as~\cite{PRA_Koch2007, APR_Krantz2019, JAP_Kwon2021}
\begin{equation}
	\omega(\Phi) \approx \sqrt{8E_{\mathrm{JJ}}E_{\mathrm{C}}\sqrt{\delta^2+\cos^2{\Big(\frac{\pi\Phi}{\Phi_0}}\Big)}}-E_{\mathrm{C}},
\end{equation}
where $\Phi_0=h/(2e)$ is the unit flux, $E_{\mathrm{JJ}}=\sqrt{I_{c_1}I_{c_2}}\Phi_0/\pi$ denotes the geometric mean of two junctions energy in the zero field and $\delta=|I_{c_1}-I_{c_2}|/(2\sqrt{I_{c_1}I_{c_2}})$ represents the junction asymmetry. Here, the total magnetic flux $\Phi$ is in direct proportion with the strength of the magnetic field threading the loop and this weak magnetic field induced by Z pulse is approximately proportional to Zpa ($\Phi\propto$~Zpa). Thus, the mapping between qubit Zpa and its frequency can be given by
\begin{equation}
	\label{eq:zpa2freq}\omega(\mathrm{Zpa}) \approx \sqrt{8E_{\mathrm{JJ}}E_{\mathrm{C}}\sqrt{\delta^2+\cos^2{(A\cdot \mathrm{Zpa}+\phi)}}}-E_{\mathrm{C}}
\end{equation}
and
\begin{equation}
	\label{eq:freq2zpa}\mathrm{Zpa}(\omega) \approx \frac{\arccos\bigg[\pm\sqrt{\frac{(\omega+E_{\mathrm{C}})^4}{(8E_{\mathrm{JJ}}E_{\mathrm{C}})^2}-\delta^2}\bigg]-\phi}{A},
\end{equation}
where $E_{\mathrm{C}}$ can be measured by the two-photon excitation experiment (double difference between two-photon excitation frequency and qubit frequency), the remaining parameters $E_{\mathrm{JJ}}$, $\delta$, $A$ and $\phi$ will be obtained by fitting the two-dimensional spectrum of qubits. Here parameter $A$ describes the efficiency of qubit bias which depends on the attenuation on the Z control line, while $\phi$ is the initial flux shift. Even if the refrigerator temperature rises and cools again, only $\phi$ may have some displacement. As long as the circuit wiring does not change, parameter $A$ will keep its value. Notice, however, that Eq.~\eqref{eq:zpa2freq} and Eq.~\eqref{eq:freq2zpa} need to be modified by the crosstalk of Z control lines if the multi-qubit case is involved.

When we design the multi-qubit levels, Eq.~\eqref{eq:freq2zpa} will be beneficial to obtain the corresponding Zpa according to the target frequency. A more accurate frequency calibration can be implemented by the Ramsey fringe experiment.
\begin{figure}[ht]
	\includegraphics[width=3.0in]{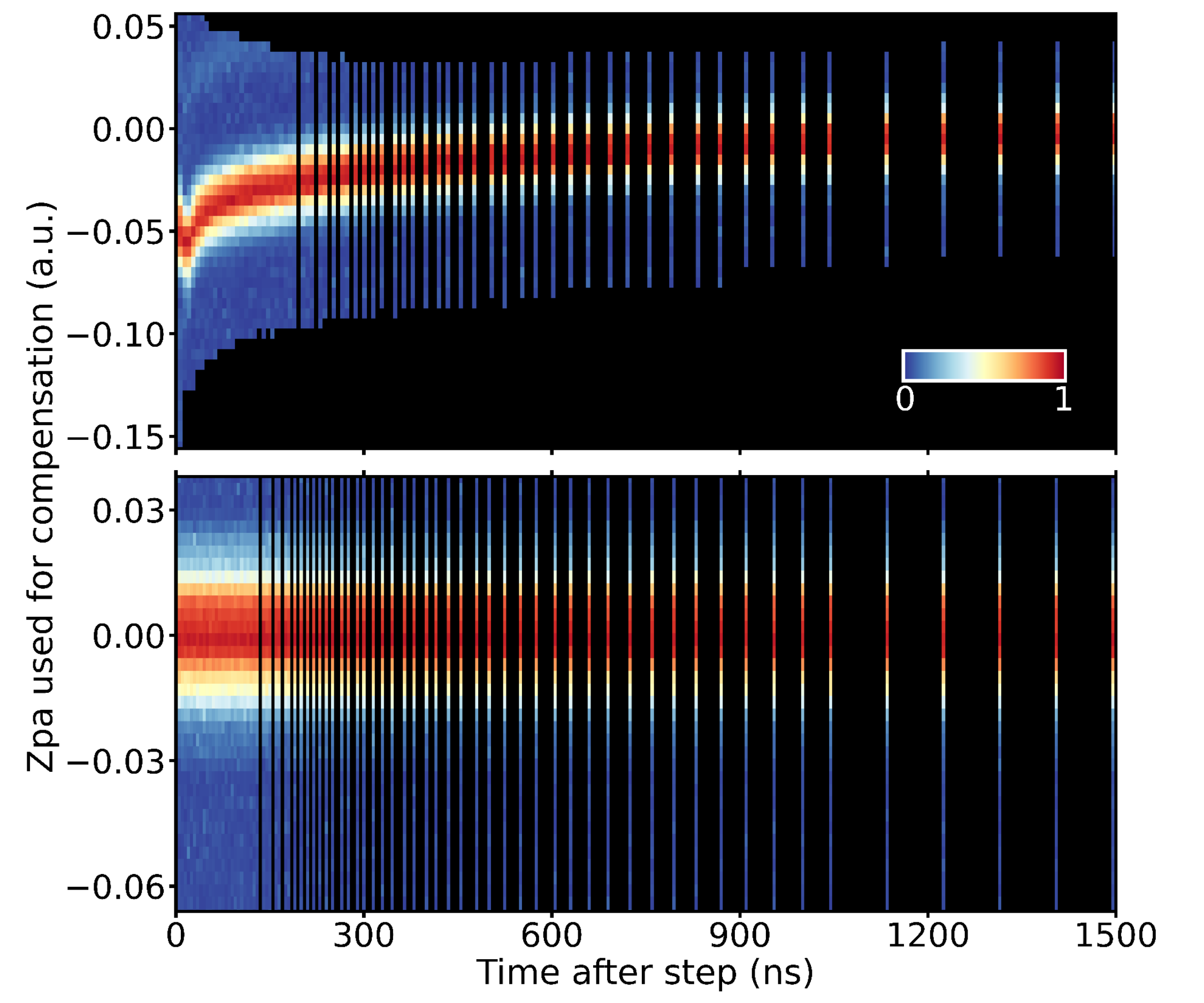}
	\caption{Experimental data of pulse shape measurement. The black area is unscanned in order to save time. Here the heatmap denotes the probability of qubit in $\ket{1}$. The uncorrected pulse (up) is distorted, while the corrected result (down) shows a stationary step response. \label{fig:pulseshape} }
\end{figure}

\begin{figure}[hb]
	\includegraphics[width=3.2in]{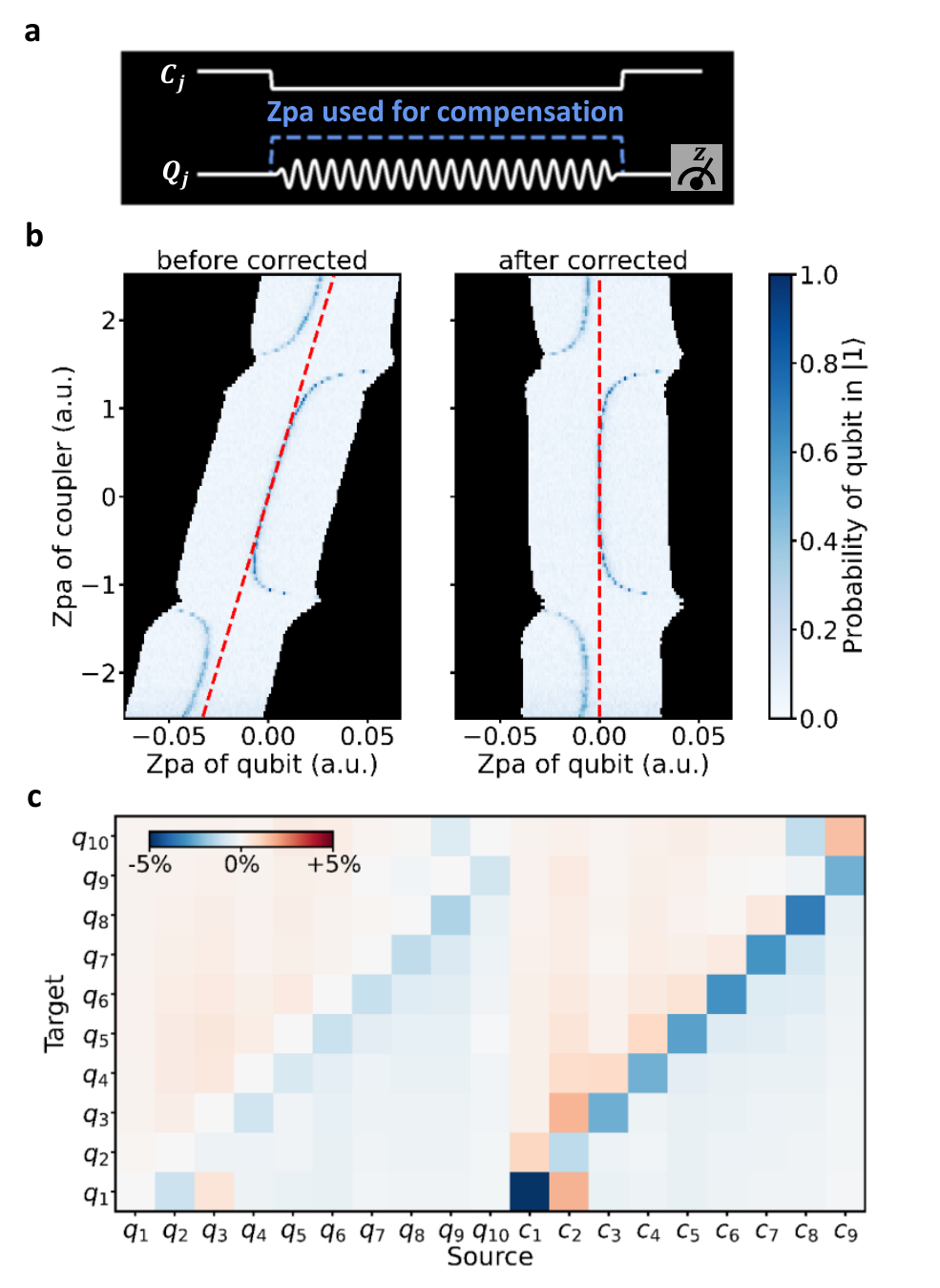}
	\caption{Experimental data of automatical Z crosstalk calibration. \textbf{a}, Pulse sequence for measurement of Z crosstalk of coupler to qubit. \textbf{b},  Before Z crosstalk is corrected, one can observe a tilted anti-crossing pattern of qubit and its nearest neighbor coupler. By constantly fine-tuning the crosstalk coefficient and applying the Zpa to compensate for crosstalk from coupler Z line, a symmetrical anti-crossing pattern will be obtained after corrected. The black area is unscanned, while the red lines are the results of the linear fitting. Here we show the experimental data of Z crosstalk calibration \textbf{c}, All the coefficients of Z crosstalk. Compared with the high crosstalk from couplers Z line to qubits, the absolute coefficients of Z crosstalk between qubits are all at a low level ($<2\%$).  \label{fig:zxtalk} }
\end{figure}

\subsection{\label{subsec:cali_auto2} Calibration of pulse distortion and Z crosstalk}

Although the Z pulse generated by AWG is designed carefully, the shape of the pulse is distorted when it interacts with the qubit. To calibrate the distortion of step response, we use several first-order infinite impulse response (IIR) filters and a finite impulse response (FIR) filter~\cite{APL_Rol2020}. Here IIR filters are designed to be an integration of several exponential functions and the FIR filter is described by a polynomial with 20 parameters. The results of pre- and post-correction are shown in Fig.~\ref{fig:pulseshape}.

For the crosstalk of Z control lines between qubits and qubits or qubits and their non-nearest neighbor couplers, a routine Z crosstalk measurement with a small scanning range is adopted, which can be used to estimate the crosstalk coefficients by measuring the frequency response to the Z control lines. However, it may be better to extend to a wider range of scanning when it comes to the Z crosstalk of couplers to qubits. If the frequency of coupler approaches the frequency of qubit, the effect of anti-crossing will be amplified due to the strong coupling between the coupler and its nearest neighbor qubit, leading to a distinctly non-linear relationship between coupler Zpa and qubit Zpa (as shown in Fig.~\ref{fig:zxtalk}\textbf{b}). To correct the crosstalk from classical flux crosstalk of Z control lines that basically meet the linear relationship, we first select a range of data away from the resonance points to use linear fitting, and constantly fine-tune the corresponding crosstalk coefficient until a symmetrical anti-crossing pattern is obtained. For a more accurate Z crosstalk calibration, we still take advantage of Ramsey fringe experiment, but proximity to the resonance points should be avoided. Here we emphasize that in our procedure of calibration, Z crosstalk of couplers to qubits must be corrected in order to more accurately measure the spectrum of coupler and coupling strengths, as explained in the following.

\subsection{\label{subsec:cali_auto3} Spectrum of coupler and anti-crossing of energy levels}

\begin{figure}[t]
	\includegraphics[width=3.2in]{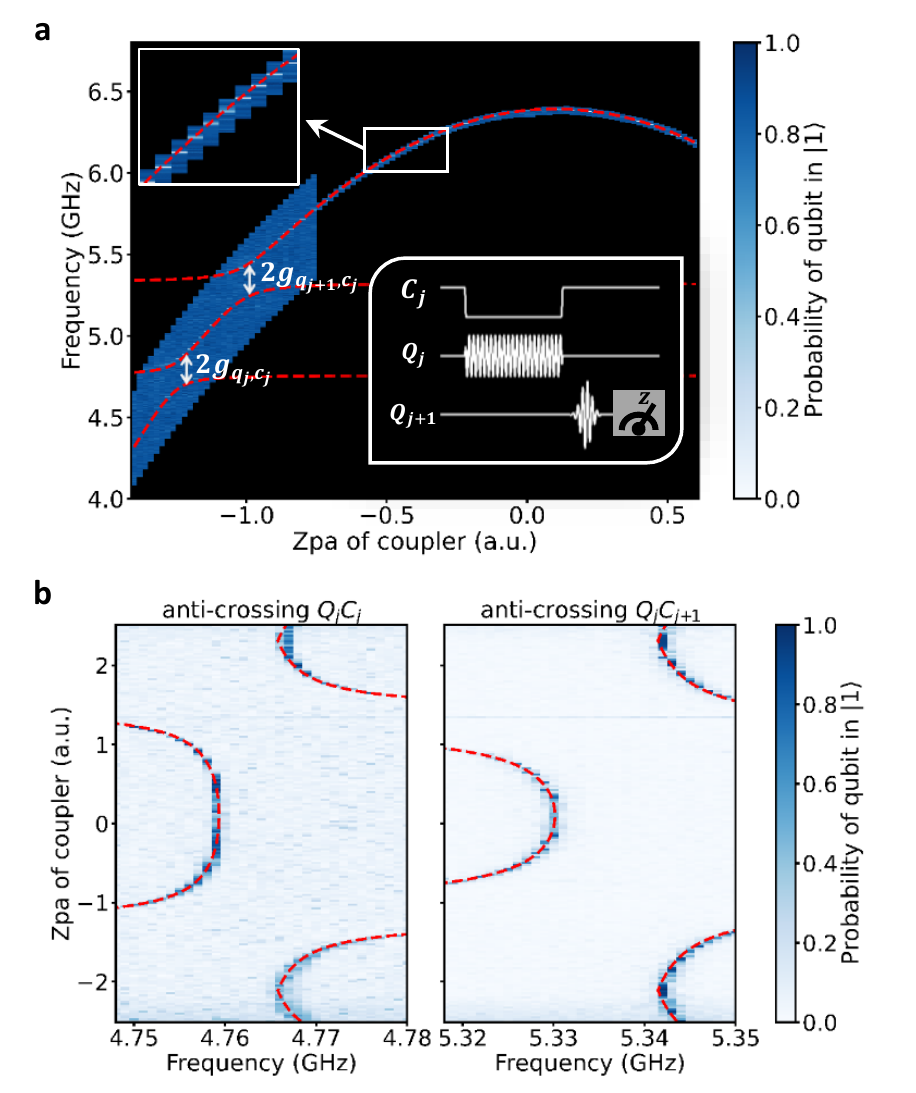}
	\caption{Experimental data of coupler automatical spectroscopy measurement. The red curves are numerical simulation results for fitting the peaks of spectroscopy data, which is based on a multi-objective optimization. \textbf{a}, The spectrum of the local $Q_jC_jQ_{j+1}$ three-body system. The black area is unscanned, while the experimental data consists of the blue area. When the frequency of coupler is far from the qubits’ frequency, we only need to scan a very narrow width like the single-qubit spectroscopy measurement. As it approaches the anti-crossing points, we increase the scan width to reduce the impact caused by predictive error, which ensures a clear three-body spectrum and saves time simultaneously. \textbf{b}, Experimental data of anti-crossing spectrums of $Q_jC_j$ (left) and $Q_{j+1}C_j$ (right). Here the results of $C_2$ are taken as an example. \label{fig:spec_c2} }
\end{figure}

As what mentioned above, it is difficult to directly excite and measure a coupler because it has no XY control line and readout resonator. Therefore, we make use of two qubits ($Q_j$ and $Q_{j+1}$) that are adjacent to the coupler ($C_j$) to perform a coupler spectroscopy measurement (after the calibration of Z crosstalk). To be specific, we apply XY excitation pulse to one of the qubits ($Q_j$) and vary the Zpa of coupler. If the coupler is excited to $\ket{1}$ by the crosstalk from $Q_j$ XY line, the frequency of another qubit ($Q_{j+1}$) will be changed due to the AC Stark effect between them. At the moment, a $\pi$ pulse calibrated before is unable to cause the perfect transition of $Q_{j+1}$ due to its variation of frequency and its population of excited state will be decreased~\cite{PRL_Xu2020,PRX_Sung2021} (see pulse sequence in the inset of Fig.~\ref{fig:spec_c2}\textbf{a}). In this way, one can obtain the spectrum of the local $Q_jC_jQ_{j+1}$ three-body system, which are actually the first three eigen-spectra (red lines in Fig.~\ref{fig:spec_c2}\textbf{a}) of the three-body Hamiltonian~\cite{PRApplied_Yan2018,PRL_Xu2020}
\begin{eqnarray}
	\hat{H}_{Q_jC_jQ_{j+1}} &=& \sum_{k}\omega_{k}\hat{b}^{\dagger}_{k}\hat{b}_{k} + \frac{\alpha_{k}}{2}\hat{b}^{\dagger}_{k}\hat{b}^{\dagger}_{k}\hat{b}_{k}\hat{b}_{k} \nonumber\\
	&~& +\sum_{k\neq l} g_{k,l}(\hat{b}^{\dagger}_{k}\hat{b}_l + \hat{b}_{k}\hat{b}^{\dagger}_l) \label{eq:H3}
\end{eqnarray}
with $k\in \{q_j,c_j,q_{j+1}\}$. In Eq.~\eqref{eq:H3}, the qubits frequencies $\omega_{q_j}$ and $\omega_{q_{j+1}}$ are fixed and all anharmonicities $\alpha = - E_{\mathrm{C}}$ can be obtained by two-photon excitation measurement. Furthermore, the coupling between qubit and coupler is much stronger than $g_{q_j,q_{j+1}}$ in our device. Thus, there leaves 5 parameters to be determined in this three-body Hamiltonian, namely 2 coupling strengths (i.e., $g_{q_j,c_j}$ and $g_{q_j,c_{j+1}}$) and 3 parameters of $\omega_{c_j}$ (i.e., $E_{\mathrm{JJ}}$, $A$, $\phi$ in Eq.~\eqref{eq:zpa2freq}).

\begin{figure}[ht!]
	\includegraphics[width=3.3in]{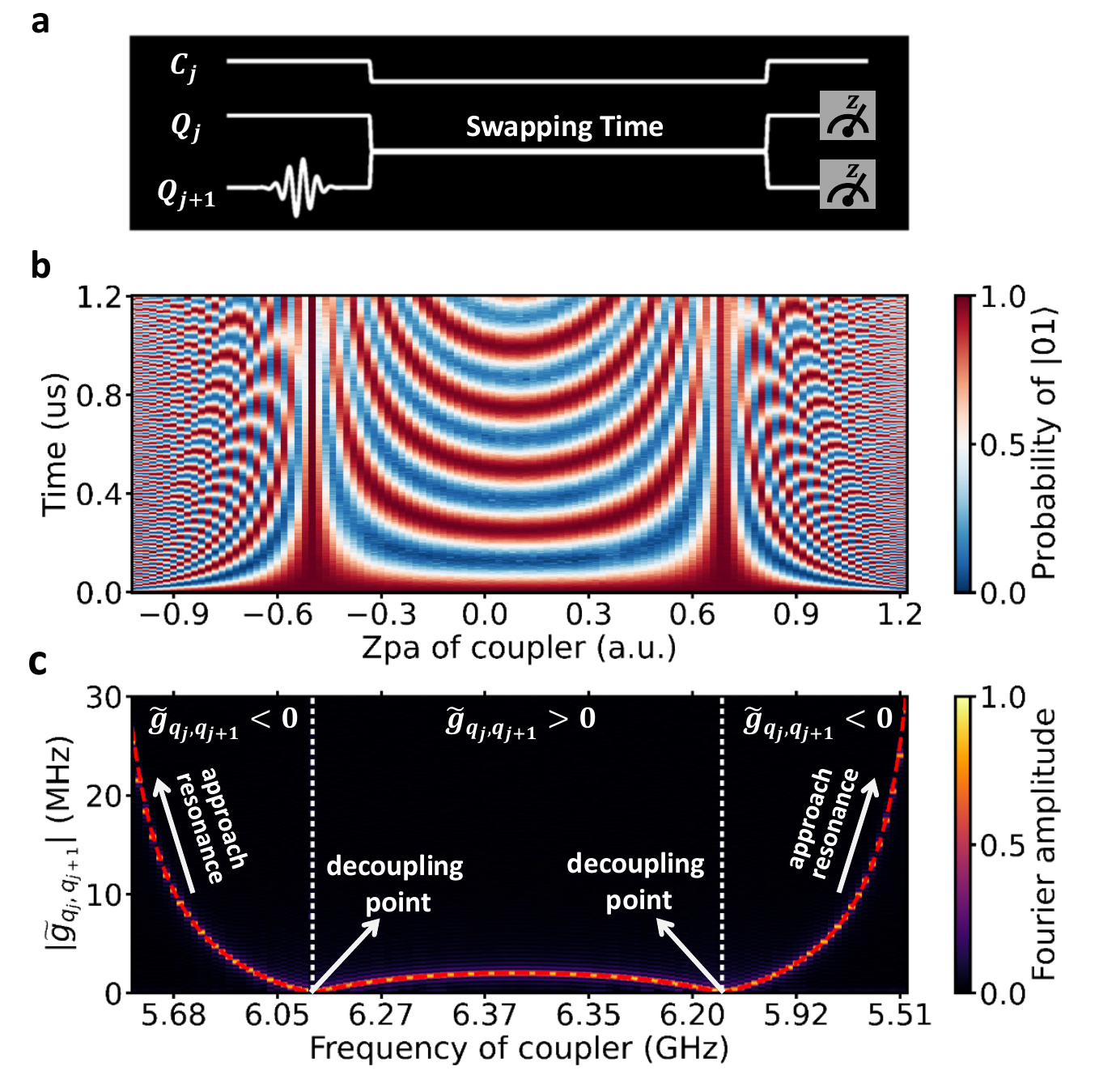}
	\caption{Experimental data of the effective coupling strength measurement. \textbf{a}, Pulse sequence for measurement of swapping between qubits while changing the Zpa of coupler. \textbf{b}, Measured joint probability $P_{01}$ of qubits vs Zpa of coupler (or corresponding frequency) and the swapping time. \textbf{c}, The Fourier
		transform of \textbf{b}, where the heatmap represents the normalized Fourier amplitude. The relation between absolute effective coupling strength $|\widetilde{g}_{q_j,q_{j+1}}|$ and coupler Zpa (or corresponding frequency) is given by each peak of normalized Fourier amplitude. The red dash line is the fitting curve of $|\widetilde{g}_{q_j,q_{j+1}}|$ by using Eq.~\eqref{eq:zpa2g}, while white dot lines denote two decoupling points ($\widetilde{g}_{q_j,q_{j+1}}=0$). As coupler frequency decreases, $\widetilde{g}_{q_j,q_{j+1}}$ decreases from positive to zero. Once it passes the decoupling point, $\widetilde{g}_{q_j,q_{j+1}}$ becomes negative and its absolute value will increase rapidly, especially approaching the resonance point of qubits.  \label{fig:g_c2} }
\end{figure}

Note that the smallest gaps of the two anti-crossing spectral lines represent twice the coupling strengths, respectively. However, it is inaccuracy to estimate the coupling strengths only by scanning the three-body spectrum due to the broadening of spectral lines and some impure peaks~\cite{PRApplied_Li2020,CPB_Xu2021}. For more accurate measurement, we scan two extra anti-crossing spectrums of two-body systems $Q_jC_j$ and $Q_{j+1}C_j$, as shown in Fig.~\ref{fig:spec_c2}\textbf{b}. Truncated to two energy levels, the Hamiltonian of a qubit $Q_j$ coupled to a coupler $C_j$ can be expressed in the subspace basis $\{\ket{10},\ket{01}\}$ as
\begin{equation}
	\label{eq:H2qc}\hat{H}_{Q_{j}C_{j}} =
	{\left(\begin{array}{cc}
			\omega_{q_j} & g_{q_j,c_j} \\
			g_{q_j,c_j} & \omega_{c_j}
		\end{array}\right)},
\end{equation}
and its eigen-energy spectra are
\begin{equation}
	\label{eq:H2spec1}\omega_{Q_{j}C_{j}}^{\pm} = \frac{\omega_{q_j}+\omega_{c_j}}{2}\pm \sqrt{g^2_{q_j,c_j}+\frac{(\omega_{q_j}-\omega_{c_j})^2}{4}}.
\end{equation}
Similarly, the eigen-energy spectra of $Q_{j+1}C_j$ are
\begin{equation}
	\label{eq:H2spec2}\omega_{Q_{j+1}C_{j}}^{\pm} = \frac{\omega_{q_{j+1}}+\omega_{c_j}}{2}\pm \sqrt{g^2_{q_{j+1},c_j}+\frac{(\omega_{q_{j+1}}-\omega_{c_j})^2}{4}}.
\end{equation}
Combining the above two equations with the diagonalization result of Eq.~\eqref{eq:H3}, one can finally determine the coupling strengths between coupler and qubits (i.e., $g_{q_j,c_j}$ and $g_{q_{j+1},c_j}$) and the mapping between coupler frequency $\omega_{c_j}$ and its Zpa. Actually, this is a multi-objective optimization problem of simultaneously fitting 3 spectroscopy results via 5 parameters. We utilize the optimization function \verb+scipy.optimize.minimize+ in the Python module SciPy to solve this problem.

\subsection{\label{subsec:cali_auto4} Measurement of the effective coupling}

To measure the effective coupling strength $\widetilde{g}_{q_j,q{j+1}}$, we measure the joint probability as a function of qubit-qubit swapping time $t$ and the Zpa of coupler~\cite{CPB_Xu2021,PRApplied_Li2020}, as shown in Fig.~\ref{fig:g_c2}\textbf{b}. Similar to Eq.~\eqref{eq:H2qc}, the swapping Hamiltonian of $Q_jQ_{j+1}$ in the subspace effective basis $\big\{\ket{\widetilde{10}},\ket{\widetilde{01}}\big\}$ is
\begin{equation}
	\hat{H}_{Q_{j}Q_{j+1}} =
	{\left(\begin{array}{cc}
			\widetilde{\omega}_{q_j} & \widetilde{g}_{q_j,q_{j+1}} \\
			\widetilde{g}_{q_j,q_{j+1}} & \widetilde{\omega}_{q_{j+1}}
		\end{array}\right)}.
	\label{eq:H2qq}
\end{equation}
If $Q_jQ_{j+1}$ is initially in state $\ket{\widetilde{01}}$, the time-dependent joint probability $P_{01}(t)=\bra{\widetilde{01}}e^{-\mathrm{i}\hat{H}_{Q_{j}Q_{j+1}}t}\ket{\widetilde{01}}$ can be expressed as
\begin{equation}
	\label{eq:P01} P_{01}(t)=\frac{1}{2}\cos{\Big[\sqrt{4\widetilde{g}^2_{q_j,q_{j+1}}+(\widetilde{\omega}_{q_j}-\widetilde{\omega}_{q_{j+1}})^2}t\Big]}+\frac{1}{2},
\end{equation}
which is reduced to
\begin{equation}
	\label{eq:P01reduce} P_{01}(t)=\frac{1}{2}\cos{(2\widetilde{g}_{q_j,q_{j+1}}t)}+\frac{1}{2}
\end{equation}
when the two qubits are resonant, namely $\widetilde{\omega}_{q_j}=\widetilde{\omega}_{q_{j+1}}$. Thus, the effective coupling strength can be calculated as half the Fourier frequency of probability $P_{01}(t)$. It needs to be emphasized that decoherence may cause the damping amplitude of swapping probability but does not affect the Fourier frequency.

For each Zpa of coupler (related to its frequency), we calculate $\widetilde{g}_{q_j,q_{j+1}}$ via measuring $P_{01}(t)$ and performing Fourier transform (as shown in Fig.~\ref{fig:g_c2}\textbf{c}). Subsequently, one can utilize Eq.~\eqref{eq:effective_g} to draw the mapping between the effective coupling strength and coupler Zpa:
\begin{equation}
	\label{eq:zpa2g}\widetilde{g}_{q_j,q_{j+1}}(\mathrm{Zpa})= g_{q_j,q_{j+1}} + \frac{g_{q_j,c_{j}}g_{q_{j+1},c_{j}}}{\omega_{c_j}(\mathrm{Zpa})-\omega},
\end{equation}
where $\omega=\omega_{q_j}=\omega_{q_{j+1}}$ is the resonant frequency of qubits, the direct coupling $g_{q_j,q_{j+1}}$ is the fitted value and the coupler frequency $\omega_{c_j}(\mathrm{Zpa})$ obeys Eq.~\eqref{eq:zpa2freq}. Hence, if the Zpa of coupler is given, the effective coupling strength can be computed via Eq.~\eqref{eq:zpa2g}; or given a target coupling, one can estimate the Zpa of coupler by
\begin{widetext}
	\begin{equation}
		\label{eq:g2zpa} \mathrm{Zpa}(\widetilde{g}_{q_j,q_{j+1}})\approx \frac{1}{A}\arccos\left[\pm\frac{1}{8E_{\mathrm{JJ}}E_{\mathrm{C}}}\sqrt{\left(\omega+E_{C}+\frac{g_{q_j,c_{j}}g_{q_{j+1},c_{j}}}{\widetilde{g}_{q_j,q_{j+1}}-g_{q_j,q_{j+1}}}\right)^2-\Big(8E_{\mathrm{JJ}}E_{\mathrm{C}}\delta\Big)^2}\right]-\frac{\phi}{A},
	\end{equation}
\end{widetext}
where $E_{\mathrm{JJ}}$ and $E_{\mathrm{C}}$ are the Josephson energy and the charging energy of coupler, respectively. Eq.~\eqref{eq:g2zpa} is a crucial foundation for engineering arbitrary coupling distribution in a superconducting circuit with tunable couplers.

\begin{figure}[ht]
	\includegraphics[width=3.0in]{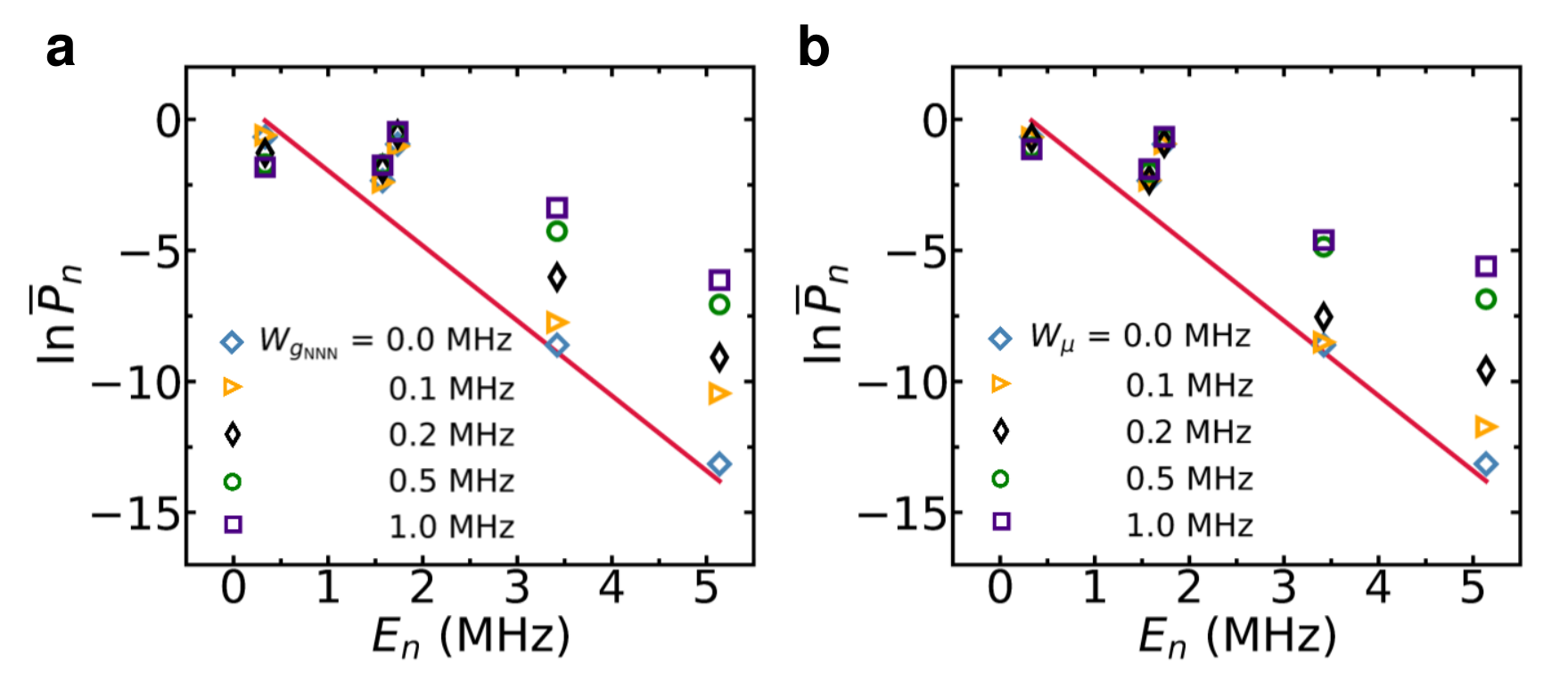}
	\caption{The effects of two typical disorders on Hawking radiation. \textbf{a}, The logarithm of average radiation probability vs. the positive energy of particle with different disorder strengths of $g_{\mathrm{NNN}}$. \textbf{b}, The logarithm of average radiation probability vs. the positive energy of particle with different disorder strengths of $\mu$. Here, the red solid line represents the theoretical result.  \label{fig:discuss_disorder} }
\end{figure}

\section{\label{sec:discussion} Additional discussion}
For further discussion, we perform additional numerical simulations to compare and supplement with our results in this paper. In the following, the effects of disorders, different coupling distribution, finite size, and continuum limit are investigated.

\begin{figure}[hb]
	\includegraphics[width=2.6in]{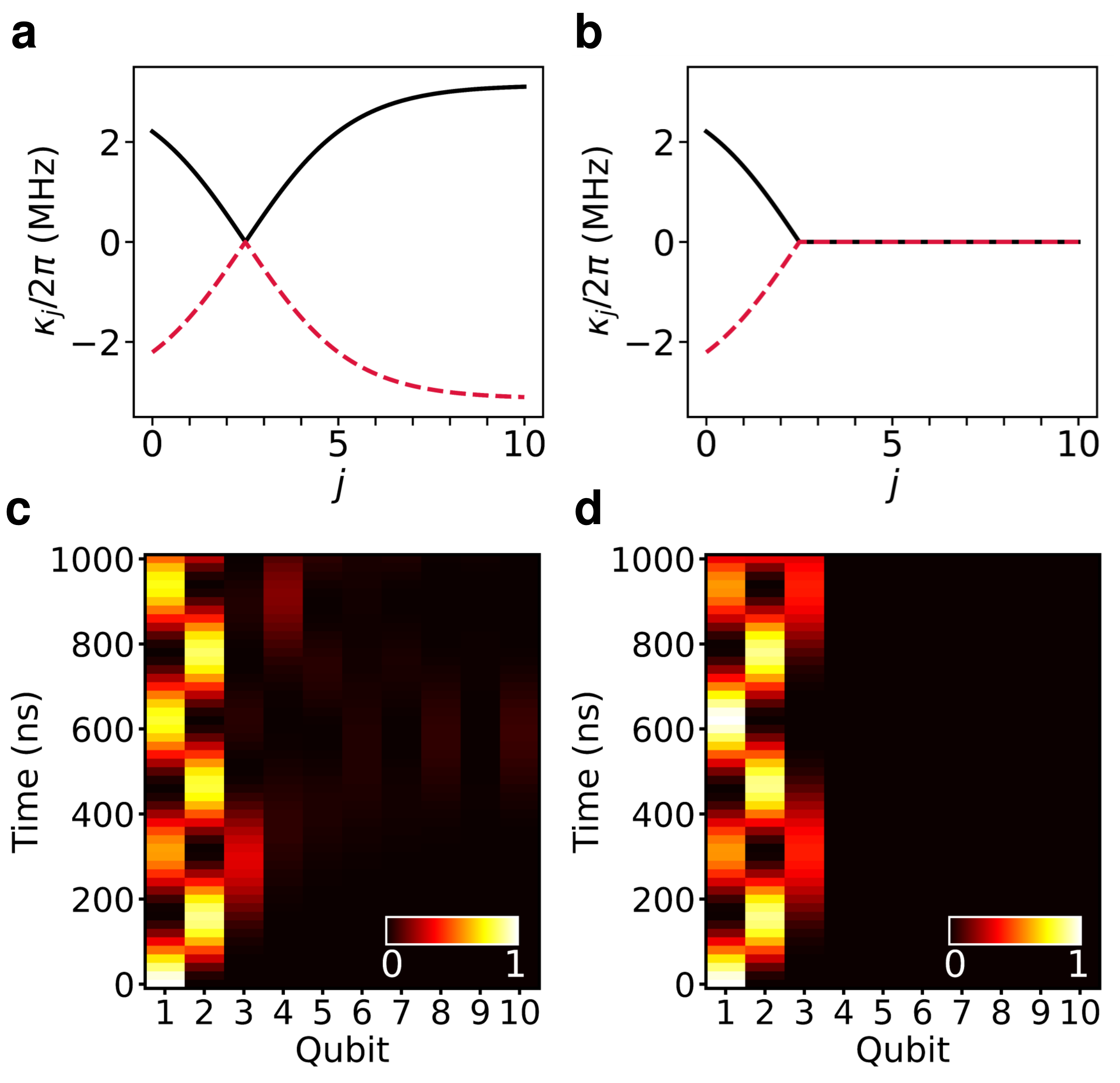}
	\caption{Quantum walks of different coupling distribution. \textbf{a}, The couplings keep the same amplitude inside and outside of the black hole, where the black solid line represents all couplings keeping positive and the red dash line denotes all couplings remaining negative. These two cases are equivalent and the same results are shown in \textbf{c}. \textbf{b}, Non-zero coupling inside the black holes (black solid line and red dash line imply positive coupling and negative coupling respectively) and zero coupling between all sites outside the black hole, where the corresponding results of quantum walks are shown in \textbf{d}. Here we take the initial state $\ket{1000000000}$ as an example. \label{fig:qw_with_different_couplings} }
\end{figure}

\subsection{\label{subsec:disorders} The effects of disorders}

\begin{figure*}[ht]
	\includegraphics[width=6.0in]{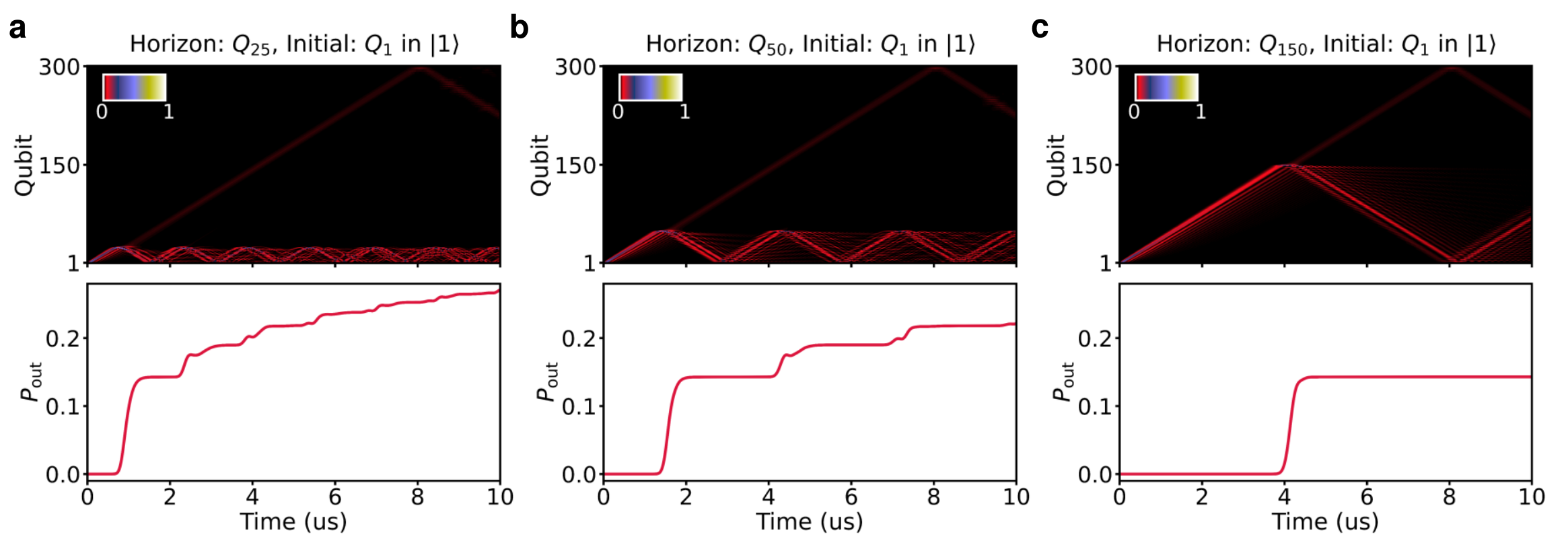}
	\caption{Simulation of a 300-qubit chain with horizons at different locations. Here the coupling $\kappa_j$ takes the form of Eq.~\eqref{eq:kappa_j}, where $d=0.35$ and $\beta/(2\pi)=4.39$ MHz. From \textbf{a} to \textbf{c}, the corresponding horizons are located at $Q_{25}$, $Q_{50}$ and $Q_{150}$, respectively. $P_{\mathrm{out}}$ is defined as the sum of probabilities of all the qubits outside the horizon. \label{fig:finite_size_effects}}
\end{figure*}

\begin{figure*}
	\includegraphics[width=6.0in]{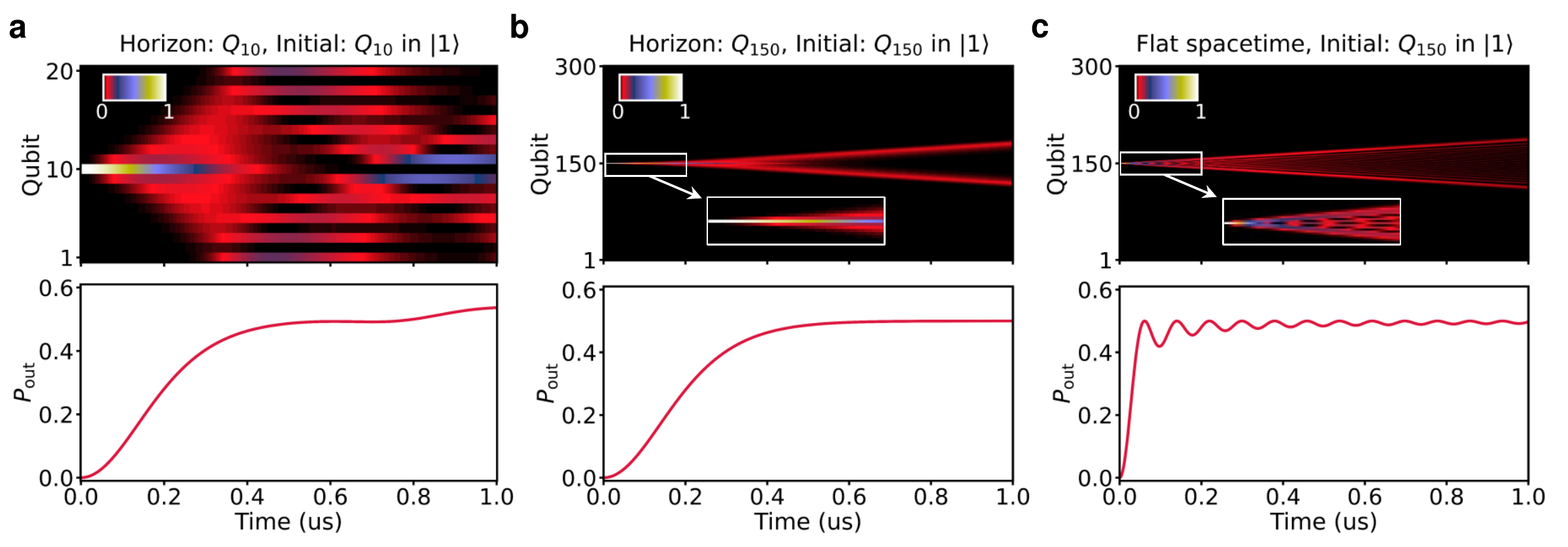}
	\caption{The finite-size effects on the horizon. The particle is initialized at the horizon. \textbf{a}, Simulation of a 20-qubit chain with the horizon located at $Q_{10}$. \textbf{b}, Simulation of a 300-qubit chain with the horizon located at $Q_{150}$. Here the coupling $\kappa_j$ takes the form of Eq.~\eqref{eq:kappa_j}, where $d=0.35$ and $\beta/(2\pi)=4.39$ MHz. \textbf{c}, Simulation of a 300-qubit chain in flat spacetime with $\kappa_j=3.14$ MHz. $P_{\mathrm{out}}$ is defined as the sum of probabilities of all the qubits outside the horizon. \label{fig:horizon_finite_size_effects} }
\end{figure*}

In reality, qubits are doomed to be disturbed by various disorders, leading to the nuance between experimental conditions and theoretical assumptions. For a 1D-array of qubits, one can consider two disorders about next-nearest-neighbor (NNN) coupling $g_{\mathrm{NNN}}$ and on-site potential $\mu$ with the corresponding disorder strengths $W_{g_{\mathrm{NNN}}}$ and $W_{\mu}$. Specifically, the Hamiltonian of disorders is
\begin{equation}
	\hat{H}_{\mathrm{dis}} = \sum_{j=1}^{8}g_{\mathrm{NNN}}^j(\hat{\sigma}^{+}_j\hat{\sigma}^{-}_{j+2}+\hat{\sigma}^{-}_j\hat{\sigma}^{+}_{j+2})
	-\sum_{j=1}^{10}\mu_j\hat{\sigma}^{+}_j\hat{\sigma}^{-}_j,
\end{equation}
where the site-dependent $g_{\mathrm{NNN}}^j$ and $\mu_j$ are in $[-W_{g_{\mathrm{NNN}}}, W_{g_{\mathrm{NNN}}}]$ and $[-W_{\mu}, W_{\mu}]$, and their distribution are assumed to be uniform. To probe the effects of disorders, we numerically model the dynamics of Hawking radiation by considering $\hat{H}_{\mathrm{dis}}$. In Fig.~\ref{fig:discuss_disorder}\textbf{a} and \textbf{b}, one can note that both disorders diverge the probability spectrum and theoretical results of Hawking radiation, especially in the condition of strong disorder. However, we measure the NNN coupling of $g^j_{\mathrm{NNN}}\approx0.1$ MHz and the frequencies difference of $|\mu_j-\omega_{\mathrm{ref}}|<0.2$ MHz with reference frequency $\omega_{\mathrm{ref}}/(2\pi)\approx5.1$ GHz. According to Fig.~\ref{fig:discuss_disorder}, such a small degree of disorders has little impact on the results of Hawking radiation in the experiment. In fact, we measure the initial density matrix of 7 qubits outside the horizon. The fidelity between the imperfect initial state in the experiment and the ideal initial state is $99.2\%$ (see Fig.~3a in the main text), which may be caused by the XY crosstalk, thermal excitation, leakage, etc. We substitute such an experimental state for the ideal initial state in the numerical simulation of Hawking radiation, then the results of the numerical simulation agree with the experimental results better (see Fig.~3d in the main text).

\subsection{\label{subsec:different_couplings} Different coupling distribution}

Admittedly, our model does not mandate flipping the sign of coupling $\kappa_j$ near the horizon. In the main text, we request that the coupling goes monotonically from negative to positive (or vice versa) from the left of the horizon to its right side. This is based on the realistic consideration of the smoothness of $f(x)$. In fact, if instead of flipping the sign of couplings they were all kept positive (or negative) inside and outside of the black hole (Fig.~\ref{fig:qw_with_different_couplings}\textbf{a}), all of the results would be similar, as shown in Fig.~\ref{fig:qw_with_different_couplings}\textbf{c}. For the case with non-zero coupling inside the black hole but zero coupling between all sites outside the black hole (Fig.~\ref{fig:qw_with_different_couplings}\textbf{b}), one can find the results inside the black hole are also similar to the results in case of flipping the sign of couplings, but it is quite different for the results outside the black hole. Due to the zero coupling between all sites outside, no particles can travel in the exterior (Fig.~\ref{fig:qw_with_different_couplings}\textbf{d}) and thus no radiation can be detected by the observer outside.

\subsection{\label{subsec:finite_size_effects} The finite-size effects}
Here we perform the numerical simulation of a 300-qubit chain to show the finite-size effects more clearly when we initialize the system by preparing a particle in the black hole. When the particle arrives at the horizon, it is going to be reflected back into the black hole in all probability but has a little chance to appear in the outside. The horizon is similar to a `membrane' with certain transmittance, see Fig.~\ref{fig:finite_size_effects}. The probability of finding the particle outside $P_{\mathrm{out}}$ shows a general upward trend due to the Hawking radiation. However, the particle will be reflected when it arrives at the boundary ($Q_1$ or $Q_{300}$) due to the finite-size effects. When the particle reflected by the boundary of the black hole reaches the horizon again, it has a certain probability to escape into the outside and $P_{\mathrm{out}}$ thus increases again (see Fig.~\ref{fig:finite_size_effects}\textbf{a} and Fig.~\ref{fig:finite_size_effects}\textbf{b}). Conceivably, if there are no boundaries, $P_{\mathrm{out}}$ will increase to a certain value and eventually the particle reaches a steady state of radiation.

In addition, the finite size can also affect the horizon. In the continuous curved spacetime, the particle initialized at the horizon is bound to the horizon forever due to the zero couplings on both sides of the horizon. However, in the finite-size lattice, the coupling strengths on both sides of the horizon are not strictly zero even though they are very small. As shown in Fig.~\ref{fig:horizon_finite_size_effects}\textbf{a} and Fig.~\ref{fig:horizon_finite_size_effects}\textbf{b} (also Fig. 2\textbf{b} in the main text), although the particle seems to be localized at the horizon for a very short time, it is doomed to escape from the constraints. When the particle is far from the horizon, its behavior is similar to that in flat spacetime (see Fig.~\ref{fig:horizon_finite_size_effects}\textbf{b} and Fig.~\ref{fig:horizon_finite_size_effects}\textbf{c}).

\begin{table*}[ht!]
	\begin{ruledtabular}
		\begin{tabular}{ccccc}
			Reference & Experimental system & Surface gravity $g_{\mathrm{h}}/(2\pi)$ & Hawking temperature $T_{\mathrm{H}}$ & Ratio of mass $M/M_{\mathrm{s}}$\\[.05cm] \hline\\[-.2cm]
			\cite{PRL_Weinfurtner2011} & Shallow water wave & $\sim10^{-1}$~Hz & $\sim10^{-12}$~K & $\sim10^{4}$\\[.09cm]
			\cite{PRL_Steinhauer2010,PRL_Isoard2020,Nature_Steinhauer2019,NP_Steinhauer2021} & Bose-Einstein condensates & $\sim10^{1}$~Hz & $\sim10^{-10}$~K & $\sim10^{2}$\\[.09cm]
			\cite{NP_Steinhauer2016} & Bose-Einstein condensates & $\sim10^{2}$~Hz & $\sim10^{-9}$~K & $\sim10^{1}$\\[.09cm]
			This work & Superconducting qubits & $\sim10^{5}$~Hz & $\sim10^{-5}$~K & $\sim10^{-3}$\\[.09cm]
		\end{tabular}
	\end{ruledtabular}
	\caption{Comparison of analogue Hawking radiation with other experiments. Considering a Schwarzschild black hole in four-dimensional spacetime with the same Hawking temperature, the mass of analogue black hole can be calculated by $M/M_{\mathrm{s}}=6.4\times10^{-8} \mathrm{K}/T_{\mathrm{H}}$, where $M_{\mathrm{s}}\approx2\times10^{30}$ kg is the solar mass.}\label{tab:hawking_compare}
\end{table*}

\subsection{Continuum limit}
To confirm that our model realizes an analogy of black hole, we now consider the wave prorogating in classical and continuum limit. To simulate the wave prorogating in the classical limit, we consider a state $|\psi\rangle$ such that
\begin{equation}\label{classicaleq1}
	\langle\psi|(\hat{c}_n-\psi_n)^\dagger(\hat{c}_n-\psi_n)|\psi\rangle\ll|\psi_n|^2
\end{equation}
where $\psi_n= \langle\psi|\hat{c}_n |\psi\rangle$. Then Eq.~\eqref{eq:Heisenberg} becomes
\begin{equation}\label{fulleoms2}
	\begin{split}
		-i\frac{\td}{\td t}\psi_n=\kappa_n \psi_{n-1}+\kappa_{n+1} \psi_{n+1}+\mu\psi_n
	\end{split}
\end{equation}
We choose the function $f(x)$ to be
$$f(x)=\tanh(\alpha x)\,.$$
with a constant $\alpha$ and set $\mu=0$ for simplicity. This choice of metric describes an asymptotically flat spacetime since $f(x)\rightarrow1$ as $x\rightarrow\pm\infty$.

In the asymptotically flat region $x\rightarrow\infty$, due to the translation symmetry, we can take a planar wave ansatz
\begin{equation}\label{twoks1}
	\psi_{n,k}=c_0\exp[-i(\omega_k t-2x_n(k -\pi/4d)]\,,
\end{equation}
with $x_n=nd$. Note that a coefficient 2 of $x_n$ appears in Eq.~\eqref{twoks1}, which is different from the usual mode expansion $e^{-i(\omega_k t-x k+\varphi_0)}$. This is because here we use Eddington-Finkelstein coordinate and need Eq.~\eqref{twoks1} to match the equation~\eqref{outmodes}. Eq.~\eqref{twoks1} leads to following dispersion relationship
\begin{equation}\label{despereq1a}
	\omega_k=\frac1{2d}\cos(2kd-\pi/2)=\frac1{2d}\sin 2kd\,.\\
\end{equation}
In the limit $d\ll1$ but keep $k$ finite, this recovers the dispersion relationship of massless particle and describes a massless particle travailing at the speed of light in the asymptotic flat regions.
\begin{figure}[hb]
	\centering
	\includegraphics[width=0.4\textwidth]{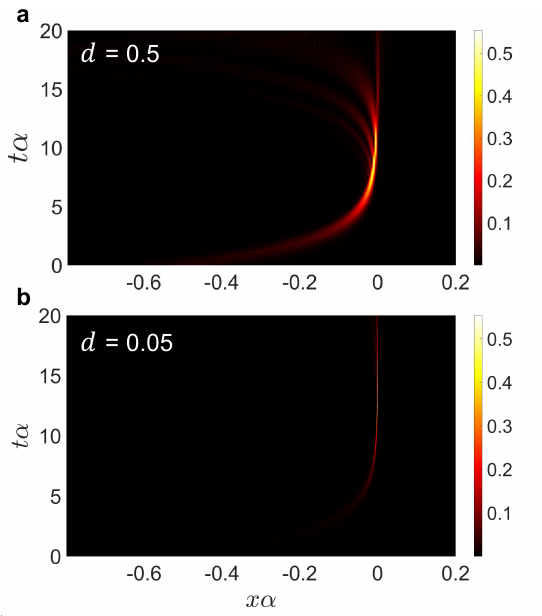}
	\caption{Numerical results of outgoing modes with different widths of wave packets. Here we set $\Delta\alpha=0.2$,  $k=0.01$, $\alpha=0.01$, $d=0.5$ (\textbf{a}) and $d=0.05$ (\textbf{b}). \label{traject1}}
\end{figure}
In the region of curved spacetime, we also require that the curvature radius of spacetime is much larger than the scale of wave packet so that the wave will propagate along trajectory of light. This requirement is easy found by using the Maxwell field in four-dimensional curved spacetime,
\begin{equation}\label{maxwellfield}
	\square A_\mu- R_{\mu\nu}A^\nu=0\,.
\end{equation}
Here $A_\mu$ is the gauge field, $\square$ is the d'Alembert operator in curved spacetime and $R_{\mu\nu}$ is the Ricci curvature tensor. In general, such an equation does not admit a plane wave solution due to the existence of the curvature term. In situations where the spacetime scale of variation of the electromagnetic field is much smaller than that of the curvature, the solutions of Maxwell's equations have a wave oscillating with nearly constant amplitude. In this case the \textit{geometric optics approximation} can be used and eletromagnetic wave will travel along null geodesic approximately.

\begin{figure*}[ht]
	\includegraphics[width=7.1in]{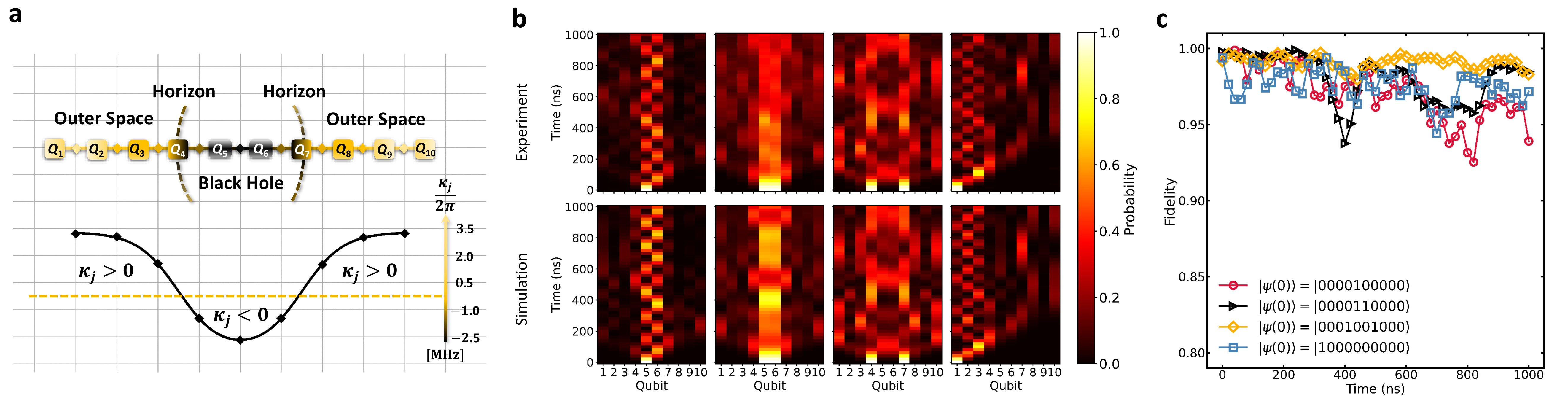}
	\caption{Quantum walks in a 1D array of 10 superconducting qubits with black hole at the center. \textbf{a}, Schematic representation of the black hole at the center and the corresponding coupling distribution. \textbf{b}, Quantum walks in such a curved spacetime. The heatmap denotes the probabilities of excited-state for $Q_i$ in time. The horizontal axis is indexed as qubit number $i$, the vertical axis is time. Here we show both the numerical simulation and experiment data. \textbf{c}, Fidelity of the experimental data
		compared to ideal numerical simulation of quantum walks. \label{fig:walk_extended}}
\end{figure*}

We can read the effective wavelength from Eq.~\eqref{twoks1} $\lambda=|2\pi/(2k-\pi/2d)|\approx d/4$. When the parameter $\alpha$ satisfies
\begin{equation}\label{alphalambda}
	\alpha\lambda\ll1\Leftrightarrow d\alpha\ll1/4\,,
\end{equation}
The variation of spacetime caused by the curvature in curved region of spacetime will be much lower than the variation caused by phase factors of wave. Then the geometric optics approximation can be used. In this case, the wave will travel as light rays.

In numerical simulation, we take the initial wave packet to be Gaussian with the following form
\begin{equation}\label{initialwave1}
	\tilde{\psi}_n(0)=\frac1{Z}e^{-\frac{(x_n-\tilde{x}_0)^2}{\Delta^2}}\psi_{n,k}\,.
\end{equation}

Here $Z$ is the normalized constant, $\Delta$ describes the width of wave packet, $\tilde{x}_0$ is the center of the initial wave packet and $\psi_{n,k}$ is defined by Eq.~\eqref{twoks1}. Since the the wavepacket is modulated by  Gaussian distribution, except Eq.~\eqref{alphalambda} we also require $\Delta \alpha\ll1$ so that geometric optical approximation is valid. On the other hand, to have a well-defined momentum, the width of wavepacket should much be larger than the wavelength, i.e. $\Delta\gg d$.

To summary, in order to check our model indeed forms an analogues of massless scalar particle in a black hole spacetime, we require the parameters to satisfy
\begin{equation}\label{contlimit}
	kd\ll1,~~d\alpha\ll1,~~\Delta\alpha\ll1,~~\text{and}~d/\Delta\ll1
\end{equation}

In the Fig.~\ref{traject1}, we set parameters  $\Delta\alpha=0.2$,  $k=0.01$, $\alpha=0.01$, $d=0.5$ (\textbf{a}) and $d=0.05$ (\textbf{b}) so that the geometric optical approximation is valid. From  Fig.~\ref{traject1} we see that the outgoing mode inside the black hole will be increasingly closer to the horizon but not pass through the horizon. For finite $d$, the wave can only be ``trapped'' into the horizon for a finite time due to finite discretization. In the limit $d\rightarrow0$, the ``light'' will be trapped into the horizon forever. This is just what we expect for the black hole horizon.

\subsection{Comparison of measured Hawking temperature with other experiments}

Here we compare the measured Hawking temperature with other previous experiments, as shown in Table~\ref{tab:hawking_compare}. As we mentioned in the main text, the significant difference in magnitude of the results is attributed to the scales of the setup in different experimental systems. In our superconducting qubits system, we set a stronger surface gravity, contributing to a higher Hawking temperature.

\section{\label{sec:extended} Extended Data}

In addition, we design an analogue black hole at the center of our 1D array superconducting qubits, see Fig.~\ref{fig:walk_extended}\textbf{a}. Here the coupling $\kappa_{j}$ is designed as $\beta\left(\tanh((j-7/2)\eta d)-\tanh((j-13/2)\eta d)+1\right)/(4\eta d)$ with $\eta d=1$ and $\beta/(2\pi)\approx13.2$ MHz. The qubits $Q_5$ and $Q_6$ are in the interior of the black hole, $Q_4$ and $Q_7$ are at the horizon, and other qubits are in outer space. In Fig.~\ref{fig:walk_extended}\textbf{b}, we prepare four initial states to show quantum walks in such a curved spacetime, i.e., $\ket{0000100000}$, $\ket{0000110000}$, $\ket{0001001000}$ and $\ket{1000000000}$, respectively. The results of quantum walks of the particle at the center of the chain are similar to the results of the particle at the leftmost boundary in the main text. However, when the black hole is at the center, the finite size effect in the experimental 10-qubit 1D array becomes obvious, for which the tunneling out particle soon reaches the boundary and is reflected back.





\providecommand{\noopsort}[1]{}\providecommand{\singleletter}[1]{#1}%